\documentclass[a4paper,aps,preprintnumbers,showpacs,twocolumn,superscriptaddress,nofootinbib]{revtex4-1}

\usepackage{amsmath}
\usepackage{amssymb}
\usepackage{graphicx}
\usepackage{epsfig}
\usepackage{color}
\usepackage{url}
\usepackage{times}
\usepackage{bm}
\usepackage{mathrsfs}
\usepackage[utf8]{inputenc}
\usepackage{hyperref}

\newcommand{\beq}{\begin{equation}}
\newcommand{\eeq}{\end{equation}}
\newcommand{\bea}{\begin{eqnarray}}
\newcommand{\eea}{\end{eqnarray}}
\newcommand{\bit}{\begin{itemize}}
\newcommand{\eit}{\end{itemize}}
\newcommand{\ben}{\begin{enumerate}}
\newcommand{\een}{\end{enumerate}}
\newcommand{\nn}{\nonumber}

\newcommand{\coord}{\tau, \sigma, \theta,\varphi}

\font\tenscr=rsfs10 scaled1100
\font\sevenscr=rsfs7 
\font\fivescr=rsfs5 
\skewchar\tenscr='177
\skewchar\sevenscr='177
\skewchar\fivescr='177
\newfam\scrfam
\textfont\scrfam=\tenscr
\scriptfont\scrfam=\sevenscr
\scriptscriptfont\scrfam=\fivescr

\def\scrI{{\fam\scrfam I}}
\def\scri{{\fam\scrfam I}^+}

\newcommand{\TPI}{\affiliation{Theoretisch-Physikalisches Institut,
    Friedrich-Schiller-Universität Jena,\\ Max-Wien-Platz 1,
          D-07743 Jena, Germany}}

\begin{document}
\title{Spectral decomposition of black-hole perturbations on hyperboloidal slices}
\author{Marcus Ansorg} 
\author{Rodrigo Panosso Macedo} 
\TPI
\date{\today}

\begin{abstract}
In this paper we present a spectral decomposition of solutions to relativistic wave equations described on horizon-penetrating hyperboloidal slices within a given Schwarzschild-black-hole background. The wave equation in question is Laplace-transformed which leads to a spatial differential equation with a complex parameter. For initial data which are analytic with respect to a compactified spatial coordinate, this equation is treated with the help of the \texttt{Mathematica}-package in terms of a sophisticated Taylor series analysis. Thereby, all ingredients of the desired spectral decomposition arise explicitly to arbitrarily prescribed accuracy, including quasinormal modes, quasinormal mode amplitudes as well as the jump of the Laplace-transform along the branch cut. Finally, all contributions are put together to obtain via the inverse Laplace transformation the spectral decomposition in question. The paper explains extensively this procedure and includes detailed discussions of relevant aspects, such as the definition of quasinormal modes and the question regarding the contribution of infinity frequency modes to the early time response of the black hole.
 \end{abstract}

\maketitle

\section{Introduction}
Since the advent of general relativity, studies within the linear regime of Einsteins's equations have played a crucial role in understanding important aspects of both mathematical and physical sides of the theory. In the particular case of black-hole perturbation theory, the work by Regge and Wheeler~\cite{Regge57} usually marks the birth of the field. Their analysis of a special class of perturbations of the Schwarzschild spacetime was later generalised by Zerilli~\cite{Zerilli70,Zerilli70a}. Still in the context of the Schwarzschild spacetime, Bardeen and Press~\cite{Bardeen73a} used the Newman-Penrose formalism~\cite{Newman62a} to derive the equations describing the propagation of scalar, electromagnetic and gravitational wave perturbations within the aforementioned background. This approach is the same as the one that guided Teukolsky towards the derivation of his equation, which takes the Kerr solution as the background spacetime~\cite{Teukolsky73}. 

First observed by Vishveshwara~\cite{Vishveshwara70b}, the time evolution of the perturbing field shows, after an initial dynamics, an intermediate phase dominated by exponentially damped oscillations, the so-called {\em quasi-normal modes} (QNMs). The remarkable feature is that the oscillation and decay time scales depend solely on the black-hole parameters, which allows one to infer essential properties of the black hole in the gravitational wave signal~\cite{Abbott:2016gw} and to determine whether the object is black hole or something else~\cite{Cardoso:2016rao,Chirenti:2016hzd}. For the late time evolution, Price showed that the dynamics is characterised by a power law decay, also referred to as {\em tail decay}~\cite{Price72}.

A complete revision of the literature related to this field is a task that goes beyond the scope of this work. Instead of overwhelming the reader with all the development that followed the seminal works mentioned previously, we would rather point out Chadrasekhar's book~\cite{Chandrasekhar83} that reviews the state-of-art during the 1980's and elucidates the connections between different formalisms. Also worth mentioning are references~\cite{Kokkotas99a} and~\cite{Nollert99}, the two reviews that appeared in the late 1990's. Finally, Berti, Cardoso and Starinets~\cite{Berti:2009kk} summarised the more recent development on black-hole perturbation theory. Apart from a very interesting and useful chronological ``roadmap" in terms of papers considered as milestones, the results presented in ~\cite{Berti:2009kk} range from astrophysical scenarios (with insights into the numerical simulations and the efforts to the detection of gravitational waves), passing by applications in gauge-gravity duality theories, up to some recent developments on quantum black holes. Among the many important milestones listed in~\cite{Berti:2009kk}, we mention here Leaver's work~\cite{Leaver85,Leaver86b,Leaver86c}, which provides one of the most accurate methods to compute the QNMs.  

Note however, that in the great majority of these works, QNMs are defined in a phenomenological way motivated by a comparison with the analysis of normal modes. (see, for an example~\cite{CardosoPhD}). 

A formal definition of the QNMs is presented in the second chapter of reference~\cite{Kokkotas99a}. It starts out with the description of {\em normal} modes as the real eigenvalues $\omega_n$ of an appropriate differential operator. This operator acts on a corresponding Hilbert-space whose measure characterizing the inner product, arises through the requirement that the operator be {\em self-adjoint}. The eigenvalues can then be used along with the associated eigenvectors $\phi_n(x^k)$ to build up the solutions $V$ of a specific nondissipative wave equation,
\footnote{In this paper, time in the wave equations is denoted by $\tau$, whereas $x^k$ ($k=1\ldots3$) stands for the collection of relevant spatial coordinates.} 
\beq\label{eq:Spectral_Decomp_1} V(\tau ,x^k)=\sum_{n=0}^\infty \eta_ne ^{ i \omega_n \tau}\phi_n(x^k).\eeq
Of course, this only works if the self-adjoint operator in question has a pure point spectrum $\sigma_p=\{\omega_n\}_{n=0}^\infty$. More generally, if in addition a continuous spectrum $\sigma_c$ is present, then the superposition needs to include {\em improper} eigenvalues and eigenvectors:
\beq
\label{eq:Spectral_Decomp_2}V(\tau,x^k)=\sum_{n=0}^\infty \eta_ne ^{ i \omega_n\tau}\phi_n(x^k)+ \int\limits_{\omega\in\sigma_c} \eta(\omega)e ^{ i \omega\tau}\phi(\omega; x^k)d\omega.
\eeq
The computation of the corresponding amplitudes, $\eta_n$ and $\eta(\omega)|_{\omega\in\sigma_c}$, amounts to projecting the initial data onto the complete orthonormal system of (proper and improper) eigenfunctions in terms of the inner product. 

As expressed in the review \cite{Kokkotas99a}, it would be clearly desirable to take this formulation to the realm of linear black-hole perturbations, that is, to write the solutions of initial value problems associated with  linear {\em dissipative} wave equations in a given black-hole background-metric as a similar superposition, with the {\em quasi}-normal modes being defined as the eigenvalues of an appropriate operator. However, the very definition of the QNMs, as  pursued in \cite{Kokkotas99a}, follows a different route. Through a Laplace transform applied to the wave equation in question, a spatial differential equation arises with an inhomogeneity formed by the initial data. The QNMs are then defined as the zeros of the corresponding Wronskian determinant, formed by specifically normalized solutions of the associated {\em homogeneous} equation. This definition is a perfectly working characterization of QNMs in many situations. However, in the context of linear perturbations in asymptotically flat black-hole spacetimes, it appears that a corresponding direct computation poses substantial technical difficulties. In this paper (Sec.~\ref{Sec:QNMs}) we provide a detailed discussion of this matter.

Until now there is no strict mathematical derivation of a spectral decomposition formula \eqref{eq:Spectral_Decomp_2} for linear waves around asymptotically flat black holes. Nevertheless, in special cases it was shown \cite{Bachelot1993} that the late time behaviour of the solutions
can be approximated in finite parts of the space by a {\em finite} sum of the form  (\ref{eq:Spectral_Decomp_1}). In~\cite{Nollert99a} it was demonstrated that for a wave equation with so-called spiked truncated dipole potential the decomposition (\ref{eq:Spectral_Decomp_1}) can be constructed; particularly the amplitudes $\eta_n$ were determined explicitly. Moreover, in several papers \cite{Leaver86c,Sun88,Andersson95b,Berti:2006wq} so-called {\em quasi-normal excitation factors} are discussed which are determined through the behaviour of $\phi_n$ at the two boundaries describing spatial infinity and the event horizon. They serve for the determination of {\em quasi-normal excitation coefficients} which are equivalent to the expressions $\eta_n\phi_n$. Rigorous mathematical results including integral representations have been obtained in the case of Cauchy problems for the massive Dirac equation as well as for the Teukolsky-equation in the nonextreme Kerr-Newman geometry outside the event horizon, see e.g.~\cite{Finster:2000jz,Finster:2001vn,Finster:2003fu,Finster:2005dg,Finster:2006bi}.

In this paper we demonstrate that a superposition of the form 
\beq
\label{eq:VSol_spectral}
V(\tau, x^k) = \sum_{n=0}^{\infty} \eta_{n} e^{s_{n}\tau}\phi_n( x^k) + \int\limits_{-\infty}^{0} \eta(s)e^{s\tau}\phi( x^k;s) ds    
\eeq
can be found for solutions of initial value problems of linear wave equations in the Schwarzschild spacetime, provided that the initial data are analytical in terms of a compactified coordinate in an appropriate hyperboloidal slice to be specified in Sec.~\ref{sec:HyperCoord}. The amplitudes $\eta_n$ and $\eta(s)$ are fixed solely by the initial data, whereas the {\em quasi-}normal modes $s_n$ as well as the functions $\phi_n(x^k)$ and $\phi( x^k;s)$ are characteristics of the particular wave equation being studied and hence independent of the initial data. We stress that (\ref{eq:VSol_spectral}) is meant to provide in a strict sense the entire solution for all coordinate times $\tau>\nu$ where $\nu$ is a mutual growth rate of the excitation coefficients $\eta_n\phi_n(x^k)$ and $\eta(s)\phi(x^k;s)$ to be defined in the sequel, (see secs.~\ref{Sec:QNM_amplitudes} and \ref{Sec:branch_cut}).

Now, in order to describe the decay of a dissipative wave field, the QNMs $s_n$ need to be complex-valued with negative real part. Hence, an associated self-adjoint operator with spectrum $\sigma=\{s_n\}_{n=0}^\infty\cup(-\infty,0)$ cannot be identified. Consequently, it is not simply possible to establish $\eta_n$ and $\eta(s)$ by some orthogonal projection of the initial data onto the functions $\phi_n(x^k)$ and $\phi( x^k;s)$. Nevertheless, in this paper we develop highly accurate numerical means, based on a detailed Taylor series analysis which is established within the \texttt{Mathematica}-environment, through which all ingredients of (\ref{eq:VSol_spectral}) are determined:
\ben
\item The QNMs $s_n$ are computed through an efficient procedure which can be considered as an extension of Leaver's method of continued fractions~\cite{Leaver85,Leaver86b}.
\item The functions $\phi_n(x^k)$ and $\phi( x^k;s)$ are constructed from the wave equation under consideration (without the need to consider specific initial data).
\item The amplitudes $\eta_n$ and $\eta(s)$, being characteristics of the initial data, are obtained through an analysis that incorporates the initial data. 
\een
In addition, we provide strong evidence that the solutions $V(\tau, x^k)$ to initial value problems of wave equations in the Schwarzschild spacetime  for analytical initial data as described above are indeed entirely given in terms of (\ref{eq:VSol_spectral}) for all coordinate times $\tau$ that exceed the growth rate $\nu$ of the excitation coefficients. 

\begin{figure}[]
\begin{center}
\includegraphics[width=8.0cm]{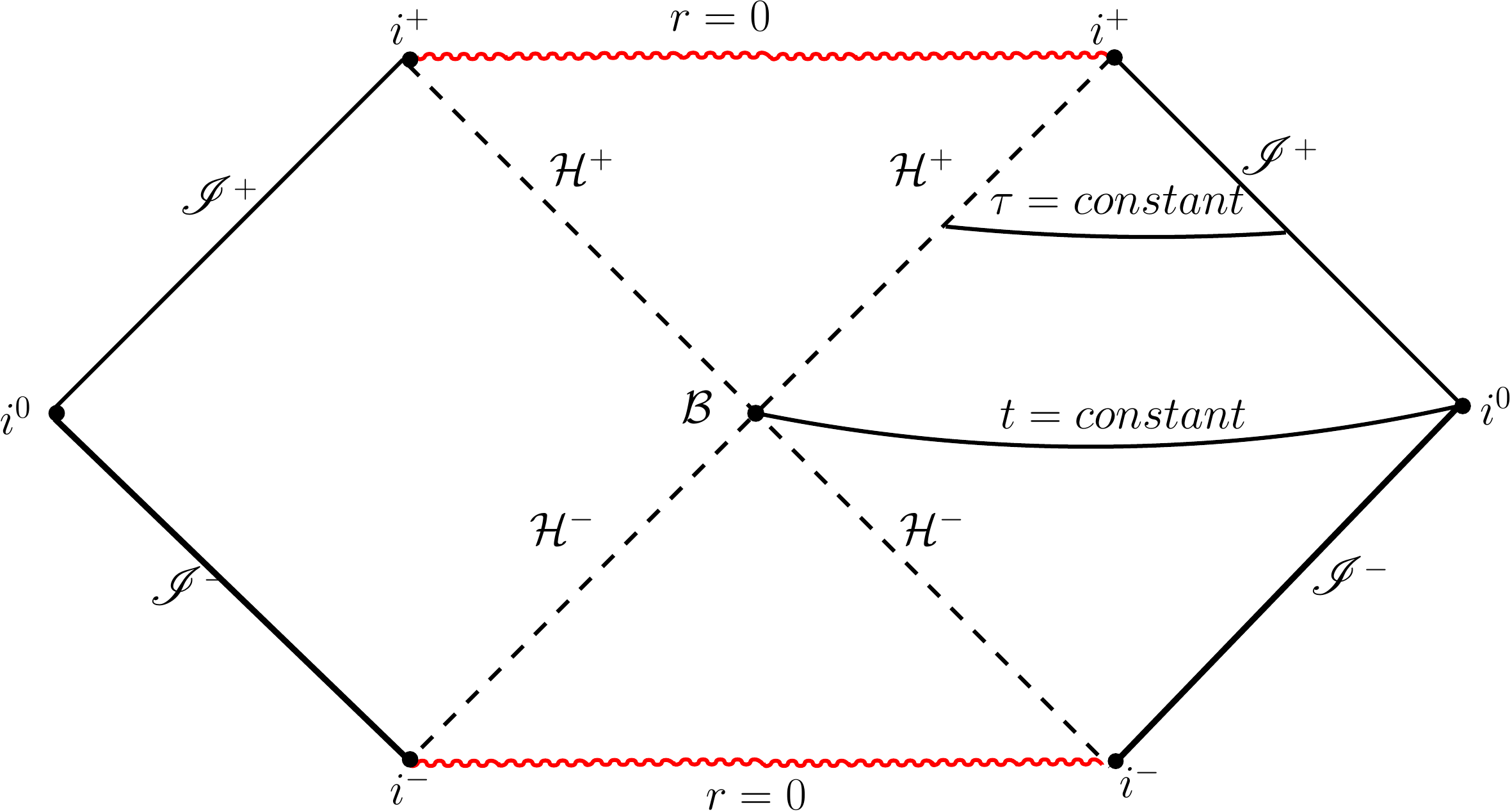}
\end{center}
\caption{Penrose-Carter conformal diagram for the extended Schwarzschild spacetime. Future and past event  horizons are denoted by ${\cal H}^+$ and ${\cal H}^-$ respectively. Likewise, future and past null infinity are specified through $\scri$ and $\scrI^-$ respectively. The bifurcation point ${\cal B}$ describes the mutual meeting point of the several horizons in this diagram. Also shown is a typical hyperboloidal slice $\tau=constant$ extending smoothly through both ${\cal H}^+$ and $\scri$, as well as a Cauchy slice $t=constant$ extending from the bifurcation point ${\cal B}$ to spatial infinity $i^0$.}
\label{fig:PenroseSchwarzschild}
\end{figure}

As mentioned above, we concentrate in this paper on a hyperboloidal formulation. However, typically the black hole perturbation theory is developed with the background metric described in terms of a coordinate system with slices of constant time extending between the bifurcation point ${\cal B}$ and spatial infinity $i^0$, see the Penrose-Carter conformal diagram in fig.~\ref{fig:PenroseSchwarzschild}. 

The most simple example is given by the Schwarzschild spacetime written in terms of the well known Schwarzschild-coordinates $\{t, r, \theta, \varphi\}$. In this context, apart from the data at the initial time slice, one needs to impose boundary conditions at ${\cal B}$ as well as at $i^0$ since the physical solution should contain only ingoing (outgoing) radiation at the horizon (spatial infinity). When treating the wave equation in terms of its Laplace transform, this framework leads to solutions of the spatial equation which grow exponentially near the boundaries. Note that the review~\cite{Nollert99} lists many difficulties for obtaining the desired representation (\ref{eq:VSol_spectral}), and the issue regarding the blow up of the solutions near the boundaries constitutes one of the main drawbacks.

To overcome these caveats, the author of~\cite{Zenginoglu2011} argues in favour of a coordinate system with time-constant surfaces extending between the future event horizon ${\cal H}$ and future null infinity $\scri$, also known as hyperboloidal slices, see~\cite{Frauendiener04} for a review. The work~\cite{Zenginoglu2011} shows that this choice resolves the issues concerning the representation of the functions associated to the QNMs. The paper emphasises the advantage of this framework in comparison with other methods and mentions that developing a black hole perturbation theory on hyperboloidal slices ``may lead to efficient numerical codes in the frequency domain"~\cite{Zenginoglu2011}. The work does, however, not advance further in this direction. In this paper, we exploit the advantages of the formulation of linear wave equations in the Schwarzschild spacetime on hyperboloidal slices. 

The paper is organized as follows. In section \ref{sec:QNMFilter}, we describe the coordinate transformation leading to the hyperboloidal slices in the Schwarzschild spacetime, and we introduce the Bardeen-Press equation that describes scalar, electromagnetic and gravitational perturbations propagating on this background. Section \ref{sec:LapTrans} is devoted to the Laplace transformation of the Bardeen-Press equation, thereby obtaining a characteristic spatial equation. Moreover, the inversion through the so-called {\em Bromwich Integral} is discussed. This section also brings a comparison with the corresponding formulation of black-hole perturbation theory on Cauchy slices. In the comprehensive section \ref{sec:TaylorExpansions} we apply sophisticated Taylor series expansions in order to get the solutions of the spatial equations. This  analysis provides us with the various ingredients of the spectral representation~(\ref{eq:VSol_spectral}) of the solution which is then derived in section \ref{sec:Spec_Decomp}. Section \ref{sec:Discussion} brings a thorough discussion, including a comparison with a similar problem in Minkowski spacetime which can be treated explicitly. The appendix comprises several sections in order to elaborate on certain aspects and issues that appear in the course of the text. Especially, the so-called algebraically special QNMs \cite{Chandrasekhar:1984} are discussed. Note that we use units such that the speed of light as well as Newton's constant of gravity are unity, $c=G=1$. 
\section{Black-hole perturbation within a Hyperboloidal foliation}\label{sec:QNMFilter}
\subsection{Hyperboloidal coordinates}\label{sec:HyperCoord}
Our starting point is a  review of the hyperboloidal coordinates used for the spacetime foliation. Following~\cite{Schinkel:2013zm,Macedo2014}, 
we write at first the Schwarzschild metric for a black hole with mass $M$ in the horizon-penetrating ingoing Eddington-Finkelstein coordinates $\{v,r, \theta, \phi \}$ and introduce then the new coordinates $(\tau, \sigma)$ via
\bea
v &=& 4M \left( \tau +   \frac{1}{\sigma}  - \log{\sigma}  \right)  \label{eq:HypCoord1} \\
r &=& \frac{2M}{\sigma}.  \label{eq:HypCoord2}
\eea
As a result, we obtain the line element 
\bea
ds^2 &=&\frac{16M^2}{\sigma^2}\Bigg[ -\sigma^2 (1-\sigma)d\tau^2 +(1+\sigma)d\sigma^2  \nn \\ &&
+(1-2\sigma^2)d\tau d\sigma 
 + \frac{1}{4}\left(d\theta^2 + \sin^2\theta d\varphi \right)\Bigg].
\eea
In the hyperboloidal coordinates $\{\tau,\sigma, \theta, \phi \}$, the horizon is given by $\sigma = 1$, while $\scri$ is fixed at $\sigma = 0$.

\subsection{Bardeen-Press equation}\label{sec:BardeenPressEq}
The equation describing the dynamics of a perturbation $U$ in a background given by the Schwarzschild solution was derived\footnote{The equation is equivalent to the Teukolsky equation~\cite{Teukolsky73} with vanishing specific angular momentum parameter, $a=0$.} by Bardeen and Press~\cite{Bardeen73a}. The equation reads in our hyperboloidal coordinates $\{ \coord \}$~\cite{Macedo2014}
\bea
\label{eq:IrregTeukolsky}
&-&  \left( 1+\sigma \right) U_{,\tau \tau}
+\left( 1-2\sigma^2 \right) U_{,\tau\sigma}  + (1-\sigma)\sigma^2 U_{,\sigma\sigma} \nn \\
 & -& \left( \frac{1+2\lambda}{\sigma} - \lambda   \frac{2-\sigma}{1-\sigma} \right) U_{,\tau} -  \sigma\left[ \sigma + \lambda(2-\sigma)\right]U_{,\sigma}\nn \\
&+ & U_{,\theta\theta} +  U_{,\theta}\cot\theta +\frac{1}{\sin^2\theta}\left(U_{,\varphi\varphi} + 2i\lambda\cos\theta U_{,\varphi} \right)  \nn \\
&+& \lambda\left(1-\lambda \cot^2\theta \right)U = 0.
\eea

The field $U(\coord)$ has spin-weight $\lambda$. A scalar field $\Phi$ propagating on the background is described by $\lambda=0$. If $\lambda=-1$ ($\lambda=+1$), $U(\coord)$ is closely related to the Newman-Penrose scalars~\cite{Newman62a} $\phi_2$ ($\phi_0$) describing outgoing (ingoing) electromagnetic waves. In the same way, gravitational waves are associated to Newman-Penrose scalars $\Psi_4$ ($\Psi_0$) with spin weight $\lambda=-2$ ($\lambda=+2$). Table \ref{tab:FieldPert} brings the relation between $U(\coord)$ and the different types of fields according to the corresponding spin $\lambda$. The table also shows the asymptotic behaviour of such fields around $\scri$ ($\sigma \to 0$) in accordance with the Peeling Theorem~\cite{Newman62a}.

\begin{table}[h]
\caption{Perturbation field and spin-weight $\lambda$}
\begin{center}
\begin{tabular}{|c|c|rcl|}
\hline
$\quad\lambda\quad$ & $\quad U\quad$ & $\quad\sigma$ &$\to$ & $0\quad$ \\
\hline
\hline
0 & $\Phi$ &  $\Phi $&=&${\cal O}(\sigma)$\\
\hline
1 & $\phi_0$ & $\phi_0$ &=&${\cal O}(\sigma^3)$ \\
\hline
-1 & $\sigma^{-2}\phi_2$ & $\phi_2$&=&${\cal O}(\sigma)$ \\
\hline
2 & $\Psi_0$ & $\Psi_0$&=&${\cal O}(\sigma^5)$ \\
\hline
-2 & $\quad \sigma^{-4}\Psi_4$ \quad& $\quad\Psi_4$&=&${\cal O}(\sigma)\quad$ \\
\hline
\end{tabular}
\end{center}
\label{tab:FieldPert}
\end{table}

Note that eq.~(\ref{eq:IrregTeukolsky}) is irregular at $\sigma=0$ and $\sigma=1$, i.e., at future null infinity and at the horizon, respectively. This property is a direct consequence of our hyperboloidal slicing. Taking into account the relation of $U(\coord)$ with the respective Newman-Penrose quantities and their asymptotic behaviour as depicted in table \ref{tab:FieldPert}, one can introduce the regular field $\tilde U$
\[
U(\coord)=\sigma^{1+2\lambda}\tilde{U}(\coord)
\] which removes the singular term going as $\sigma^{-1}$ from (\ref{eq:IrregTeukolsky}). Moreover, as noted in~\cite{Zenginoglu2011}, the singular term at $\sigma=1$ is removed by the further re-scaling
\beq
U(\coord)=\sigma^{1+2\lambda}(1-\sigma)^{-\lambda} V(\coord).
\eeq
We expand the field $V(\coord)$ into the spin-weighted spherical harmonics ${}_\lambda Y_{\ell m}(\theta, \varphi)$ basis~\cite{Goldberg:1967} 
\[
V(\coord)  = \sum_{\ell = |\lambda|}^{\infty} \sum_{m=-\ell}^{\ell} V_{\ell m}(\tau, \sigma)\,{}_\lambda Y_{\ell m}(\theta, \varphi),
\]
thus obtaining a specific wave equation for each mode $V_{\ell m}(\tau, \sigma)$: 
\footnote{For simplicity, we omit the indices $\ell m$ in $V_{\ell m}(\tau, \sigma)$ from now on.}
  \bea
  \label{eq:ReqTeukEq}
 & - &  (1+\sigma) V_{,\tau \tau}  +  \left( 1 - 2\sigma^2 \right)V_{,\tau \sigma} 
  + \sigma^2\left( 1-\sigma  \right)V_{,\sigma \sigma}   \notag \\ 
  &+& \sigma\left[ 2 - 3\sigma +\lambda(2-\sigma) \right]V_{,\sigma}  -  \left[ 2\sigma - \lambda(1-\sigma) \right] V_{,\tau} \notag \\ 
  &+&\left[   - \ell(\ell+1)  -(\sigma-\lambda)(1+\lambda) \right] V 
   = 0.
  \eea
  
Given initial data $V_0(\sigma) = V(0,\sigma)$ and $\dot V_0(\sigma) = V_{,\tau}(0, \sigma)$, eq.~(\ref{eq:ReqTeukEq}) is to be solved in the domain $(\tau, \sigma)\in [0,\tau_{\rm final}]\times[0,1]$. 

Note that the transition from the field $U$ to $V$ has not removed the degeneracies of the wave equation at the two boundaries $\sigma=0$ and $\sigma=1$. These degeneracies provide boundary conditions that guarantee that the characteristics of the wave equation always point outward the numerical domain and hence no further boundary conditions at the horizon nor at future null infinity $\scri$ are allowed to be imposed. Consequently, equation (\ref{eq:ReqTeukEq}) has to be solved as an {\em initial value problem}.


\section{Laplace transformation}\label{sec:LapTrans}
\subsection{Definition}\label{sec:LapTrans_Def}
Given initial data $V_0(\sigma)$ and $\dot V_0(\sigma)$, we follow~\cite{Kokkotas99a} and introduce the Laplace transformation
\beq
\label{eq:LaplaceTransf}
\hat V(\sigma; s) := {\cal L}[V(\tau,\sigma)](s) =  \int_0^\infty e^{-s\tau} V(\tau, \sigma) d\tau.
\eeq
As the field $V$ is strictly  bounded for all times $\tau$, it follows that $\hat V(\sigma; s)$ is complex-holomorphic in the right half-plane $\Re(s)>0$ (see figure~\ref{fig:BromwichInt}). Note that the following relation is a particular consequence of the fact that the wave-field $V(\tau,\sigma)$  is real: \footnote{Throughout this paper we use an upper asterisk ${}^*$ to denote {\em complex conjugation}.}
\beq
	\label{eq:Symmetry_hat_V}
	\hat V(\sigma;s^*)=[\hat V(\sigma;s)]^*.
\eeq
Applying now the Laplace transformation to both sides of the dynamical equation (\ref{eq:ReqTeukEq}) and considering that
\bea
	\nn {\cal L}[V_{,\tau}(\tau,\sigma)](s)&=&s\hat V(\sigma; s) - V_0(\sigma), \\
	\nn {\cal L}[V_{,\tau\tau}(\tau,\sigma)](s)&=&s^2\hat V(\sigma; s) - sV_0(\sigma)-\dot V_0(\sigma), 
\eea
we obtain an inhomogeneous ordinary differential equation (referred to as ``ODE'' in the sequel) \footnote{For addressing the function $\hat V$ defined for $\sigma\in[0,1]$ (rather than a particular function value within that interval), we simply write $\hat V(s)$. The same applies to the right hand side $B$.}
\bea
\label{eq:LaplaceTransfTeukEq}
{\mathbf A}(s) \hat{V}(s)  &=& B(s)
\eea
with the second order differential operator given by
\bea
\label{eq:OpertorA}
{\mathbf A}(s)&=&\sigma^{2}(1-\sigma)\partial_{\sigma\sigma} \nn \\
&&+\left\{ s(1-2\sigma^2)+\sigma\left[2-3\sigma+ \lambda(2-\sigma)\right]\right\}\partial_{\sigma} \nn \\
&& -\left\{s^{2}(1+\sigma) + s[2\sigma+\lambda(\sigma-1)] \right.\nn \\ 
&&\left. + \ell(\ell+1) + (\sigma-\lambda)(\lambda+1)\right\}.
\eea 
The degeneracies of the wave equation (\ref{eq:ReqTeukEq}) (see discussion at the end of Sec.~\ref{sec:BardeenPressEq}) implies that of the operator ${\mathbf A}(s)$ at the surfaces $\sigma = 0$ and $\sigma=1$. 
The inhomogeneity $B(s)$ in (\ref{eq:LaplaceTransfTeukEq}) is given in terms of the initial data $V_0=V_0(\sigma)$ and $\dot{V}_0=\dot{V}_0(\sigma)$:
\bea
B( s)  &=& (1-2\sigma^2)V_{0}{}_{,\sigma}-(1+\sigma)\dot{V}_{0}\nn \\
&&-\left[2\sigma-\lambda(1-\sigma)\right]V_{0} -s(1+\sigma)V_{0}. \label{eq:Source_b}
\eea

\subsection{Inversion}\label{sec:LapTrans_Inv}
\begin{figure}[t!]
\begin{center}
\includegraphics[width=8.0cm]{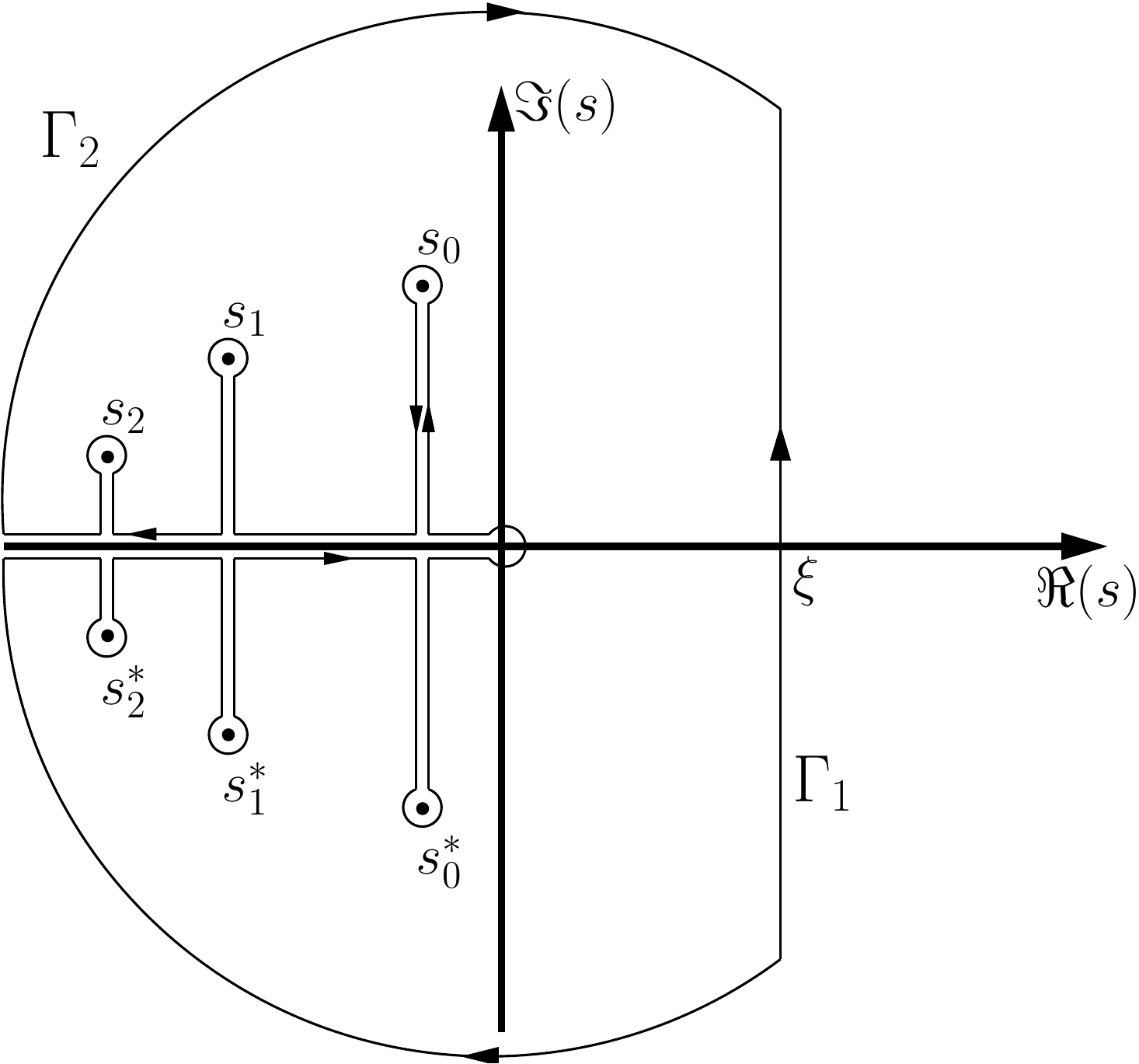}
\end{center}
\caption{Integration paths for the inverse Laplace transformation. The Bromwich integral (\ref{eq:BromInt}) is evaluated along the line $\Gamma_1$ in the half-plane $\Re(s)>0$. 
In order to arrive at the spectral decomposition formula \eqref{eq:VSol_spectral} it is essential to deform the integration path and to obtain the solution via integration along the curve $\Gamma_2$. The several ingredients, i.e.~integration around the QNMs $s_n$, along the branch cut $\mathbb{R}^-$ as well as along infinitely extended circular sections, are discussed in detail in Sec.~\ref{sec:Spec_Decomp}.}
\label{fig:BromwichInt}
\end{figure}

The solution of equation (\ref{eq:LaplaceTransfTeukEq}) with the help of a sophisticated Taylor series expansion will be depicted in Sec.~\ref{sec:TaylorExpansions}. Once established $\hat{V}(\sigma;s)$ for values $s$ on some vertical line  in the right half-plane $\Re(s)>0$ (i.e.~for $\Re(s)=\xi$ with some fixed $\xi>0$), we may write the solution of eq.~(\ref{eq:ReqTeukEq}) in terms of the inverse Laplace transformation, also known as the Bromwich Integral
\beq
\label{eq:BromInt}
V(\tau, \sigma) = \frac{1}{2\pi i}\int_{\Gamma_1}\hat{V}(\sigma;s)e^{s\tau} ds,
\eeq
with the integration path (see fig.~\ref{fig:BromwichInt})
\beq
\label{eq:Gamma_1}
\Gamma_1=\left\{s\in {\mathbb C}\,|\, s=\xi + i\chi ,\, \xi>0, \, \chi\in(-\infty, +\infty)\right\}.
\eeq
Note that \eqref{eq:BromInt} provides us with the solution $V(\tau, \sigma)$ only if $\tau>0$ (for $\tau<0$ the integral vanishes). 

We remark that the formula \eqref{eq:BromInt} is the starting point for the spectral decomposition (\ref{eq:VSol_spectral}) which comes about through an appropriate deformation of the integration path (see figure~\ref{fig:BromwichInt}) to be discussed in Sec.~\ref{sec:Spec_Decomp}.

\subsection{Comparison with Cauchy foliation}\label{sec:CauchyCompLaplace}
We end this section with a brief discussion on some relations between the Laplace transformation performed in the context of hyperboloidal foliation and the corresponding formulation of perturbation theory in the original Schwarzschild coordinates. 

Note that typically, perturbations on the Schwarzschild background are described by the Regge-Wheeler-Zerilli formalism, while here we focus on the Bardeen-Press-Teukolsky approach. Both formalisms are known to be equivalent and, in particular, the equations coincide for $\lambda=0$ (scalar perturbation). Therefore, along the paper, the discussion between the different foliations of the spacetime will always be made for the particular case $\lambda=0$. The conclusions, however, are general and should be valid also for $\lambda \ne 0$.

Let us first introduce the dimensionless coordinates $x = r^*/(2M)$ and $\bar{t} = t/(2M)$ where $r^*$ is the so-called tortoise coordinate. They are related to $\{\tau, \sigma\}$ via (cf.~eqs.~\eqref{eq:HypCoord1} and \eqref{eq:HypCoord2})
\bea
x&= & \frac{1}{\sigma} - \ln(\sigma) + \ln(1-\sigma), \label{eq:Coord_x}\\
\bar{t} &=& 2\tau + \frac{1}{\sigma} - \ln\left[\sigma(1-\sigma)\right]. \label{eq:Coord_t}
\eea
The horizon is described by $x\rightarrow -\infty$ and spatial infinity is given by $x\rightarrow +\infty$. These points correspond to the bifurcation point ${\cal B}$ and $i^0$ in the Penrose diagram (see fig.~\ref{fig:PenroseSchwarzschild}). 

Then, eq.~\eqref{eq:ReqTeukEq} is equivalent to the well-known wave equation
\beq
\label{eq:WaveEqCauchy}
- f_{,\bar{t}\bar{t}} + f_{,xx} - {\cal P}\, f = 0,
\eeq
with $f(\bar{t}, x) = V(\tau(\bar{t},x), \sigma(x))$ and 
\bea
{\cal P} &=& \left( 1- \frac{2M}{r}\right)\left(\frac{2M}{r}\right)^2\left[\frac{2M}{r} + \ell(\ell+1) \right] \nn \\
&=& \left( 1- \sigma \right)\sigma^2\left[\sigma + \ell(\ell+1) \right].
\eea

Now, we apply the Laplace transformation
\beq
\label{eq:LapCauchy}
\hat{f}(x;\bar{s})  := {\cal L}[f(\bar{t}, x)](\bar{s}) =  \int_0^\infty e^{-\bar{s}\bar{t}} f(\bar{t}, x) d\bar{t}.
\eeq
to eq.~\eqref{eq:WaveEqCauchy} and obtain the equation
\beq
\label{eq:LapTranSchwCoord}
\hat{f}_{,xx} - \left[ \bar{s}^2 + {\cal P}\right]\hat{f} = \dot{f}_0(x) + \bar{s}f_0(x),
\eeq
where $f_0(x)=f(0,x)$ and $\dot{f}_0(x)=f_{,\bar{t}}(0,x)$ are the initial data and $\bar{s}$ the Laplace parameter in the Cauchy formulation.

Considering the homogeneous equations, we find the following relation between the Cauchy and the hyperboloidal formulation. Given a function $\Omega(s)$ that satisfies the ODE ${\mathbf A}(s) \Omega(s)=0$ (cf.~\eqref{eq:LaplaceTransfTeukEq} and \eqref{eq:OpertorA}). Then the function
\beq
\label{eq:Rel_f_V}
F(x(\sigma);\bar{s}) = 2\,e^{-\bar{s}/\sigma}\sigma^{\bar{s}}(1-\sigma)^{\bar{s}} \,\Omega(\sigma; 2\bar{s})
\eeq
obeys the equation
\beq
\label{eq:HomLaplCauchy}
F_{,xx} - \left[ \bar{s}^2 + {\cal P}\right]F = 0
\eeq
where  the Laplace parameter $s$ (from our hyperboloidal foliation) is related to the Cauchy Laplace parameter $\bar{s}$ via \beq\label{eq:s_bar}s = 2\,\bar{s}.\eeq 

Note that (up to constants) the term 
\beq
\label{eq:HypCauchyFactor}
\Xi(\sigma; \bar{s}) = 2\, e^{-\bar{s}/\sigma}\sigma^{\bar{s}}(1-\sigma)^{\bar{s}}
\eeq 
corresponds exactly to the factor introduced by Leaver~\cite{Leaver85} accounting for the correct behaviour of $F$ at the boundaries $x\rightarrow \pm \infty$. Here, the relevant factor \eqref{eq:HypCauchyFactor} is motivated by the term $e^{-\bar{s}\bar{t}(\tau, \sigma)}$ in \eqref{eq:LapCauchy}, i.e.,~it follows directly from the hyperboloidal coordinates as a consequence of the fact that $\tau = constant$ surfaces extend smoothly between the black horizon ${\cal H}^+$ and future null infinity $\scri$.

\section{Taylor series expansions}\label{sec:TaylorExpansions}

\subsection{Solutions to the homogeneous Laplace transformed equation and the asymptotics of their Taylor coefficients }\label{sec:Homogeneous_Tayl_asymptotics}

Let us exclude, for the time being, nonpositive integer values of the Laplace parameter $s$, i.e.~$s\notin\mathbb{Z}^-_0$. \footnote{In Sec.~\ref{App:Neg_Lapl_Par} we extend our results to negative integer values $s$.\label{foot:Neg_s}} We then start with the homogeneous Laplace transformed equation
\bea
\label{eq:HomogLaplaceTransfTeukEq_phi}
{\mathbf A}(s) \phi(s)  &=& 0, \qquad \mbox{$\phi(\sigma;s)$ analytic for $\sigma\in(0,1]$}
\eea
and expand $\phi(s)$ in terms of a Taylor series
\beq
\label{eq:Series_phi}
\phi(\sigma;s) = \sum_{k=0}^{\infty} H_{k}(1-\sigma)^k.
\eeq
With this ansatz we follow Leaver~\cite{Leaver85} and concentrate, in a first step, on solutions which are {\em analytic} in a neighbourhood about the horizon ${\cal H}$, i.e.~at $\sigma=1$. Inserting the ansatz \eqref{eq:Series_phi} into \eqref{eq:HomogLaplaceTransfTeukEq_phi}, we obtain the following recurrence relation:
\bea
	\alpha_k H_{k+1} + \beta_k H_k + \gamma_k H_{k-1} 	&=& 0 \label{eq:RecRel_phi}
\eea
with the coefficients:
\beq\label{eq:RecRelCoef_albega}
\begin{array}{ccl}
\alpha_k &=& (k+1)( k+1 + s - \lambda),   \\
-\beta_k &=& 2(k+s)( k+1 + s) + \ell(\ell+1) - \lambda^2 + 1,  \\
\gamma_k &=& (k+s)(k+s+\lambda).
\end{array}
\eeq
The coefficients $H_k$ can now be obtained for $k\ge 1$ via 
\beq
	H_{k+1}=-\frac{1}{\alpha_k} \Big(\beta_k H_k + \gamma_k H_{k-1}\Big) = 0, \qquad H_0\stackrel{!}{=}1,\label{eq:UpRecRel_phi}
\eeq
where we have chosen $H_0=1$ as convenient scaling condition. Note that for $k\ge 1$  we have $\alpha_k\ne 0$ as long as $s\notin\mathbb{Z}^-_0$, see~\eqref{eq:RecRelCoef_albega}. We study now the asymptotics of $H_k$ for large indices $k\to\infty$.

The singular points of the ODE (\ref{eq:HomogLaplaceTransfTeukEq_phi}) are given by $\sigma=0$, $\sigma=1$, and $\sigma=\infty$. Consequently, we expect the series (\ref{eq:Series_phi}) to be convergent within the unit circle 
\beq\label{eq:circle_C} 
\mathbf{C}=\{\sigma\in\mathbb{C}: |1-\sigma|<1\},
\eeq
with analyticity breaking down at $\sigma=0$, as this represents  an essentially singular point of the ODE. We may therefore conclude that the domain of convergence of (\ref{eq:Series_phi}) does not exceed the unit circle $\mathbf{C}$, implying that
\beq\label{eq:lim_ratio}
	\lim_{k\to\infty} \left|\frac{H_{k+1}}{H_k}\right|=1.
\eeq
An asymptotic estimate was found by Leaver~\cite{Leaver85}
\[\frac{H_{k+1}}{H_k}=1\pm\sqrt{\frac{s}{k}}+\frac{\lambda + s-\frac{3}{4}}{k}+{\cal O}(k^{-3/2}),\]
from which we derive:
\bea\nn
	\log H_{k+1} &=& \log H_{k} \pm \sqrt{\frac{s}{k}} + \frac{\lambda + \frac{s}{2} -\frac{3}{4}}{k} + {\cal O}\left(k^{-3/2}\right). 
\eea
Applying this formula successively from some $k=k_0$ on,  we obtain the asymptotic formula
\bea
	\nn\log H_{k+1} &=& \pm\sqrt{s}\sum_{j=k_0}^k \frac{1}{\sqrt{j}}+
\left(\lambda + \frac{s}{2}-\frac{3}{4}\right)\sum_{j=k_0}^k\frac{1}{j} + {\cal O}(1) \\
  \nn &=& \pm\sqrt{s}\left[H^{(1/2)}_k-H^{(1/2)}_{k_0-1}\right] \\ &&\nn
    + \left(\lambda + \frac{s}{2}-\frac{3}{4}\right)\left[H^{(1)}_k-H^{(1)}_{k_0-1}\right] + {\cal O}(1) 
\eea
with 
\beq\label{eq:Harmonic_Numbers}
H^{(\nu)}_k=\sum_{j=1}^k \frac{1}{j^\nu} 
\eeq
being {\em generalized harmonic numbers}. Now these numbers possess the following asymptotics:
\beq\label{eq:Harm_Num_asymptotics}
H^{(1/2)}_k=2\sqrt{k}+ {\cal O}(1) ,\qquad H^{(1)}_k=\log k + {\cal O}(1) ,
\eeq
from which it follows that for sufficiently large $k$ :
\beq\label{eq:Hk_Asymptotics} 
	H_k=k^\zeta\left(A^+_k e^{\kappa\sqrt{k}} + A^-_k e^{-\kappa\sqrt{k}}\right),
\eeq
with
\beq\label{eq:kappa_zeta}
 \kappa=2\sqrt{s},\quad\Re(\kappa)>0,\qquad\zeta=\lambda + \frac{s}{2}-\frac{3}{4},\eeq
and where the coefficients $A^\pm_k$ tend to a finite value as $k\to\infty$, to be denoted by $A^\pm_\infty$. 
Again, this result corresponds to the  findings by Leaver \cite{Leaver86b}. 

If we now investigate the behaviour of the corresponding solution $\phi$ of \eqref{eq:HomogLaplaceTransfTeukEq_phi} we observe that, for $\Re(s)>0$, it possesses an essential singularity at $\sigma=0$, somewhat similar to the function $e^{s/\sigma}$.\footnote{Indeed, it can be shown that the Taylor series (\ref{eq:Series_phi}) of $e^{s/\sigma}$ possesses the dominant asymptotics $H_k=e^{2\sqrt{sk}}k^{-3/4} A^+_k$ with $A^+_k\to A^+_\infty(s)$, see appendix, Sec.~\ref{App:Examples_Taylor_Asymptotics}.} A more sophisticated analysis (to be conducted in Sec.~\ref{App:Examples_Taylor_Asymptotics}) reveals that 
\beq\label{eq:Phi}
	\Phi(\sigma;s) := \sigma^{s+2\lambda} e ^{-s/\sigma} \phi(\sigma;s) 
\eeq
is, for $\Re(s)>0$, analytic at any $\sigma\in(0,1]$ and still $C^\infty$ at $\sigma=0$. However, just as the function $e^{s/\sigma}$, the behaviour of $\phi$ changes, when moving into the left half-plane $\Re(s)<0$, as here $\phi$ becomes $C^\infty$ for all $\sigma\in[0,1]$.  
This fact has the interesting consequence that for $\Re(s)<0$ the solutions $\phi(s)$ of \eqref{eq:HomogLaplaceTransfTeukEq_phi} can be taken as regular $C^\infty$ initial data,
\beq\label{eq:Example_ID}
V(0,\sigma)=\Re[\phi(\sigma, s)],\qquad 
V_{,\tau}(0,\sigma)= \Re[s \phi(\sigma, s)]
\eeq
which imply the regular $C^\infty$-solution 
\beq\label{eq:Example_V} V(\tau,\sigma)=\Re[\phi(\sigma, s)e^{s\tau}]\eeq
to our wave equation (\ref{eq:ReqTeukEq}). 

\begin{figure}[t!]
\begin{center}
\includegraphics[width=8.0cm]{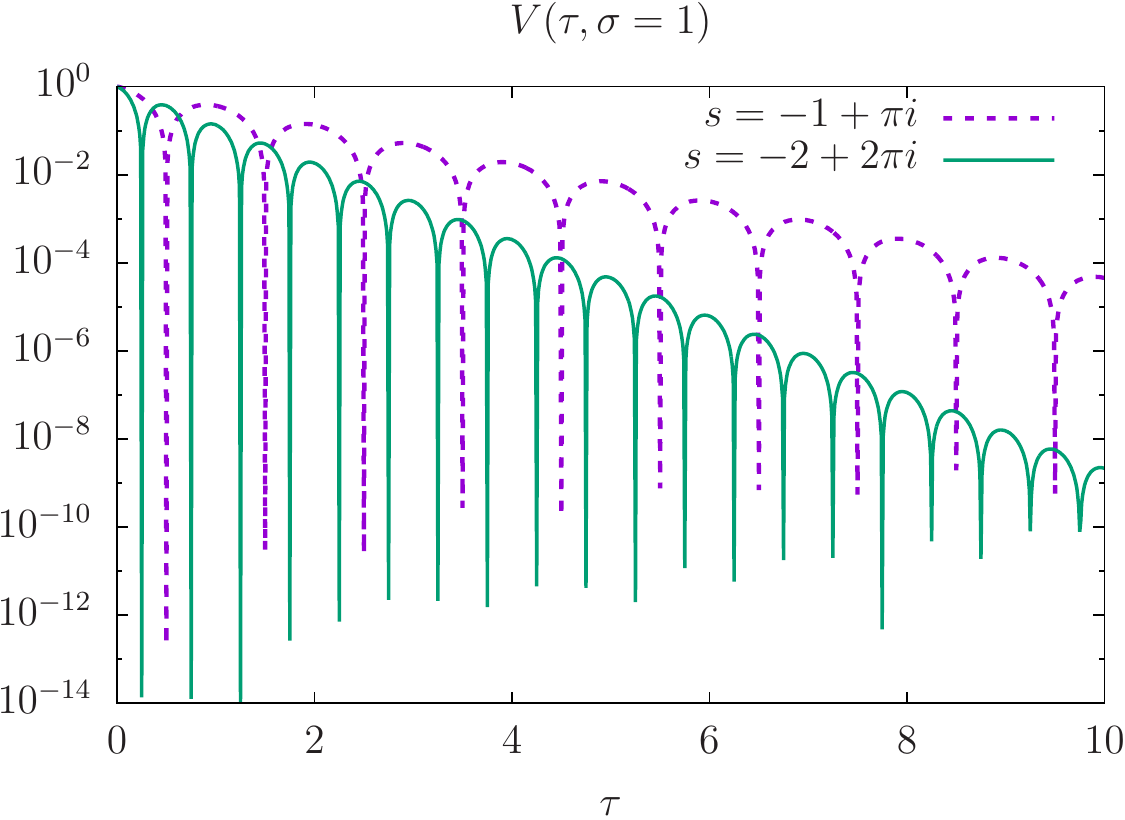}
\end{center}
\caption{Time evolution for initial data given by~\eqref{eq:Example_ID}. For arbitrarily prescribed $s$-values with $\Re(s)<0$, the evolution shows exponentially damped oscillations (cf.~\eqref{eq:Example_V}). Here, the examples are obtained for $s = -1+ \pi i $  and $s = -2 + 2\pi i$ in the case $\lambda=\ell=0$. The field is measured at $\sigma=1$. }
\label{fig:noQNMEvolution}
\end{figure}

We observe that the solution \eqref{eq:Example_V} behaves like a purely quasi-normal ringing, with damping $\Re(s)$ and frequency $\Im(s)$. Fig.~\ref{fig:noQNMEvolution} exemplifies this behaviour by showing the time evolution (obtained through the algorithm presented in \cite{Macedo2014}) for this type of initial data. In particular, in the case $\lambda=\ell=0$ the values $s = -1 + \pi i$ and $s= -2 + 2\pi i$ have been taken. However, these arbitrarily chosen $s$-values in the left plane are to be distinguished from the QNMs to be discussed in Sec.~\ref{Sec:QNMs}.
\footnote{Note that if $s$ is not a QNM, then the solution \eqref{eq:Example_V} cannot be expressed in terms of the desired spectral decomposition formula, see discussion in section \ref{Sec:Conclusion}.    
}

Let us now turn to a specific second solution $\psi(s)$ obeying the homogeneous Laplace transformed equation 
\bea
\label{eq:HomogLaplaceTransfTeukEq_psi}
{\mathbf A}(s) \psi(s)  &=& 0.
\eea
We describe this solution by a sequence $\{I_k\}_{k=-\infty}^\infty$ of coefficients satisfying
\bea
	\alpha_k I_{k+1} + \beta_k I_k + \gamma_k I_{k-1} 	&=& 0,\quad \lim_{k\to\infty} I_k e^{\kappa\sqrt{k}}k^{-\zeta}=1.\qquad\label{eq:RecRel_psi}
\eea
Here we have chosen the asymptotics \eqref{eq:Hk_Asymptotics} (which holds for any solution to the homogeneous recursion relation \eqref{eq:RecRel_phi}) with $A^+_\infty=0$ and, moreover, $A^-_\infty=1$ as convenient scaling condition.

Let us discuss the corresponding solution $\psi$ in some detail. The sequence $\{I_k\}_{-\infty}^\infty$, which is uniquely defined by \eqref{eq:RecRel_psi}, does not provide us with a decent Laurent representation of $\psi$. An analysis of the asymptotics of $I_k$ as $k\to-\infty$, performed along the same lines as the above investigation in the case $k\to+\infty$, reveals that 
\beq\label{eq:Ik_negative_asymptotics}
\lim_{m\to \infty} \Big(e^{-2\sqrt{-sm}}m^{-\zeta}I_{-m}\Big)=B^+_\infty \qquad \mbox{with\quad $|B^+_\infty|<\infty$}.
\eeq
Here, the square root in the exponent is again to be taken such that its real part is positive. If we now consider principal part $\{I_k\}_{-\infty}^{-1}$ and secondary part $\{I_k\}_{0}^\infty$ separately,
\beq\label{eq:psi_pm}\psi_-=\sum_{m=1}^\infty \frac{I_{-m}}{(1-\sigma)^m},\quad \psi_+=\sum_{k=0}^\infty I_k(1-\sigma)^k,\eeq
we find that $\psi_-$ corresponds to a function which is analytic {\em outside} $\bar{\mathbf{C}}$ (the closure of the circle $\mathbf{C}$, cf.~\eqref{eq:circle_C}), while $\psi_+$ is analytic {\em inside} $\mathbf{C}$. Hence, there is no annulus in the complex $\sigma$-plane about the point $\sigma=1$ within which the formal Laurent series $\psi=\psi_-+\psi_+$ would converge. Nevertheless, $\psi_-$ and $\psi_+$ can be extended analytically into exterior and interior of $\mathbf{C}$, respectively, and there $(\psi_-+\psi_+)$ describes the complex continuation of a function $\psi$ that satisfies \eqref{eq:HomogLaplaceTransfTeukEq_psi}. Note that the complex extension of $\psi_-$ is a function with singularities at $\sigma\in\{0,1\}$ and a branch cut discontinuity along the real interval $\sigma\in(0,1)$. 
Likewise, $\psi_+$ is a function with singularities at $\sigma\in\{0,\infty\}$ and a branch cut discontinuity along the real interval $\sigma\in(-\infty,0)$. Thus, the resulting function $\psi$ possesses a branch cut discontinuity along the real interval $\sigma\in(-\infty,1)$.

As a solution to \eqref{eq:HomogLaplaceTransfTeukEq_psi}, the function $\psi=\psi_-+\psi_+$ can be written in terms of $\phi$ as
\beq
	\label{eq:ansatz_psi}\psi(\sigma;s)=\phi(\sigma;s)[C(s)\Pi(\sigma;s)+c(s)]
\eeq
with $s$-dependent constants of integration $C(s)$ and $c(s)$ and where  
\bea
	\label{eq:chi}
	\Pi(\sigma;s)&=&\int\limits_{1/2}^\sigma e^{s/\tilde\sigma}
	\frac{[\tilde\sigma(1-\tilde\sigma)]^{\lambda-s-1} d\tilde\sigma}{\tilde{\sigma}^{3\lambda+1}[\phi(\tilde\sigma;s)]^2}\\
	&=&\int\limits_{1/2}^\sigma e^{-s/\tilde\sigma}
	\frac{[\tilde\sigma(1-\tilde\sigma)]^{\lambda-s-1} d\tilde\sigma}{\tilde{\sigma}^{1-2s-\lambda}[\Phi(\tilde\sigma;s)]^2}.\nn
\eea
With the convenient lower integration bound $1/2$,  the function $\Pi$ can be analytically extended for $0<\sigma<1$ from the right half-plane $\Re(s)>0$ onto the left half-plane $\Re(s)<0$. We remark that based on the aforementioned considerations it appears difficult to assess with certainty that the function $\psi$, when considered at {\em real} $\sigma\gtrsim 0$, is $C^\infty$-regular at $\sigma=0$ for $\Re(s)>0$. This property would be essential to qualify $\psi$ as one of the two linearly independent homogeneous solutions to \eqref{eq:HomogLaplaceTransfTeukEq_psi} that, according to the definition in \cite{Kokkotas99a},  constitute a Wronskian determinant whose zeros determine the set of QNMs. To show that $\psi$ is $C^\infty$-regular at $\sigma=0$ would mean that we have to perform a complex continuation of $\psi_-$ from the annulus $\{\sigma\in\mathbb{C}: |1-\sigma|>1\}$ to {\em real} $\sigma\gtrsim 0$. As values of $\psi_-$ anywhere inside the annulus are only given numerically, it appears extremely difficult to provide a decent analytical expansion of $\psi_-$ as required. In Sec.~\ref{Sec:QNMs_Cauchy}, we return to this issue but remark here that instead of the function $\psi$ being considered, we concentrate in the following on the corresponding sequence $\{I_k\}$ which is defined through the specific asymptotics given in \eqref{eq:RecRel_psi}. 

\begin{figure*}[t!]
\begin{center}
\includegraphics[width=8.5cm]{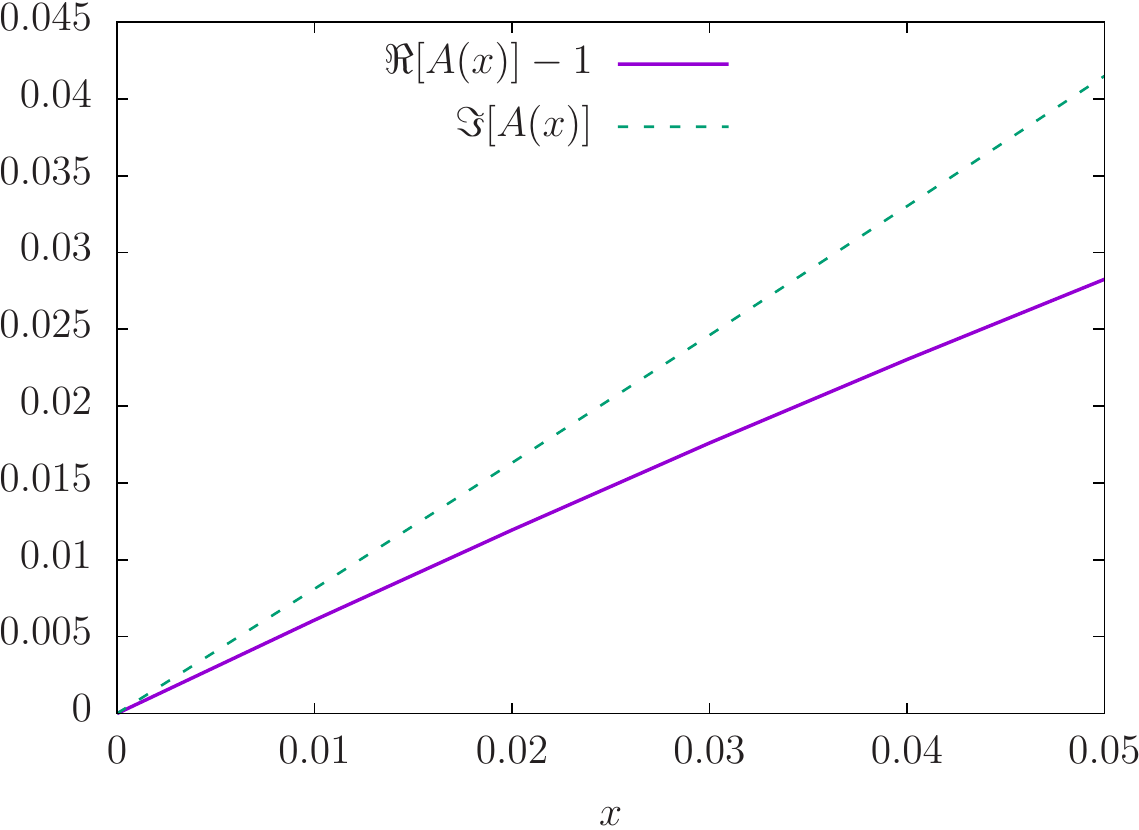}
\includegraphics[width=8.5cm]{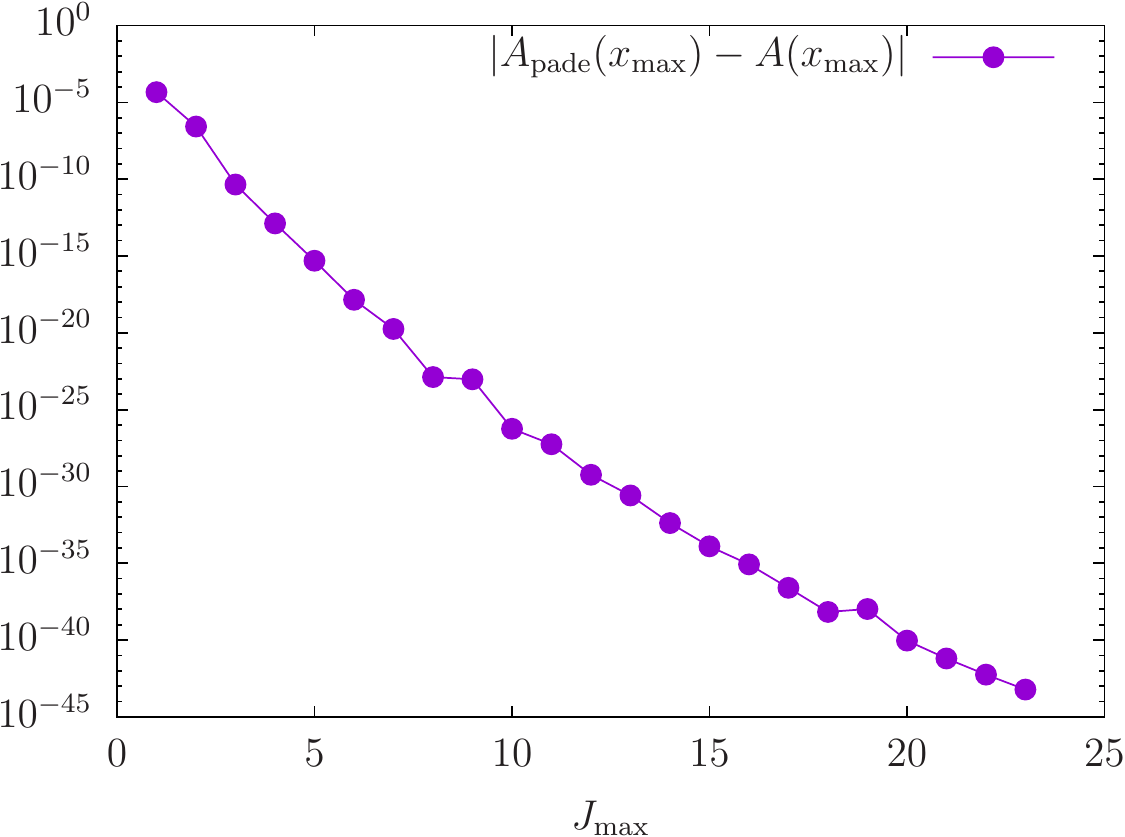}
\end{center}
\caption{Left panel: within the interval $x\in[0,x_{\rm max}]$ with $x_{\rm max}=0.05$, the function $A=A(x)$ (cf.~(\ref{eq:A_k}), (\ref{eq:A_expansion}), (\ref{eq:A_Pade})) is shown for $\ell=\lambda=0$ and the Laplace parameter $s=1+i$. The values $x\le 0.05$ correspond to the indexes $k\ge k_{\rm max}=x_{\rm max}^{-2}=400$. Right panel: maximal deviation, obtained for $x=x_{\rm max}$, of the approximation function $A_{\rm Pade}$ (cf. (\ref{eq:A_Pade}), (\ref{eq:djA})) from the function $A$, in dependence of the expansion order $J_{\rm max}$.}
\label{fig:A_Pade}
\end{figure*}

We now develop a higher-order approximation of the asymptotic expansion $(k\to+\infty)$ of the coefficients $I_k$, as they will be an essential ingredient in the solution of the Laplace transformed equation to be derived in Sec.~\ref{sec:Homogeneous_Tayl_asymptotics}. For $k\gg 1$, we write 
\bea\label{eq:A_k}
I_k&=&e^{-\kappa\sqrt{k}}k^{\zeta}A_k
\eea
where the coefficients $A_k$ are given in terms of a regular function $A$ defined on an $\varepsilon$-neighbourhood about the origin,
\beq\label{eq:A_expansion}
A_k = A\left(\frac{1}{\sqrt{k}}\right),\qquad A(x) = 1+\sum_{j=1}^\infty \mu_j x^j.
\eeq
Now, the expression \eqref{eq:A_k} may be inserted into a slightly rearranged version of the recurrence relation \eqref{eq:RecRel_psi}, thus obtaining:
\bea
	&&\alpha_k \left(1+\frac{1}{k}\right)^\zeta e^{-\kappa\left(\sqrt{k+1}-\sqrt{k}\right)}A_{k+1} + \beta_k A_k \nn \\
	&+& \gamma_k \left(1-\frac{1}{k}\right)^\zeta e^{-\kappa\left(\sqrt{k-1}-\sqrt{k}\right)} A_{k-1} 	= 0. \label{eq:RecRel_rearranged}
\eea
If we now consider \eqref{eq:A_expansion} and expand (\ref{eq:RecRel_rearranged}) in terms of $1/\sqrt{k}$ about $k=\infty$, we can successively determine the $\mu_j$'s through the method of equating the coefficients. The first two terms amount to:
\bea 
	\nn\mu_1&=&\frac{1}{48\sqrt{s}}[8 s (3+2 s) - 9-48 \ell (\ell+1)], \\
	\label{eq:mu_coeffs}\mu_2&=&\frac{1}{4608 s}\Big[-135 (1+16 s)\\
					\nn &&+32 \Big(9\ell (\ell+1) (-1+8 \ell (\ell+1))-72 (\ell+\ell^2-\lambda) s\\
	       \nn && -3 (21+16\ell (\ell+1)-48 \lambda) s^2+60 s^3+8 s^4\Big)\Big].
\eea
In this manner, all coefficients $\mu_j$ (and hence the function $A$) are completely fixed. A sample of the function $A$ for $\lambda=\ell=0$ and $s=1+ i $ is displayed in the left panel of fig.~\ref{fig:A_Pade}.

In our numerical computations we determine, for given values of $\lambda$, $\ell$ and $s$, the coefficients $\{\mu_j\}_{j=1}^{2 J_{\rm max}}$ for some integer $J_{\rm max}$ with high numerical precision. Thanks to the \texttt{Series}-command in \texttt{Mathematica}, values of $J_{\rm max}$ around 40 can easily be chosen. After the computation of the $\mu_j$'s, the function $A$ is approximated by a diagonal Pad\'e approximant
\beq\label{eq:A_Pade} A_{\rm Pade}(x)= \frac{\sum\limits_{j=0}^{J_{\rm max}}p_j x^j}{1+\sum\limits_{j=1}^{J_{\rm max}}q_jx^j} \eeq
where the coefficients $p_j, q_j$ are determined such that 
\beq\label{eq:djA}\frac{d^j A_{\rm Pade}}{dx^j}(0)= \frac{d^j A}{dx^j}(0)=j!\,\mu_j\quad \mbox{for $j=0,\ldots,2J_{\rm max}$},\eeq
with $\mu_0=1$, cf.~(\ref{eq:A_expansion}).
In this way we create a highly accurate approximation $A_{\rm Pade}$ of the function $A$, 
from which we may obtain approximative values of the two coefficients $I_{k_{\rm max}+1}$ and $I_{k_{\rm max}}$ for some large index $k_{\rm max}$ from the relation \eqref{eq:A_k}. The high order and the corresponding extreme accuracy of the Pad\'e approximation allows us to get away with rather moderate values of e.g.~$k_{\rm max}\sim 400$, see the right panel of fig.~\ref{fig:A_Pade}. 

We proceed with the determination the coefficients $I_k$ for $k\in\{-1,\ldots,k_{\rm max}-1\}$ from the backwards recurrence relation 
\beq
	I_{k-1}=-\frac{1}{\gamma_k} \Big(\alpha_k I_{k+1} + \beta_k I_k\Big) , \quad k=k_{\rm max},\ldots,0.\label{eq:DownRecRel_psi}
\eeq
Note that for $s\notin\mathbb{Z}$, the coefficients $\gamma_k$ do not vanish, cf.~\eqref{eq:RecRelCoef_albega}. The $I_k$'s for $k\le -2$ are not needed in the sequel, although they also follow from \eqref{eq:DownRecRel_psi}.

A note at the end of this section seems appropriate. The aforementioned asymptotics \eqref{eq:Hk_Asymptotics}, \eqref{eq:A_k} and \eqref{eq:A_expansion} arise as the result of a particular {\em ansatz} which appears to work by means of the asymptotic expansion of \eqref{eq:RecRel_phi} and \eqref{eq:RecRel_psi} and the subsequent application of the method of equating the coefficients. That is to say that we do not intend to provide a full proof of the aforementioned statements, but rather describe a particular route towards the solution of the ODE (\ref{eq:LaplaceTransfTeukEq}) in terms of a Taylor expansion analysis.

\subsection{Unique solutions of the Laplace 
transformed wave equation} \label{Sec:Unique_solutions}

For the construction of solutions to (\ref{eq:LaplaceTransfTeukEq}) for prescribed values $\ell,\lambda$ and $s\in\mathbb{C}$, we first consider two notations which will be useful in the sequel. For a sequence $\{a_k\}_{-\infty}^\infty$ we define the {\em discrete derivative} via the difference 
\beq\label{eq:discr_deriv} a'_k:=a_{k+1}-a_k.\eeq
Moreover, for two sequences $\{a_k\}_{-\infty}^\infty$ and $\{b_k\}_{-\infty}^\infty$ we introduce the {\em discrete Wronskian determinant} by:
\beq\label{eq:discr_Wronskian_ab}
W_k\left(\{a_k\},\{b_k\}\right):=a'_k b_k-a_k b'_k=a_{k+1}b_k-a_k b_{k+1}.
\eeq
In the following we discuss $W_k\left(\{H_k\},\{I_k\}\right)$ with the sequences $\{H_k\}_{-\infty}^\infty$ and $\{I_k\}_{-\infty}^\infty$ considered in the previous section, which we abbreviate by simply writing $W_k$. From \eqref{eq:RecRel_phi} and \eqref{eq:RecRel_psi} we find:
\bea
	 \nn 0&=&I_k\left(\alpha_k H_{k+1} + \beta_k H_k + \gamma_k H_{k-1}\right) \\ \nn && 
	 - H_k\left(\alpha_k I_{k+1} + \beta_k I_k + \gamma_k I_{k-1}\right)\\
	 &=& \alpha_k W_k-\gamma_k W_{k-1}\nn,
\eea
i.e.:
\beq
	\label{eq:discr_Wronskian_HI}
	 W_k= \frac{\gamma_k}{\alpha_k}W_{k-1}.
\eeq
We concentrate again on $s\notin\mathbb{Z}^-_0$ (see footnote~\ref{foot:Neg_s}). Also, we assume here that $s$ is not a QNM; these are just the values at which the construction to be described will fail, see Secs.~\ref{Sec:QNMs} and \ref{Sec:QNM_amplitudes}. Then we obtain nonvanishing regular values $W_k$ for $k\ge -1$:
\beq
	\label{eq:discr_Wronskian_HI_generic_s}
	 W_{-1}=I_{-1},\quad W_k= I_{-1}\prod_{j=0}^{k}\frac{\gamma_j}{\alpha_j} \quad\mbox{for $k\ge 0$}
\eeq
since $\alpha_j\ne 0$ and $\gamma_j\ne 0$ for $j\ge 0$. Moreover, we have $H_k=0$ for $k<0$ and hence $W_k=0$ for $k<-1$. 
With the scaling condition $H_0=1$, cf.~\eqref{eq:UpRecRel_phi}, we get $W_{-1}=I_{-1}$.
Note that in the following, the term
\bea
\nn\frac{1}{\alpha_j}\prod_{m=0}^{j}\frac{\alpha_m}{\gamma_m}&=&\frac{j!\, \Gamma (s) \Gamma (s+\lambda ) \Gamma (j+s-\lambda +1)}
{\Gamma(j+s+1) \Gamma (s-\lambda +1) \Gamma (j+s+\lambda +1)}\\
&=&j^{-(2 \lambda +s)} \frac{\Gamma (s) \Gamma (s+\lambda)}{\Gamma (s-\lambda +1)}\left[1+O(j^{-1})\right]\label{eq:alj_Wj}
\eea
occurs frequently.   
   Now we can address the solution to the inhomogeneous Laplace transformed wave equation (\ref{eq:LaplaceTransfTeukEq}). Let us start with the assumption that the initial data be of polynomial form, that is, for some integer $K_{\rm max}$ we may write:
\beq
\label{eq:SerieV0}
\begin{array}{lcccc}
V(0,\sigma)&=& V_0(\sigma) &=& 
{\displaystyle \sum\limits_{k=0}^{K_{\rm max}-1} v_k(1-\sigma)^k }\\[5mm]
V_{,\tau}(0,\sigma)&=&\dot{V}_0(\sigma) &=& 
{\displaystyle \sum\limits_{k=0}^{K_{\rm max}-1} w_k(1-\sigma)^k.}
\end{array}
\eeq
Later in the text, we will relieve this restriction to allow for initial data that are analytic for $\sigma\in[0,1]$.

Writing accordingly the Laplace transform $\hat V$ as
\beq
\label{eq:Series_hat_V}
\hat V(\sigma;s) = \sum_{k=0}^{\infty} a_{k}(1-\sigma)^k,
\eeq
we arrive via eq.~(\ref{eq:LaplaceTransfTeukEq}) at the recursion relation:
\beq\label{eq:RecRel}
	\alpha_k a_{k+1} + \beta_ka_k + \gamma_k a_{k-1}=B_k,\quad a_k=0\quad\mbox{for $k<0$},
\eeq
with
\bea	 
	B_k&=& (k+1) v_{k+1} -2\,(2k+s+1) v_k  \nn \\ && + (2k+\lambda +s) v_{k-1} -2  w_k +  w_{k-1}\label{eq:B_k}
\eea
and $B_k=0$ for $k<0$ and $k>K_{\rm max}$. It turns out that 
\beq\label{eq:ansatz_ak}
	a_k=c_k H_k + C_k I_k
\eeq
with the condition
\beq\label{eq:cond_ck_Ck_1}
	c'_k H_k + C'_k I_k = 0,
\eeq
where the coefficients $c_k$ and $C_k$ are to be determined, is a well-functioning ansatz for all $k\in\mathbb{Z}$ for which $W_k\ne 0$, i.e.~for $k\ge -1$. Note that from \eqref{eq:ansatz_ak} we learn that $C_k=0$ for $k<0$ since then $a_k=H_k=0$ and  $I_k\ne 0$. Moreover, the coefficients $c_k$ are arbitrary and undetermined for $k<0$. In addition, when looking at \eqref{eq:cond_ck_Ck_1} for $k=-1$ and taking $H_{-1}=C_{-1}=0\ne I_{-1}$ into account, we see that also $C_0=0$. 

If we now insert \eqref{eq:ansatz_ak} into \eqref{eq:RecRel}, thereby considering \eqref{eq:RecRel_phi}, \eqref{eq:RecRel_psi} and \eqref{eq:cond_ck_Ck_1}, we get for $k\ge -1$:
\beq\label{eq:cond_ck_Ck_2}\alpha_k\left(H_{k+1}c'_k + I_{k+1} C'_k\right)=B_k.\eeq
As $\alpha_{-1}=B_{-1}=0$, this equation is trivially satisfied for $k=-1$. For $k\ge 0$ the equations \eqref{eq:cond_ck_Ck_1} and \eqref{eq:cond_ck_Ck_2} can be written as 
\beq\label{eq:cond_ck_Ck_3}
	\left(\begin{array}{cc}
		H_k & I_k   \\ H_{k+1} & I_{k+1}
	\end{array}\right)
	\left(\begin{array}{c}
		c'_k \\ C'_k
	\end{array}\right)
	=
	\left(\begin{array}{c}
		0 \\ \frac{B_k}{\alpha_k} 
	\end{array}\right)	
\eeq
with the solution
\beq\label{eq:sol_ck_Ck_1}
	c'_k=\frac{I_kB_k}{\alpha_k W_k},\qquad C'_k=-\frac{H_k B_k}{\alpha_k W_k},
\eeq
i.e.~by virtue of $C_0=0$:
\beq\label{eq:sol_ck_Ck_2}
	c_k = c_0+\sum_{j=0}^{k-1}\frac{I_j B_j}{\alpha_j W_j},\quad C_k = -\sum_{j=0}^{k-1}\frac{H_j B_j}{\alpha_j W_j}.
\eeq
In order to determine the constant $c_0$ appearing in this context, we require that $a_k\sim I_k$ as $k\to\infty$ 
for polynomial initial data. This ensures that for $\Re(s)>0$ the Laplace transform $\hat V$ is $C^\infty$ at $\sigma=0$ and does not blow up exponentially there. We remark that only such solutions $\hat V$ can be taken to build up the wave field $V$ via the Bromwich integral \eqref{eq:BromInt}.

From \eqref{eq:sol_ck_Ck_2} we learn that for polynomial initial data \eqref{eq:Series_hat_V} we have for $k>K_{\rm max}$:
\bea\label{eq:sol_ck_Ck_3}
	c_k &=& c_0+\sum_{j=0}^{K_{\rm max}}\frac{I_j B_j}{\alpha_j W_j}=:c^*={\rm const.},\\ 
	C_k &=& -\sum_{j=0}^{K_{\rm max}}\frac{H_j B_j}{\alpha_j W_j}={\rm const.}
\eea
Now, the requirement $a_k\sim I_k$ as $k\to\infty$ means that $c^*$ must vanish (cf.~\eqref{eq:ansatz_ak}), which leads us with \eqref{eq:discr_Wronskian_HI_generic_s} to the final solution
\bea
	a_k&=&-\frac{1}{I_{-1}}\left(H_k\sum_{j=k}^{K_{\rm max}}\frac{I_jB_j}{\alpha_j}\prod_{m=0}^j\frac{\alpha_m}{\gamma_m} \nn \right.\\ 
	&&\left. \hspace{12mm}+I_k\sum_{j=0}^{k-1}\frac{H_jB_j}{\alpha_j}\prod_{m=0}^j\frac{\alpha_m}{\gamma_m}\right). \label{eq:sol_ak}
\eea
In this expression, the limit $K_{\rm max}\to \infty$ is easily performed, allowing for initial data whose complex extension is analytic within the circle $\mathbf{C}$. Indeed, if we write
\beq\label{eq:sol_ak_Green}a_k=\sum_{j=0}^\infty G_{kj}B_j\eeq
with
\beq\label{eq:discr_Green}
	G_{kj}=-\frac{1}{I_{-1}}\times\left\{
	\begin{array}{cc}
	\dfrac{I_k H_j}{\alpha_j}\prod\limits_{m=0}^j\dfrac{\alpha_m}{\gamma_m} & \mbox{for $j< k$}\\[4mm]
	\dfrac{H_k I_j}{\alpha_j}\prod\limits_{m=0}^j\dfrac{\alpha_m}{\gamma_m} & \mbox{for $j\ge k$} 
	\end{array}
	\right.,
\eeq
we obtain the implication:
\beq\label{eq:epsilon_for_B}\lim_{j\to\infty}\left|\frac{B_{j+1}}{B_j}\right|=\epsilon\quad\Longrightarrow \quad
		\lim_{j\to\infty}\left|\frac{G_{k(j+1)}B_{j+1}}{G_{kj}B_j}\right|=\epsilon,\eeq
and hence convergence in \eqref{eq:sol_ak_Green} for all $k\ge 0$ in the case of initial data which are  analytic within the circle $\mathbf{C}$ and thus satisfy $\epsilon<1$. Note that \eqref{eq:sol_ak_Green} presents a representation of the solution $a_k$ of \eqref{eq:RecRel} in terms of the {\em discrete Green's function} $G_{kj}$ given in \eqref{eq:discr_Green}. This name is justified because of the relation 
\beq\label{eq:RecRel_Green}
	\alpha_k G_{(k+1)m} + \beta_k G_{km} + \gamma_k G_{(k-1)m}=\delta_{km}, 
\eeq
which follows from \eqref{eq:sol_ak_Green} and \eqref{eq:RecRel} for $B_j=\delta_{jm}$.

Let us now discuss the case in which the initial data are analytic for all $\sigma\in[0,1]$ but whose complex extension is {\em not} analytic within the circle $\mathbf{C}$, i.e.~$\epsilon>1$ in \eqref{eq:epsilon_for_B}. Then the series \eqref{eq:sol_ak_Green} does not converge. Yet, in order to determine a definite limit even in this case we introduce
\beq\label{eq:ak_of_x_Green}a_k(x)=\sum_{j=0}^\infty G_{kj}B_j x^j\eeq
which is a Taylor series that converges for $|x|<1/\epsilon$. The corresponding function $a_k(x)$ may, however, be defined on the entire interval $x\in[0,1]$, and we obtain a good approximation again via a diagonal Pad\'e approximant
\beq\label{eq:ak_of_x_Pade} a_k^{\rm Pade}(x)= \frac{\sum\limits_{j=0}^{j_{\rm max}}p_j x^j}{1+\sum\limits_{j=1}^{j_{\rm max}}q_jx^j} \eeq
where the coefficients $p_j, q_j$ (different from the ones in \eqref{eq:A_Pade}) are determined such that 
\beq\label{eq:djak}\frac{d^j a_k^{\rm Pade}}{dx^j}(0)= \frac{d^j a_k}{dx^j}(0)=j!\,G_{kj}B_j\eeq
for $j=0,\ldots,2j_{\rm max}$. The values $a_k^{\rm Pade}(x=1)$ serve us as good approximations for the $a_k$, and we thus obtain the Taylor series \eqref{eq:Series_hat_V} which describes a unique solution $\hat V$ to the inhomogeneous ODE (\ref{eq:LaplaceTransfTeukEq}). As the coefficients $B_j$ of the initial data appearing in \eqref{eq:LaplaceTransfTeukEq} are characterized by a specific radius $1/\epsilon<1$ of convergence of the associated Taylor series, we expect that also the coefficients $a_k$ are subject to that convergence radius, meaning that \eqref{eq:Series_hat_V} is valid only for $|1-\sigma|<1/\epsilon$. Similar to the treatment above, we establish $\hat V(\sigma; s)$ in the entire range $\sigma\in[0,1]$ by utilizing once more a diagonal Pad\'e-approximant (here with respect to the coordinate $\sigma$):
\beq\label{eq:hat_V_Pade} \hat V_{\rm Pade}(\sigma; s)= \frac{\sum\limits_{j=0}^{k_{\rm max}/2}p_j (1-\sigma)^j}
{1+\sum\limits_{j=1}^{k_{\rm max}/2}q_j(1-\sigma)^j} ,
\eeq
where the coefficients $p_j, q_j$ (different from the ones in \eqref{eq:A_Pade} and \eqref{eq:ak_of_x_Pade}) are determined such that 
$\hat V_{\rm Pade}$ agrees with $\hat V$ to the order $k_{\rm max}$, which amounts to the conditions
\bea\nn\frac{d^k \hat V_{\rm Pade}}{d\sigma^k}(1;s)
&=& \frac{d^k \hat V}{d\sigma^k}(1;s)=(-1)^kk!\,a_k\\ \label{eq:djV}&&\mbox{for $k=0,\ldots,k_{\rm max}$}.
\eea

\begin{figure*}[t!]
\begin{center}
\includegraphics[width=8.5cm]{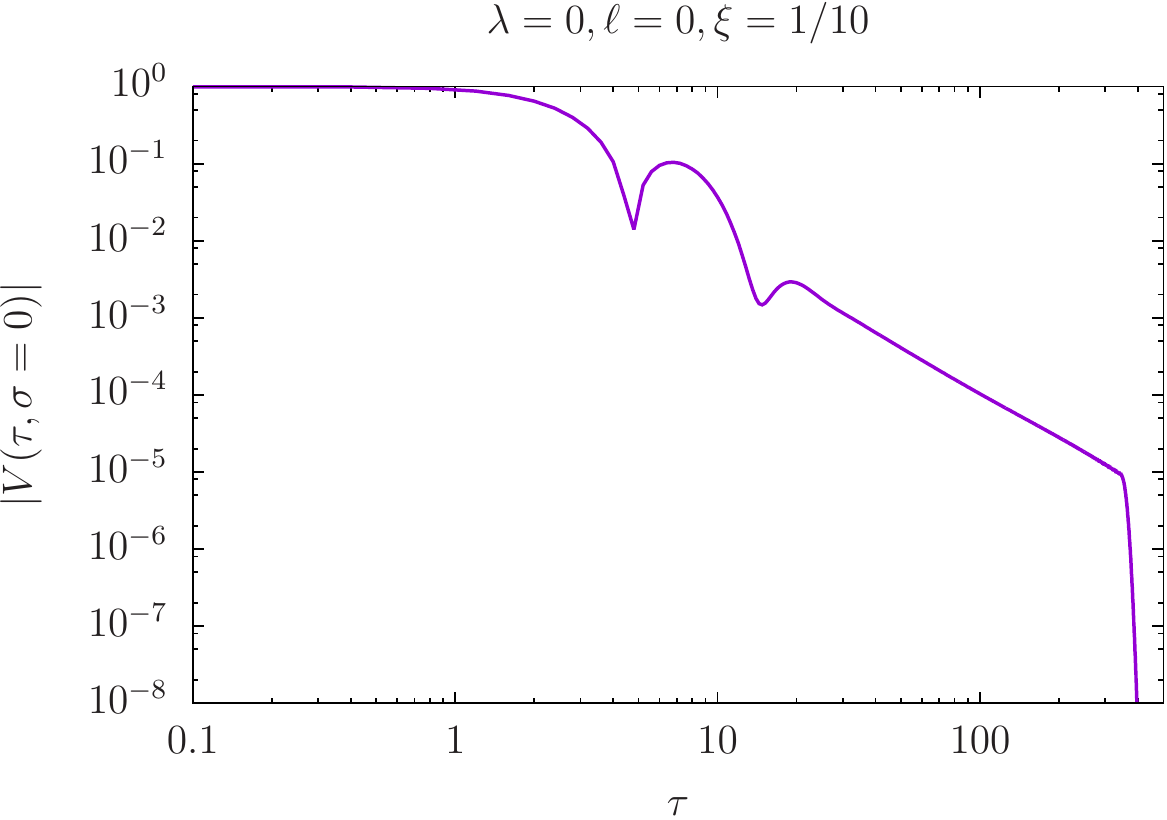}
\includegraphics[width=8.5cm]{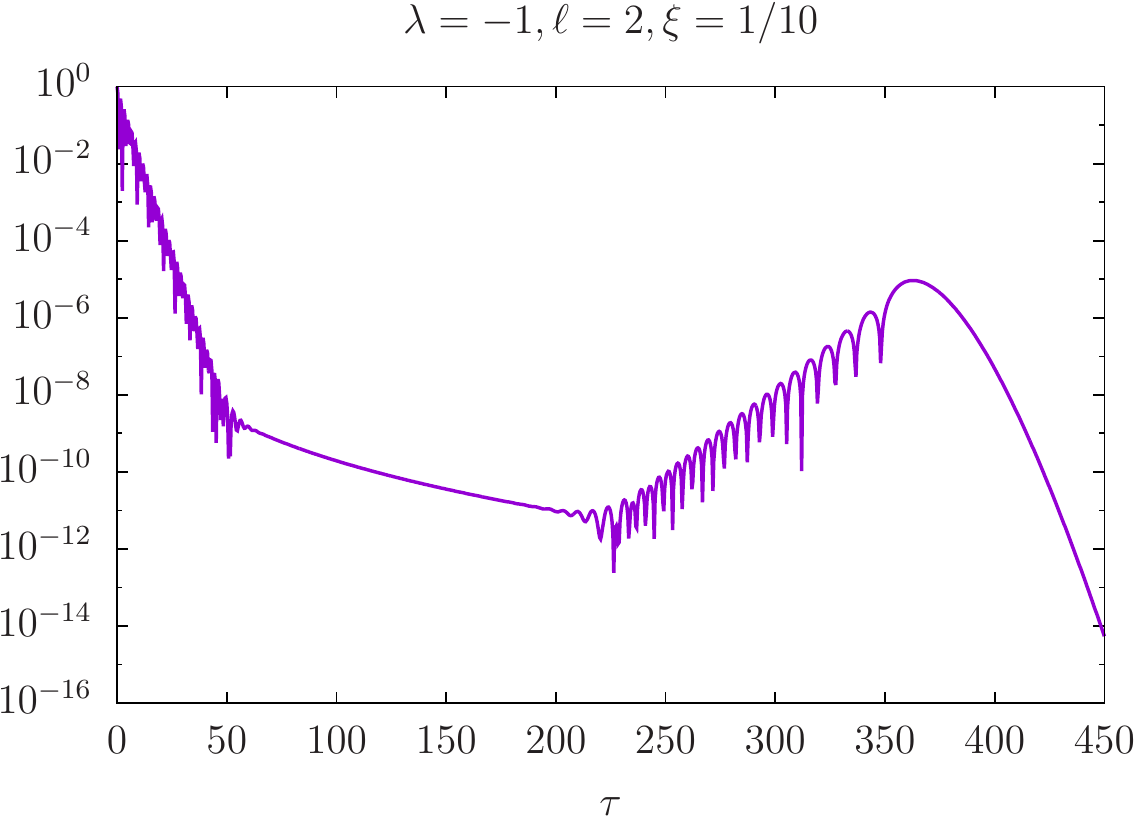}
\end{center}
\caption{Solution of the Barden-Press equation (\ref{eq:ReqTeukEq}) via the Bromwich integral method (\ref{eq:BromInt}) for the initial data $V_0(\sigma)=1, \dot{V}_0(\sigma)=0$.\\ Left panel: time evolution of a scalar perturbation with parameters $\lambda=0$ and $\ell = 0$.  Right panel: electromagnetic perturbation with parameters $\lambda = -1$ and $\ell = 2$. Both fields are evaluated at $\scri$, i.e.~at $\sigma = 0$, and the Bromwich integral is performed along the line $\Re(s) = 1/10$ (see fig.~\ref{fig:BromwichInt}).}
\label{fig:BromwitchIntegral}
\end{figure*}

The corresponding solutions $\hat V(\sigma; s)$ are analytic for $\sigma\in(0,1]$ and still $C^\infty$ at $\sigma=0$. Computed at the line (\ref{eq:Gamma_1}), they are perfectly suited to express the wave-field $V(\tau,\sigma)$ via the Bromwich integral (\ref{eq:BromInt}), as is done in Secs.~\ref{sec:BromIntResults} and \ref{App:BromwichInt}. 

\subsection{Numerical evaluation of the Bromwich integral}\label{sec:BromIntResults}

Before proceeding further on the way towards the desired spectral decomposition of the wave field, we investigate as a first application and test of the procedure described in Sec.~\ref {Sec:Unique_solutions} the numerical evaluation of the inverse Laplace transform. To this end, we compute the function $\hat V(s)$ for values of $s$ located on the path $\Gamma_1$ (see Fig.~\ref{fig:BromwichInt} and eq.~(\ref{eq:Gamma_1})) and construct from these the solution to the wave eq.~(\ref{eq:ReqTeukEq}) in the form of the Bromwich integral (\ref{eq:BromInt}). While appendix \ref{App:BromwichInt} contains more details regarding these calculations, we concentrate here on the discussion of the numerical behavior of this solution technique.

Fig.~\ref{fig:BromwitchIntegral} brings two examples of the time evolution for initial data prescribed by $V_0(\sigma) = 1$ and $\dot{V}_0(\sigma) = 0$. The first one represents a scalar perturbation with parameters $\lambda = 0$ and $\ell = 0$ while the second one corresponds to an electromagnetic perturbation  with $\lambda = -1$ and $\ell = 2$. The Bromwich integral is evaluated along the line $\Re(s) = 1/10$, and we plot $V(\sigma=0,\tau)$, i.e.~the dynamics of the wave field at $\scri$. The time evolution of the first example shows an early tail decay whereas the second one possesses a long-lasting ring-down phase. 

In the figures it becomes apparent that the Bromwich integral method, when computed along the path $\Gamma_1$,  does not seem to be well suited for the study of the wave's late time behavior. This observation is a consequence of the systematic errors introduced by the discretisation of the integral. As discussed in appendix VII A, within our approach the numerical solution tends to zero exponentially which prohibits the resolution of the tail at very large times. In contrast, the spectral decomposition of the wave field, to be derived in the sequel, does provide in principle an arbitrarily accurate description of the very late time tail. Note that all the results in this section were obtained with the fixed Taylor resolution $k_{\rm max} = 500$, $J_{\rm max} = 10$. Moreover, the resolution for the numerical evaluation of the integral~(\ref{eq:BromInt}) was set to $n_{\chi} = 200$ (see appendix \ref{App:BromwichInt}).

\subsection{quasinormal modes} \label{Sec:QNMs}

\begin{figure*}[t]
\begin{center}
\includegraphics[height=5.0cm]{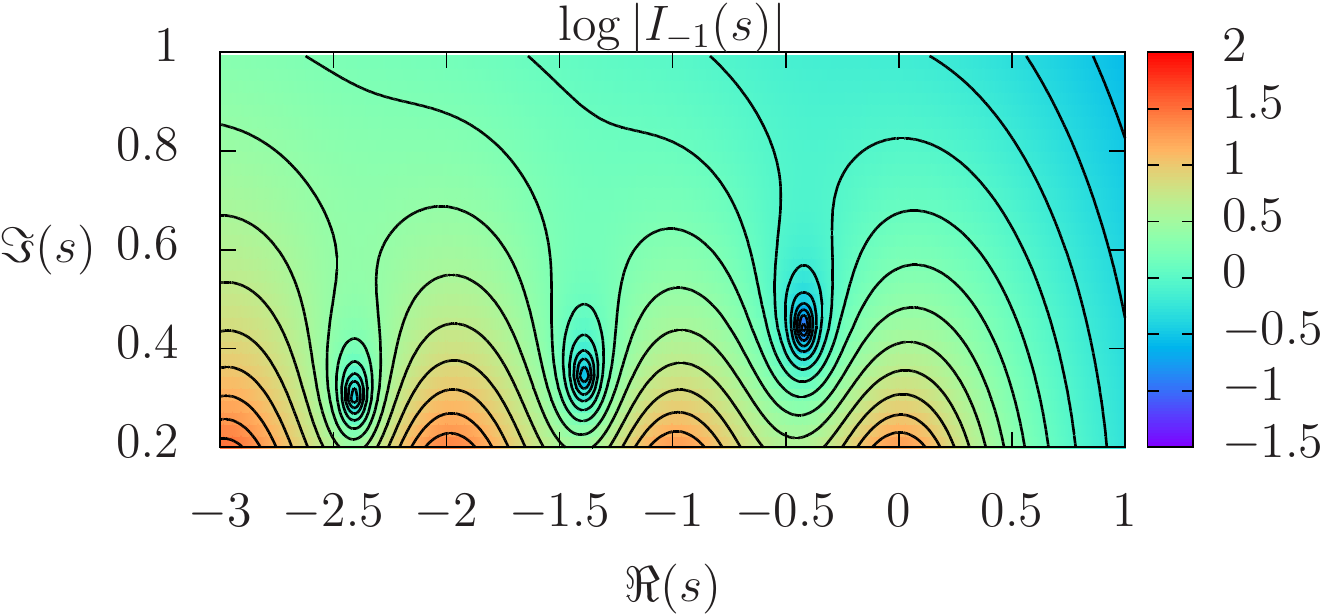}
\end{center}
\caption{Contour plot of the function $\log|I_{-1}(s)|$ in the complex $s$-plane for $\lambda=\ell=0$. The centres of the concentrically  arranged closed curves are the locations of the quasinormal modes $s_n$, shown here for $n\in\{0,1,2\}$. At these points, $I_{-1}(s)$ vanishes.}
\label{fig:QNM_contourplot}
\end{figure*}

\subsubsection{Approach within the Cauchy formulation}\label{Sec:QNMs_Cauchy}

We start by reviewing the definition of QNMs as usually expressed in the literature (see, for instance,~\cite{Kokkotas99a}) and describe why a direct computation according to this characterization seems to pose essential technical difficulties when analysed in the context of perturbations in asymptotically flat black-hole spacetimes.

Working with the Cauchy coordinates $\{\bar{t}, x\}$ introduced in sec.~\ref{sec:CauchyCompLaplace}, one first considers in the region $\Re{(\bar{s})}>0$ the homogeneous equation \eqref{eq:HomLaplCauchy}
which has two linearly independent solutions $F^\pm(x;\bar{s})$, chosen in such a way that they stay bounded as $x\rightarrow \pm \infty$. More specifically, the solutions satisfy
\beq
\label{eq:RegCondCauchy}
\lim_{x\rightarrow \pm \infty} \left| e^{\pm x\bar{s}}F^\pm(x;\bar{s})\right| = 1.
\eeq

Then one constructs the Wronskian determinant\footnote{For equations of the form \eqref{eq:HomLaplCauchy}, the Wronskian determinant does not depend on the coordinate $x$~\cite{Nollert99}.}
\beq
\label{eq:WronskianCauchy}
\bar{\cal W}(\bar{s}) = F^+_{,x}(x;\bar{s})F^-(x;\bar{s}) 
- F^-_{,x}(x;\bar{s})F^+(x;\bar{s})
\eeq
and analytically extends it onto the half-plane $\Re(s)<0$. The QNMs are then defined as the values $\bar{s}_n$ for which $\bar{\cal W}(\bar{s}_n)=0$.

In order to compute $\bar{\cal W}$ explicitly as defined above\footnote{As mentioned in sec.~\ref{sec:CauchyCompLaplace}, the comparison between both approaches is performed for $\lambda=0$. We furthermore recall that the Laplace parameters are related by $s=2\bar{s}$.}, we relate the two functions $F^\pm$ to solutions $\Omega(s)$ of the homogeneous Laplace transformed equation ${\mathbf A}(s) \Omega(s)=0$ (cf.~\eqref{eq:HomogLaplaceTransfTeukEq_phi} and \eqref{eq:HomogLaplaceTransfTeukEq_psi}) appearing in our hyperboloidal framework. From the exponential fall-off of $F^-$ when $x\to-\infty$ we conclude that $F^-$ can be written in terms of $\phi$. Concretely, with the help of eq.~\eqref{eq:Rel_f_V} and the regularity condition \eqref{eq:RegCondCauchy} we find
\beq
	F^-(x(\sigma); \bar{s}) = \frac{e^{2\bar s}}{2}\,\Xi(\sigma; \bar{s}) \, \phi(\sigma; 2\bar{s}) \label{eq:f_minus_phi} 
\eeq
with the function $\Xi$ defined in \eqref{eq:HypCauchyFactor}. 
For the second solution $F^+$ that needs to satisfy the regularity condition \eqref{eq:RegCondCauchy} we have to take
\beq\label{eq:f_plus}
	F^+(x(\sigma); \bar{s})= \Xi(\sigma; \bar{s}) \Psi(\sigma; 2\bar s),
\eeq
where 
\beq\label{eq:Psi} 
\Psi(\sigma; s)=\frac{s}{2}\Phi(0;s)\phi(\sigma;s)\int\limits_{0}^\sigma e^{s/\tilde\sigma}
	\frac{[\tilde\sigma(1-\tilde\sigma)]^{-(s+1)} d\tilde\sigma}{\tilde{\sigma}[\phi(\tilde\sigma;s)]^2} 
\eeq
is a solution to \eqref{eq:HomogLaplaceTransfTeukEq_psi} that is $C^\infty$-regular at $\sigma=0$ for all $\Re(s)>0$ (recall that $\lambda=0$  here). Note that once more the function $\Phi$ appears which was defined in \eqref{eq:Phi}. 

We now can write the Wronskian determinant $\bar{\cal W}(s)$ in \eqref{eq:WronskianCauchy} as
\beq
\bar{\cal W}(\bar{s}) =\frac{e^{2\bar s}}{2} \,\sigma^2 (1-\sigma) \, \Xi(\sigma; \bar{s})^2\, {\cal W}(\sigma; 2\bar{s})
\eeq
with ${\cal W}(\sigma; s)$ being the Wronskian determinant formed from  $\phi$ and $\Psi$:
\bea
\label{eq:Hyp_Wrosk}
{\cal W}(\sigma; s) &=& \phi_{,\sigma}(\sigma;s) \, \Psi(\sigma;s) - \Psi_{,\sigma}(\sigma;s) \, \phi(\sigma;s) \nn \\
			   &=&  \frac{s}{2}\Phi(0;s) \, e^{s/\sigma} \sigma^{-(s+2)}(1-\sigma)^{-(s+1)}.
\eea
We finally find
 \beq
\bar{\cal W}(\bar{s}) = 2\bar{s}\, e^{2\bar{s}}\,\Phi(0;2\bar{s}), \label{eq:WroskCauchy}
\eeq
which depends solely on $\bar{s}$, as expected. Following the discussion made after the introduction of $\Phi(\sigma; s)$ in \eqref{eq:Phi}, we see that $\bar{\cal W}(\bar{s})$ is well defined for $\Re(\bar{s})>0$. 
However, the values $\Phi(\sigma=0;2\bar s)$ are involved which appear to be determinable only numerically to some accuracy. Now, they are supposed to constitute a function that is well-defined on the right complex half-plane $\Re(\bar{s})>0$. In order to determine the set of QNMs it would then be necessary to continue this function analytically to the left half-plane $\Re(\bar{s})<0$ and identify there its set of zeros. As an example, if we simply had $\Phi(\sigma; 2\bar s)=e^{-\bar s/\sigma}+1+\bar s$ then we would get $\Phi(\sigma=0;2\bar s)=1+\bar s$ for $\Re(\bar{s})>0$, and this can of course trivially be analytically expanded to the entire $\bar s$-plane, despite the fact  that $\Phi(0; 2\bar s)$ does not exist for $\Re(\bar{s})<0$.  On the technical side it appears extremely complicated to provide a reasonable and sound, sufficiently accurate continuation of the numerically determined function values $\Phi(\sigma=0;2\bar s)$ onto the left half-plane. Hence we feel that the concrete determination according to the usual definition of QNMs given in \cite{Kokkotas99a} needs to be performed in some different, indirect manner.

In the next section we provide a working definition of the QNMs which results in the same expressions as the ones used by Leaver in his approach utilizing continued fractions ~\cite{Leaver85}. In particular, we derive that the QNMs can be characterized through the vanishing of the Wronskian determinant 
\beq
\label{eq:Wronski_phi_psi}
\mathbb{W}(\sigma; s) = \phi_{,\sigma}(\sigma;s) \, \psi(\sigma;s) - \psi_{,\sigma}(\sigma;s) \, \phi(\sigma;s)
\eeq
of the two solutions $\phi$ and $\psi$ 
to \eqref{eq:HomogLaplaceTransfTeukEq_psi} that were discussed extensively in Sec.~\ref{sec:Homogeneous_Tayl_asymptotics}. 
Note that in order to analyse whether this approach is equivalent to definition of QNMs given in \cite{Kokkotas99a} it would be necessary to show that
$\Psi$ is proportional to $\psi$, i.e.~to prove that 
$\psi$, when considered for {\em real} $\sigma\gtrsim 0$, is $C^\infty$-regular at $\sigma=0$  for $\Re(s)>0$. Even a mere numerical check of this equivalence would pose again substantial technical difficulties, see discussion in Sec.~\ref{sec:Homogeneous_Tayl_asymptotics}.

\subsubsection{quasinormal modes as zeros of the discrete Wronskian determinant}\label{Sec:QNMs_from_discr_Wronskian}

\begin{figure*}[t!]
\begin{center}
\includegraphics[width=8.0cm]{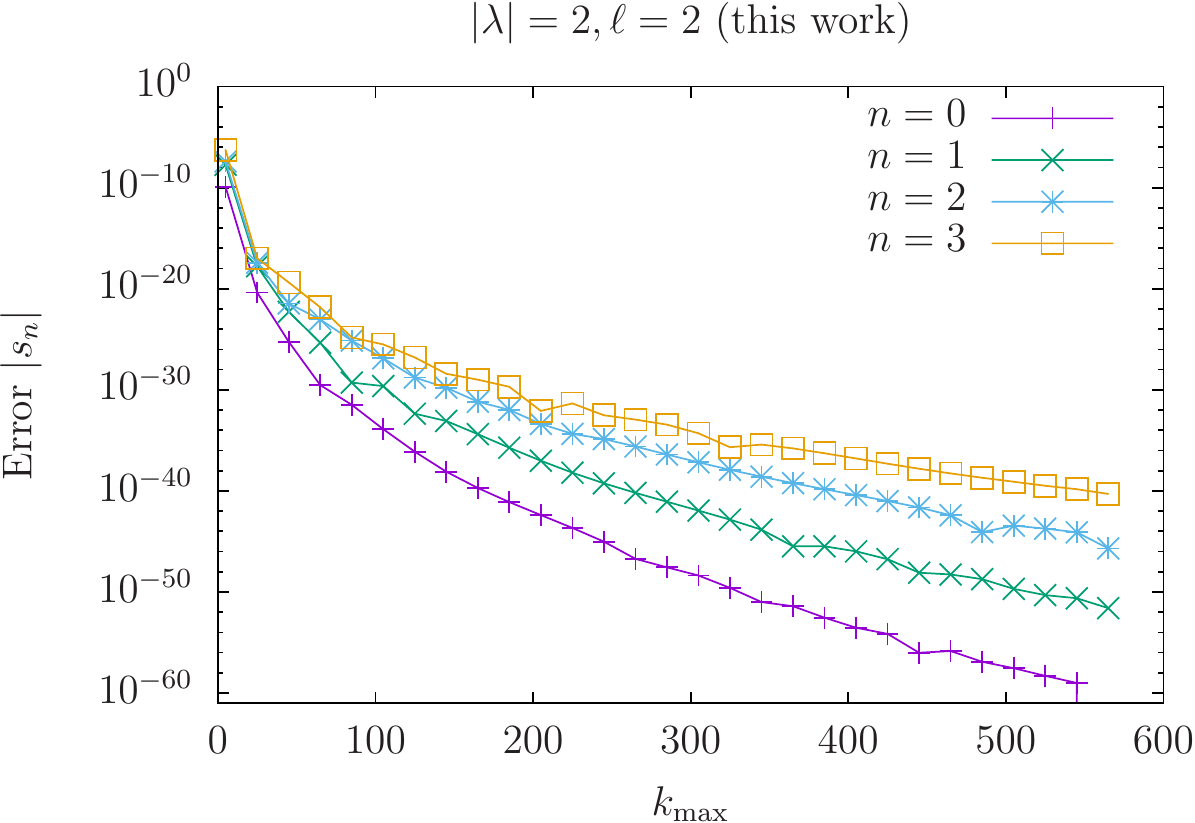}
\includegraphics[width=8.0cm]{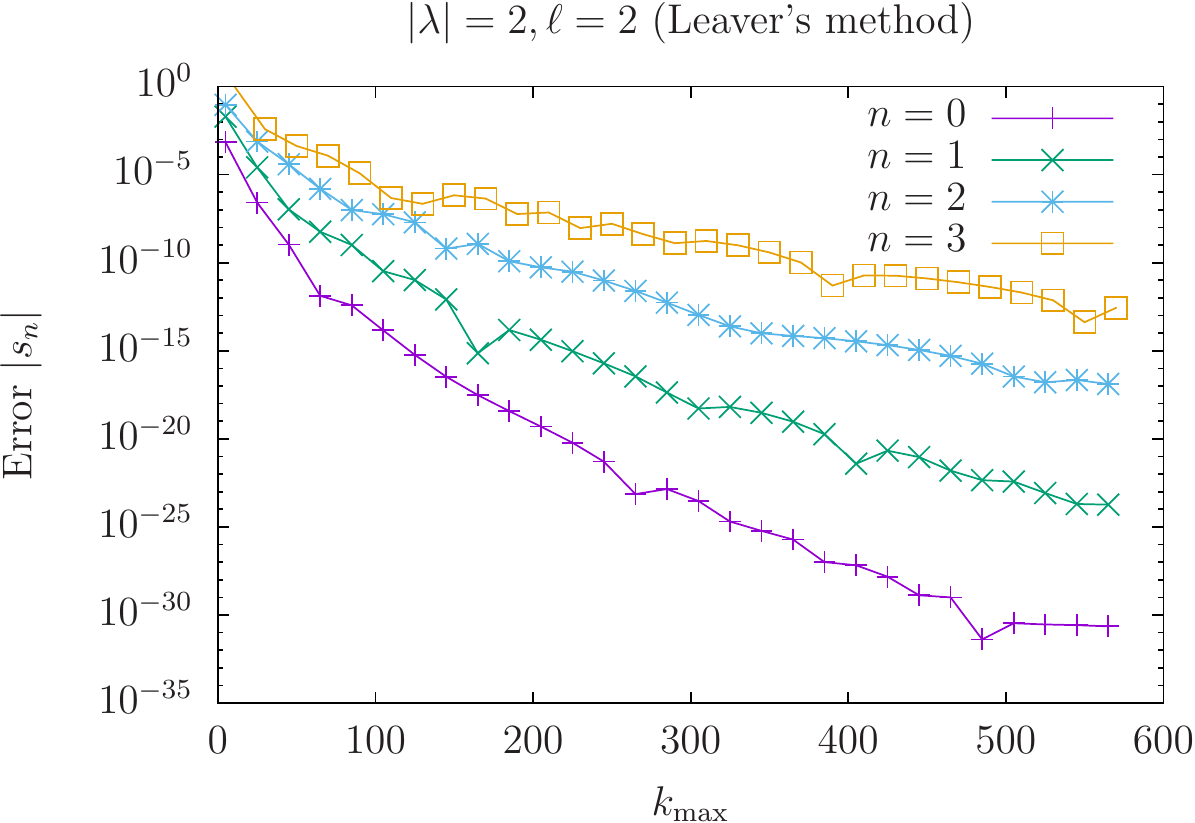}
\end{center}
\caption{Errors in the numerical determination of the quasi-normal modes $s_n$, displayed as functions of the truncation number $k_{\rm max}$ for a gravitational perturbation with parameters $|\lambda|=2$ and $\ell=2$. The left panel shows the results according to the algorithm presented in this work (with $J_{\rm max} = 10$), while the right panel brings the corresponding value obtained with Leaver's method of continued fractions. It becomes apparent that the errors are drastically reduced by taking the asymptotic behaviour (\ref{eq:A_k}), \eqref{eq:A_expansion} of the coefficients $I_k$ into account.}
\label{fig:QNM_ComparisionLeaver}
\end{figure*}

quasinormal modes $s_n$ are specific values in the complex $s$-plane for which the procedure, described in Sec.~\ref{Sec:Unique_solutions} to determine a unique solutions of the Laplace transformed wave equation, fails. 
The reason for the failure is given by the fact that 
\beq\label{eq:QNM_I_m1_=0}
I_{-1}(s_n)=0,
\eeq
from which it follows that the construction \eqref{eq:sol_ak_Green}, \eqref{eq:discr_Green} cannot be performed. The zeros of the function $I_{-1}(s)$ have always been found to be distributed discretely in the left half-plane $\Re(s)<0$. Fig.~\ref{fig:QNM_contourplot} shows a contour plot of $I_{-1}(s)$ for $\lambda = \ell = 0$. 

We immediately arrive at several equivalent characterizations:
\ben
\item {\em For $s=s_n$ being a QNM, the discrete Wronskian determinant $W_k\left(\{H_k\},\{I_k\}\right)$ vanishes for all $k\in\mathbb{Z}$.} 

This follows directly from \eqref{eq:discr_Wronskian_HI_generic_s}.
\item {\em For $s=s_n$ being a QNM, the Wronskian determinant $\mathbb{W}(\sigma;s_n)$, as defined in \eqref{eq:Wronski_phi_psi} in terms of the two solutions $\phi$ and $\psi$ introduced in Sec.~\ref{sec:Homogeneous_Tayl_asymptotics}, vanishes for all $\sigma\in[0,1]$.}

With \eqref{eq:QNM_I_m1_=0} we obtain that $I_k=0$ for all $k<0$ (see \eqref{eq:DownRecRel_psi} with $\alpha_{-1}=0$, cf.~\eqref{eq:RecRelCoef_albega}) and hence $I_k=I_0 H_k$, i.e.~the two sequences $\{H_k\},\{I_k\}$ are linearly dependent. As a consequence, the associated functions $\phi$ and $\psi$ satisfy $\psi=I_0\phi$, and hence $\mathbb{W}(\sigma;s_n)=0$, cf.~\eqref{eq:Wronski_phi_psi}.

\item\label{QNM_charct_3} {\em A QNM $s_n$ is defined by the existence of a nontrivial solution $\phi=\phi_n$ to \eqref{eq:HomogLaplaceTransfTeukEq_phi}
that possesses a Taylor expansion \eqref{eq:Series_phi} with  rapidly decreasing coefficients. That is to say that there is for each $\nu\in\mathbb{N}$ a positive constant $C_\nu$ such that for all $k\in\mathbb{N}$:
\beq\label{eq:p_k_faster_than_algebraic}
	|H_k(s_n)|<\frac{C_\nu}{k^\nu}.
\eeq
}
This formulation follows directly from the fact that $H_k=I_k/I_0 \sim e^{-\kappa\sqrt{k}}k^{\zeta}$ for $k \to\infty$, see \eqref{eq:RecRel_psi}.
\een
Among these three characterizations, point No.~3 plays a preferred role. Points 1.~and 2.~make use of the two sequences $\{H_k\}$ and $\{I_k\}$ of coefficients which cannot be defined in its entirety for $s\in\mathbb{Z}^-$, see discussion in Sec.~\ref{App:Neg_Lapl_Par}. This means that the negative integer, so-called  {\em algebraically special} QNMs $s_{(\ell)}$ \cite{Chandrasekhar:1984} (to be treated in Sec.~\ref{Sec:AlgSpecial}) is excluded in points 1.~and 2 (in the definition of $H_k$ and $I_k$ in Sec.~\ref{sec:Homogeneous_Tayl_asymptotics} we have set $s\notin\mathbb{Z}_0^-$). However, point No.~3 still applies since the corresponding solutions $\phi_{(\ell)}$ are polynomials and satisfy therefore trivially this formulation. We conclude that No.~3 should be regarded as generically valid definition of QNMs for perturbations in the asymptotically flat Schwarzschild spacetime.

For the numerical computation of the QNMs $s_n\notin\mathbb{Z}_0^-$, Leaver~\cite{Leaver85} has looked at the recursion relation 
\beq
\label{eq:RecRel_b2}
b_{k-1}=- \frac{\gamma_k}{\beta_k+\alpha_k b_k}\qquad\mbox{with}\quad b_k:=\frac{I_{k+1}}{I_k},
\eeq
through which the coefficients $b_k$ can be written in terms of continued fraction expressions. Starting with $b_{k_{\rm max}}=1$ for some very large value $k_{\rm max}$ and climbing down via \eqref{eq:RecRel_b2} one obtains through the condition 
\[0=\frac{1}{I_0}\left(\alpha_0I_1+\beta_0 I_0+\gamma_0I_{-1}\right)=\alpha_0 b_0+\beta_0\]
 an equation that determines the QNMs $s_n$. In practice, this equation can be solved numerically using the \texttt{FindRoot}-command provided within \texttt{Mathematica} (see e.g.~\cite{CardosoWebSite}). This method achieves in principle arbitrary precision. However, for very accurate calculations, extremely large vales $k_{\rm max}$ and a high internal working precision of the \texttt{Mathematica} notebook need to be chosen. The costs of this numerical calculation can be reduced drastically by making use of the approximation function $A_{\rm Pade}$ introduced in Sec.~\ref{sec:Homogeneous_Tayl_asymptotics} above. As described therein, given a suitable expansion order $J_{\rm max}$ we can choose a moderate number $k_{\rm max}$ (say about 400) to obtain extremely accurate values $I_k$, in particular for $I_{-1}$. The zero $s_n$ is then found by a Newton-Raphson scheme
\beq
	s_n=\lim_{j\to\infty} s^{(j)}_n,\qquad s^{(j+1)}_n=s^{(j)}_n-\frac{I_{-1}(s^{(j)}_n)}{(\partial_s I_{-1})(s^{(j)}_n)},
\eeq
where a suitable initial guess $s^{(0)}_n$ is needed. Note that for the numerical computations, the derivative $\partial_s I_{-1}$ can be approximated through a finite difference expression. An illustrative example demonstrating the accuracy and performance of this calculation in comparison with Leaver's method is shown in fig.~\ref{fig:QNM_ComparisionLeaver}.

As an additional note, we remark that \eqref{eq:HomogLaplaceTransfTeukEq_psi} has, for $s=s_n$ being a QNM, a second solution $\Lambda$ which is linearly independent of $\phi_n$. This solution $\Lambda$ can be described similarly as $\psi$ in Sec.~\ref{sec:Homogeneous_Tayl_asymptotics} by a series $\{L_k\}_{-\infty}^\infty$ with
\bea\nn
	\alpha_k L_{k+1} + \beta_k L_k + \gamma_k L_{k-1} 	&=& 0,\\ 
	 \lim_{k\to\infty} L_k e^{-\kappa\sqrt{k}}k^{-\zeta}&=&1,\label{eq:RecRel_xi}
\eea
as $\Lambda=\Lambda_+ + \Lambda_-$, where
\beq\label{eq:xi_pm}
		\Lambda_-=\sum_{m=1}^\infty \frac{L_{-m}}{(1-\sigma)^m},\quad \Lambda_+=\sum_{k=0}^\infty L_k(1-\sigma)^k.\eeq
Observe the diverging asymptotics of the $L_k$ in \eqref{eq:RecRel_xi}, and hence $\Lambda$ does not present a solution of the kind described in No.~\ref{QNM_charct_3} of the above characterization. Furthermore, as $L_k\ne 0$ for $k<0$, $\Lambda$ is singular at $\sigma=1$. We conclude that the entirety of solutions to \eqref{eq:HomogLaplaceTransfTeukEq_phi} is given by ${\rm span}_\mathbb{C} (\phi_n)$. This point will be relevant below, (see Sec.~\ref{Sec:QNM_amplitudes}).

We finally mention that, as a consequence of the symmetry relation (\ref{eq:Symmetry_hat_V}) we find that the quasinormal modes $s_n$ come in complex-conjugated pairs, $s_n$ and $s_n ^*$.

\subsection{quasinormal mode amplitudes} \label{Sec:QNM_amplitudes}

What happens to the solutions $\hat V=\hat V(s)$ of the inhomogeneous ODE (\ref{eq:LaplaceTransfTeukEq}), computed in sec.~\ref{Sec:Unique_solutions}, if we approach a QNM, i.e.~if $s\to s_n$ (here again $s_n\notin\mathbb{Z}^-$)? As then $I_{-1}\to 0$, eqn.~\eqref{eq:sol_ak}
suggests that $\hat V(s)$ will diverge in this limit. Assuming that $I_{-1}$ has merely a {\em single} zero at $s=s_n$ \footnote{This assumptions has always been found to be realized.}, the corresponding $\hat V(s)$ will possess a single pole at $s=s_n$. Accordingly, we write for values $s$ close to $s_n$:
\beq\label{eq:ak_Pole}
	a_k(s)=\frac{h_k}{s-s_n}+g_k(s),
\eeq
where $h_k$ does not depend on $s$ (in contrast to $a_k$ and $g_k$). With the single polelike singularity of $\hat V$, the coefficient $h_k$ is supposed to be the residue of $a_k$ at $s_n$ whereas $g_k(s)$ presents the secondary part of its Laurent series. That is to say that the coefficients $g_k(s)$ are supposed to be analytic in the vicinity of $s_n$. As will become clear below, the residues $\{h_k\}$ turn out to be proportional to $\{H_k(s_n)\}$, and the corresponding proportionality factor $\eta_n$ will play a crucial role as {\em quasinormal mode amplitude} in the desired spectral decomposition \eqref{eq:VSol_spectral}.

\begin{figure*}[t!]
\begin{center}
\includegraphics[width=7.8cm]{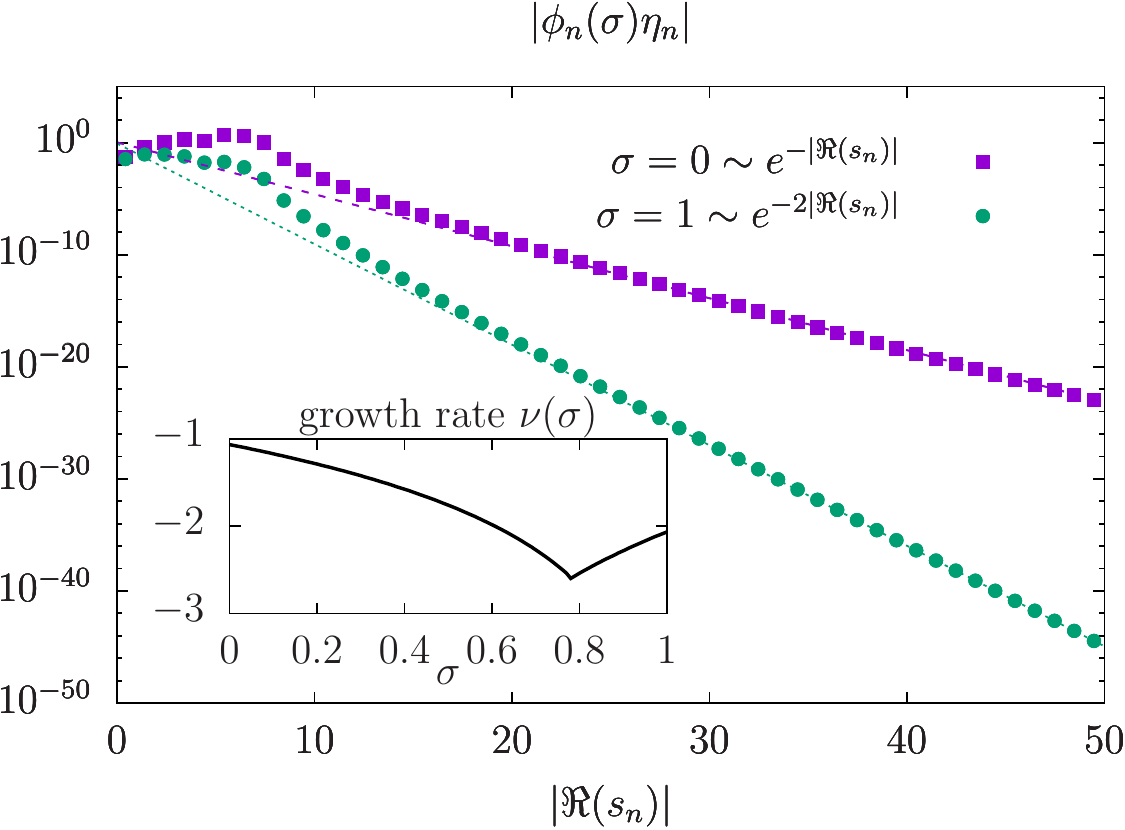}
\includegraphics[width=8.0cm]{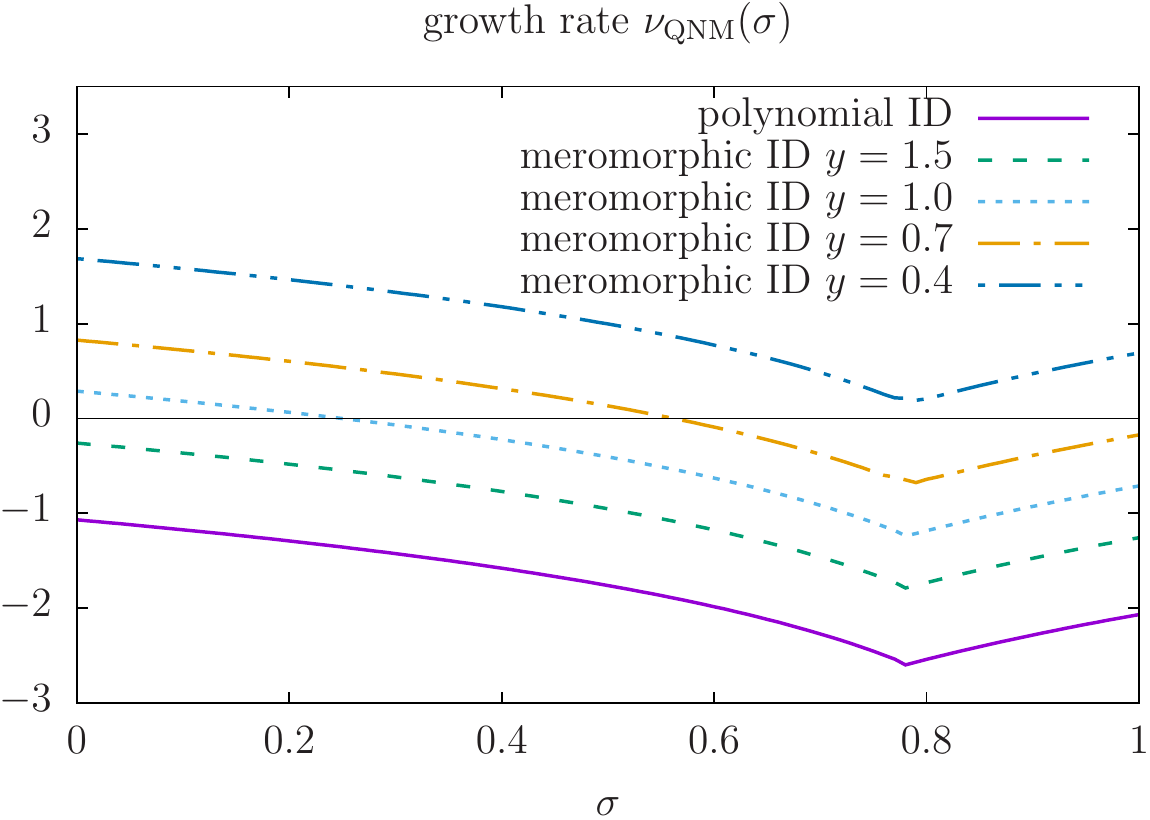}
\end{center}
\caption{Behaviour of the residues $\eta_n\phi_n(\sigma)$ of the Laplace transform $\hat V(\sigma;s)$ at the quasinormal modes $\{s_n\}$ with prescribed parameters $\lambda=l=0$. Left panel: for the polynomial initial data $V_0(\sigma) = \sigma^2 (1-\sigma)^4$, $\dot V_0(\sigma)=0$ the decay is shown for $\sigma=0$ (at $\scri$) and for $\sigma=1$ (at the horizon). The inset brings the {\em growth rate} $\nu_{\rm QNM}(\sigma)$ for all $\sigma\in[0,1]$. Right panel: comparison of the grow rate for the polynomial initial data against meromorphic initial data with single poles in the complex $\sigma$-plane. Here,  $V_0(\sigma) = 2\, \Re\left[ (\sigma - \sigma_{0})^{-1}\right]$, $\dot V_0(\sigma)=0$, with $\sigma_{0} = 0.5 + i\, y$. While polynomial initial data show a universal behaviour, more generic initial data introduce a constant shift in the profile.
}
\label{fig:Amplitude_s0l0_Inset}
\end{figure*}

For the computation of $\eta_n$, let us start with initial data that are analytic within the circle $\mathbf{C}$ (i.e.~for which $\epsilon<1$ in \eqref{eq:epsilon_for_B}) and generalize the corresponding expression in a subsequent step. We formulate the recursion relation (\ref{eq:RecRel}) as
\beq\label{eq:Aa=B}
{\cal A}(s)\cdot\{a_k\} =\{B_k\}
\eeq
with the operator ${\cal A}(s)$ defined by
\beq \big[{\cal A}(s)\cdot\{a_k\}\big]_k:=\alpha_k a_{k+1} + \beta_ka_k + \gamma_k a_{k-1}.\eeq
We now insert \eqref{eq:ak_Pole} into the relation \eqref{eq:Aa=B}, thereby rewriting \eqref{eq:Aa=B} by using another operator ${\cal C}_n$ which is given through
\beq {\cal A}(s) = {\cal A}(s_n) +(s-s_n){\cal C}_n(s), \eeq
i.e.:
\bea\nn\big[{\cal C}_n(s)\cdot\{a_k\}\big]_k&=&(k+1)a_{k+1} -2 (2k+1 + s + s_n)a_k \\ 
\label{eq:Operator_C} &&+ (2 k + s + s_n + \lambda) a_{k-1}.
\eea
We obtain
\beq\label{eq:Aa=B_expanded}
\Big[{\cal A}(s_n) +  (s-s_n){\cal C}_n(s)\Big] \cdot
\frac{\{h_k\}}{s-s_n}+{\cal A}(s)\cdot\{g_k\} = \{B_k\},
\eeq
which provides us in the limit $s\to s_n$ with the condition
\[{\cal A}(s_n)\cdot\{h_k\}\stackrel{\displaystyle !}{=}0.\]
In the vicinity of the QNM $s_n$, i.e.~for $0<|s-s_n|<\varepsilon$ with some $\varepsilon\ll 1$, eq.~\eqref{eq:ak_Pole} describes the solution to \eqref{eq:Aa=B_expanded} with vanishing and rapidly decreasing coefficients $\{h_k\}, \{g_k\}$ for $k<0$ and $k\to\infty$ respectively. Hence, according to the fact that $\phi_n$ is the {\em only} $C^\infty$-solution on the interval $\sigma\in[0,1]$ (cf.~discussion at the end of Sec.~\ref{Sec:QNMs}), we have
\beq\label{eq:hk=eta_Hk}
	h_k = \eta_n H_k(s_n)
\eeq
with the QNM amplitude $\eta_n$ as proportionality factor. We thus obtain
\[\Big[{\cal A}(s_n) +  (s-s_n){\cal C}_n(s)\Big] \cdot \frac{\{h_k\}}{s-s_n} =\eta_n {\cal C}_n(s)\cdot\{H_k(s_n)\} \]
Equation \eqref{eq:Aa=B_expanded} reads now:
\beq\label{eq:ODE_inh_sn}{\cal A}(s) \cdot\{g_k\}(s)
= \{B_k\}(s)-\eta_n{\cal C}_n(s)\cdot\{H_k(s_n)\}.\eeq
For $s\ne s_n$, the solution of \eqref{eq:ODE_inh_sn} can be found using the formula \eqref{eq:sol_ak}. In particular, we obtain 
\beq\label{eq:coeff_g0} g_0(s)=-\frac{1}{I_{-1}}\sum_{j=0}^{\infty}\frac{I_j(B_j-\eta_n C_j)}{\alpha_j}\prod_{m=0}^j\frac{\alpha_m}{\gamma_m} \eeq
with
\bea\label{eq:coeffs_Ck}
C_j&:=&\big[{\cal C}_n(s)\cdot\{H_k(s_n)\}\big]_j\,.
\eea
In the limit $s\to s_n$ we have $I_{-1}\to 0$, and \eqref{eq:coeff_g0} provides us with a {\em finite} value $g_0$ only if
\beq
	\label{eq:eta_n}
	\eta_n=\left[
	\dfrac{\sum\limits_{j=0}^{\infty}\dfrac{H_jB_j}{\alpha_j}\prod\limits_{m=0}^j\dfrac{\alpha_m}{\gamma_m}}
               {\sum\limits_{j=0}^{\infty}\dfrac{H_jC_j}{\alpha_j}\prod\limits_{m=0}^j\dfrac{\alpha_m}{\gamma_m}}\,\,\right]_{s=s_n}
\eeq
where we utilized that $I_j(s_n)=I_0(s_n)H_j(s_n)$. Note that the sum in the denominator is assured to converge since the addends are rapidly decreasing as $j\to\infty$ (see \eqref{eq:RecRel_psi}, \eqref{eq:alj_Wj}, \eqref{eq:Operator_C}). The numerator can be written as $N=\sum_{j=0}^\infty G_j B_j$, and since $|G_{j+1}B_{j+1}/(G_{j}B_{j})|\to\epsilon$ as $j\to\infty$ for our initial data that are analytic within the circle $\mathbf C$, we can again conclude convergence. 

Now the case in which the initial data are analytic for all $\sigma\in[0,1]$ but whose complex extension is not analytic within  $\mathbf{C}$, i.e.~$\epsilon>1$ in \eqref{eq:epsilon_for_B}, is treated as in Sec.~\ref{Sec:Unique_solutions}. We introduce
$N(x)=\sum_{j=0}^\infty G_jB_j x^j$ and expand it to $x=1$ with the help of an associated diagonal Pad\'e approximant.

As an important quantity for the time range, within which the desired spectral decomposition formula \eqref{eq:VSol_spectral} is valid, we now define the {\em growth rate} of the quasinormal mode amplitudes,
\beq\label{eq:growth_rate_snk}
	\nu_{\rm QNM}(\sigma)= \lim_{n\to\infty}\frac{\ln|\eta_n\phi_n(\sigma)|}{|\Re(s_n)|}.
\eeq
The following statements hold:
\ben
	\item As 
	\[\nu_{\rm QNM}(\sigma)=\underbrace{ \lim_{n\to\infty}\frac{\ln|\eta_n|}{|\Re(s_n)|}}_{=\nu_{\rm QNM}(1)} + \lim_{n\to\infty}\frac{\ln|\phi_n(\sigma)|}{|\Re(s_n)|},\]
	the profile of $\nu_{\rm QNM}(\sigma)$ is universally given by the second term, while the first one is $\nu_{\rm QNM}(1)$ 
	(which follows from the scaling condition $H_0=1$) and presents a constant shift depending on the	initial data.
	\item For polynomial initial data \eqref{eq:SerieV0} we obtain a universal growth rate $\nu_{\rm QNM}(\sigma)$. This follows from the fact that the numerator $N$ in
	 \eqref{eq:eta_n} is algebraic in $s_n$. Hence, we get:
	\bea\nn \nu_{\rm QNM}(1)&=& - \lim_{n\to\infty}\frac{\ln\left|\sum\limits_{j=0}^{\infty}\dfrac{H_jC_j}{\alpha_j}\prod\limits_{m=0}^j\dfrac{\alpha_m}{\gamma_m}\right|_{s=s_n}}{|\Re(s_n)|}\\ \nn &=&-2,\eea
	i.e.~an expression  that is independent of the particular choice of the polynomial initial data. Note that in our numerical investigations, this limit has been observed to be always -2, independently of the choices for $\lambda$ and $\ell$.
\item Looking at \eqref{eq:VSol_spectral} we see that the addends in the sum fall off exponentially for $\tau>\nu_{\rm QNM}(\sigma)$. Adding a similar argument with respect to the continuous integral part, we will be able to argue in Sec.~\ref{sec:Discussion}, that this means full validity of the spectral decomposition formula \eqref{eq:VSol_spectral} for such times. 
\een

\subsection{The Laplace transform along the branch cut}\label{Sec:branch_cut}

\begin{figure*}[t!]
\begin{center}
\includegraphics[width=8.0cm]{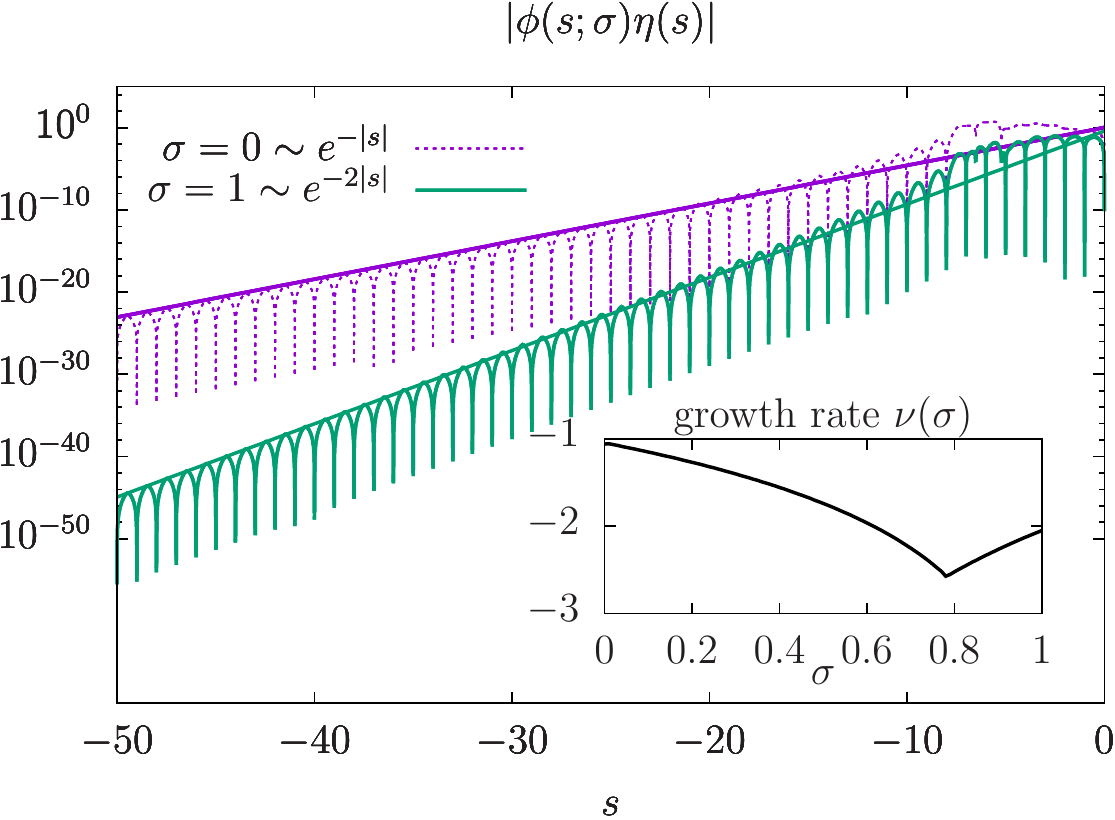}
\includegraphics[width=8.2cm]{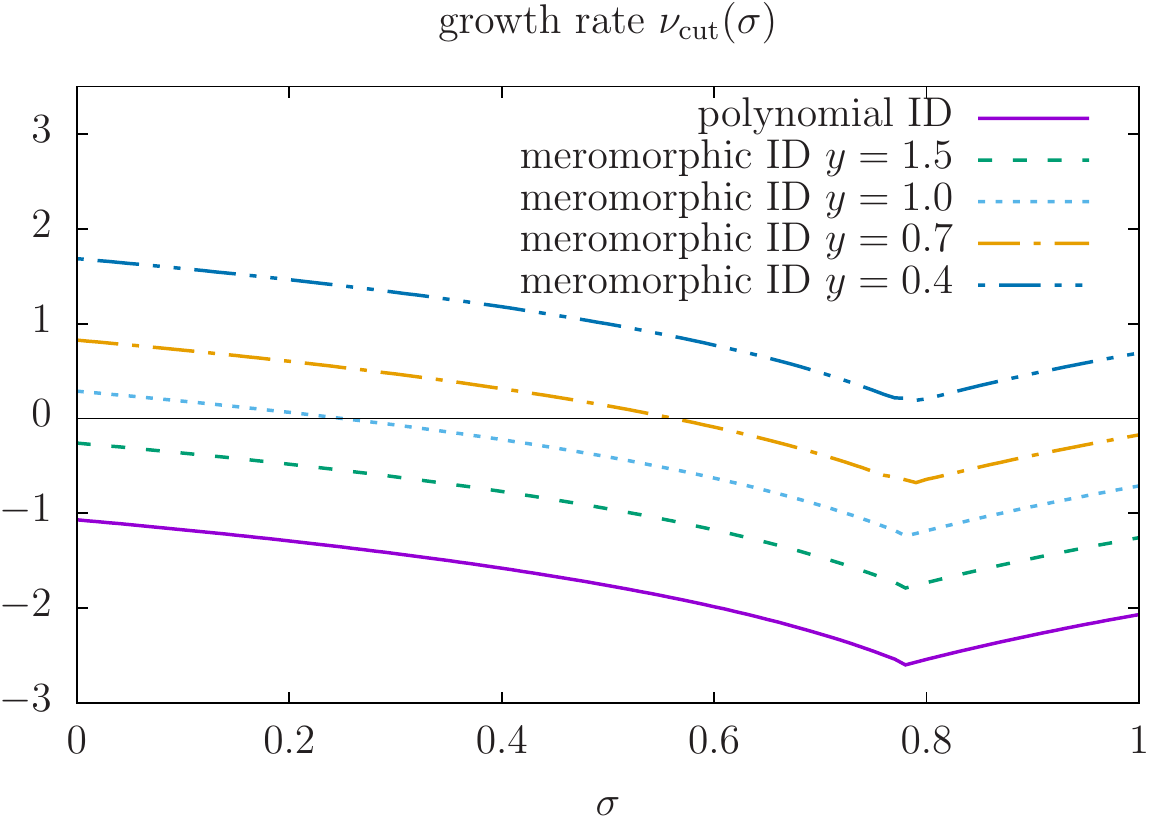}
\end{center}
\caption{
Jump $\eta_n\phi_n(\sigma)$ of the Laplace transform $\hat V(\sigma;s)$ along the negative axis $s\in\mathbb{R}^-$ with prescribed parameters $\lambda=l=0$. Left panel: for polynomial initial data $V_0(\sigma) = \sigma^2 (1-\sigma)^4$, $\dot V_0(\sigma)=0$ the decay is shown for $\sigma=0$ (at $\scri$) and for $\sigma=1$ (at the horizon). The inset brings the {\em growth rate} $\nu_{\rm cut}(\sigma)$ for all $\sigma\in[0,1]$. Right panel: comparison of the grow rate for polynomial initial data against meromorphic initial data with single poles in the complex $\sigma$-plane. Here,  $V_0(\sigma) =2\,\Re\left[ (\sigma - \sigma_{0})^{-1}\right]$, $\dot V_0(\sigma)=0$, with $\sigma_{0} = 0.5 + i\, y$. We observe exactly the same behaviour for the growth rate as in the quasi-normal mode case (see fig.~\ref{fig:Amplitude_s0l0_Inset}), i.e., $\nu_{\rm QNM}(\sigma) = \nu_{\rm cut}(\sigma)$.}
\label{fig:BranchAmplitude_s0l0}
\end{figure*}

Let us now consider the situation in which we approach the negative real axis within the complex $s$-plane. Again we exclude, for the time being, integer $s$-values, i.e.~$s\in \mathbb{R}^-\backslash\mathbb{Z}^-$. (As mentioned in footnote \ref{foot:Neg_s}, $s\in\mathbb{Z}^-$ will be treated in Sec.~\ref{App:Neg_Lapl_Par}.) Towards this axis we can get from above or from below. In the first case, $\arg(s)\to \pi$, the corresponding $\kappa=2\sqrt{s}$ with $\Re(\kappa)>0$ tends to the positive imaginary axis, while in the latter one, $\arg(s)\to -\pi$, it runs towards the negative imaginary axis. Consequently, we obtain via the method presented in Sec.~\ref{Sec:Unique_solutions} for $s\in\mathbb{R}^-\backslash\mathbb{Z}^-$ the two solutions
\[\hat V^\pm(s) = \lim_{\varepsilon\to 0} \hat V(s\pm i|\varepsilon|)\]
with
\[\hat V^\pm(\sigma; s) = \sum_{k=0}^\infty a_k^\pm(1-\sigma)^k.\]
In the computation of the $a_k^\pm$ according to steps described in Sec.~\ref{Sec:Unique_solutions}, the coefficients $B_k$, $H_k$ as well as $I_k^\pm$ are involved. Only the latter ones are different when getting to the negative real axis from above or from below, see asymptotics in \eqref{eq:RecRel_psi}. 

Now, the symmetry condition (\ref{eq:Symmetry_hat_V}) implies that
\[\hat V^-(s)=[V^+(s)]^*,\qquad a_k^-=[a_k^+]^*,\qquad 
		I_k^-=[I_k^+]^*,\]
and the asymptotics \eqref{eq:RecRel_psi} tells us that these quantities have, in general, both real- and imaginary parts. With nonvanishing imaginary part of $\hat V^\pm(s)$ we conclude that the Laplace transform $\hat V(s)$ possesses, besides the simple poles at the QNMs $s_n$, a {\em jump} along the negative real axis, $s\in\mathbb{R}^-$:
\[\lim_{\varepsilon\to 0} \left[\hat V(s+ i|\varepsilon|)-\hat V(s- i|\varepsilon|)\right] 
= \sum_{k=0}^\infty \underbrace{(a_k^+-a_k^-)}_{=:d_k}(1-\sigma)^k.\]
Therefore, the negative real axis appears as a branch cut with respect to $\hat V(s)$. 
Now, as both $a_k^+$ and $a_k^-$ satisfy the recursion relation \eqref{eq:RecRel}, the coefficients $d_k=a_k^+-a_k^-$ satisfy the homogeneous recurrence relation \eqref{eq:RecRel_phi} and, moreover, $d_{-1}=0$. Hence we have $d_k=d_0 H_k$ and thus: 
\beq\label{eq:jump_height}\lim_{\varepsilon\to 0} \left[\hat V(s+ i|\varepsilon|)-\hat V(s- i|\varepsilon|)\right] =-2\pi i\eta(s)\phi(s), \eeq
where we introduced $\eta(s):=-d_0/(2\pi i)$ which will appear in the spectral decomposition formula \eqref{eq:VSol_spectral} as a {\em branch cut amplitude}. The computation of $d_0$ in terms of \eqref{eq:sol_ak} yields
\[d_0=a_0^+-a_0^-=-\sum_{j=0}^\infty \frac{B_j}{\alpha_j}\prod_{m=0}^j\frac{\alpha_m}{\gamma_m}
\underbrace{\left(\frac{I_j^+}{I_{-1}^+}-\frac{I_j^-}{I_{-1}^-}\right)}_{=:D_j}\]
where the coefficients $D_j$ satisfy again the homogeneous recurrence relation \eqref{eq:RecRel_phi} and, moreover, $D_{-1}=0$. We conclude  that $D_j=D_0 H_j$ and obtain finally:
\beq\label{eq:eta_of_s}
\eta(s)=\Im\left(\frac{I_0^+}{\pi\, I_{-1}^+}\right)\sum_{j=0}^\infty B_j\underbrace{\frac{H_j}{\alpha_j}\prod_{m=0}^j\frac{\alpha_m}{\gamma_m}}_{=:G_j}.
\eeq
Again, this formula is valid as it stands for initial data that are analytic within the circle $\mathbf{C}$. For initial data that are analytic for all $\sigma\in[0,1]$ but whose complex extension is not analytic within  $\mathbf{C}$, i.e.~$\epsilon>1$ in \eqref{eq:epsilon_for_B}, we introduce once more $N(x)=\sum_{j=0}^\infty G_jB_j x^j$ and expand it to $x=1$ with the help of an associated diagonal Pad\'e approximant. 

For representative sample initial data, the magnitude $|\eta(s)\phi(\sigma;s)|$ is shown in Fig.~\ref{fig:BranchAmplitude_s0l0}. 

Finally, let us define as in Sec.~\ref{Sec:QNM_amplitudes} the {\em growth rate} of the branch cut amplitudes,
\beq\label{eq:growth_rate_s}
	\nu_{\rm cut}(\sigma)=-\lim_{n\to\infty}\left[\max_{-(n+1)\le s<-n}\frac{\ln|\eta(s)\phi(\sigma;s)|}{|s|}\right].
\eeq
Note that the three statements at the end of Sec.~\ref{Sec:QNM_amplitudes} can be transferred to the growth rate of the branch cut amplitudes, with the second point modified by:
\[ -\lim_{n\to\infty}\left[\max_{-(n+1)\le s<-n} \,\frac{1}{|s|}\ln\left|\Im\left(\frac{I_0^+}{\pi\, I_{-1}^+}\right)\right|\,\right]=-2.\]
It is interesting to note that in our numerical investigations of analytic initial data we observe that the growth rates $\nu_{\rm QNM}$ and $\nu_{\rm cut}$ of quasinormal mode amplitudes and branch cut amplitudes coincide. We believe that this is a consequence of the fact that these quantities arise from similar expressions, cf.~\eqref{eq:eta_n} and \eqref{eq:eta_of_s}. Henceforth we will simply use $\nu(\sigma)$ to denote the mutual growth rates $\nu_{\rm QNM}$ and $\nu_{\rm cut}$.

\subsection{Asymptotic expansion of the Laplace Transform}\label{sec:Asymptotic}

\begin{figure}[t!]
\begin{center}
\includegraphics[width=8.0cm]{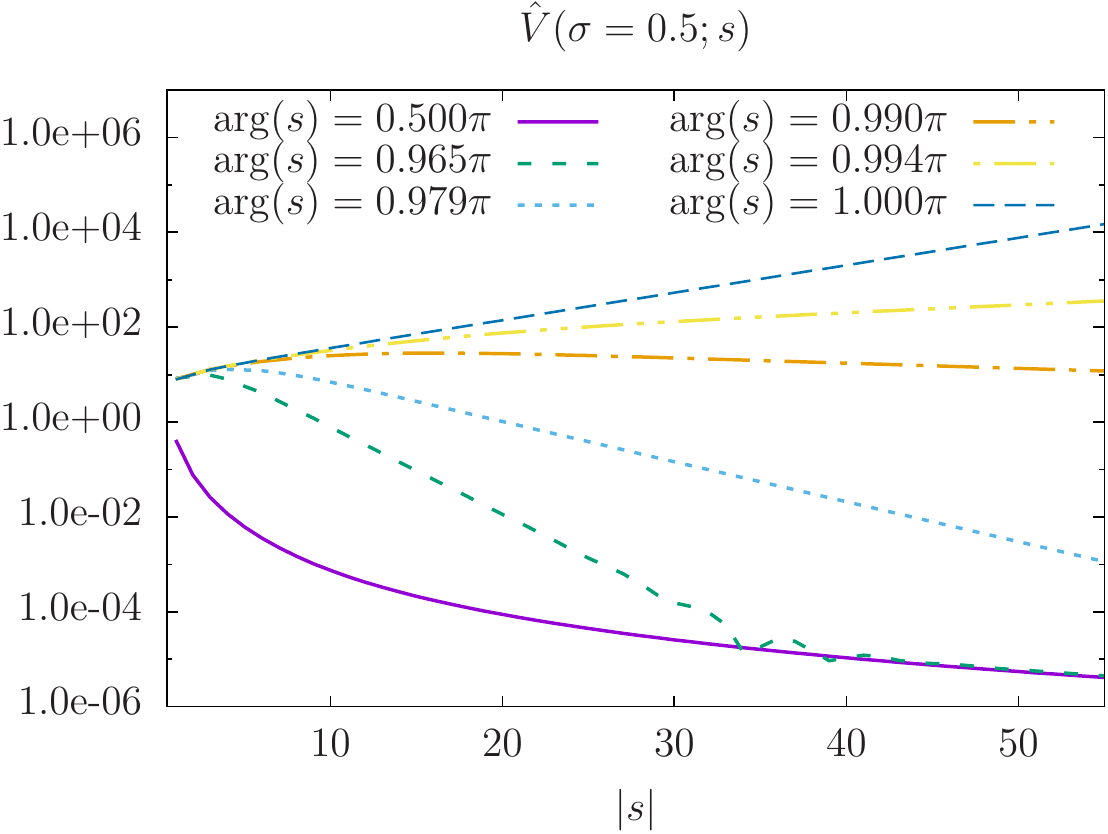}
\end{center}
\caption{Behaviour of the Laplace transform $\hat V$ at the spatial location $\sigma=0.5$ in the half-plane $\Re(s)<0$ for meromorphic initial data $V_0(\sigma) = 2\,\Re\left[ (\sigma - \sigma_{0})^{-1}\right]$, $\dot V_0(\sigma)=0$ with $\sigma_{0} = 0.5 + i\, 0.7$. It becomes apparent that $\hat{V}$ grows exponentially for $|\arg(s)| > 0.99 \pi$. The strongest divergence is obtained for $|\arg(s)| = \pi$ and coincides there with the growth rate $\nu$ (here, $\nu\approx0.13$).}
\label{fig:HalfCircle_s0l0_AnlID_020_sig050}
\end{figure}

In order to arrive at the spectral decomposition (\ref{eq:VSol_spectral}) in Sec.~\ref{sec:Spec_Decomp} via an appropriate deformation of the Bromwich integration path (see fig.~\ref{fig:BromwichInt}), we have to consider the contribution of the integral over the semicircular portion of $\Gamma_2$ in the complex $s$-plane for $\tau>0$. The two cases, positive and negative growth rates $\nu$, have to be discussed separately.

Let us approach the matter with the auxiliary function
\beq\label{eq:W}W(\tau)=\int_{-\infty}^0 e^{s(\tau-\nu)}\sin(\omega s) d s
=\Im\left[\frac{1}{\tau-\nu-i\omega}\right], \eeq
where we took in the desired formula (\ref{eq:VSol_spectral}) only the continuous integral part and, moreover, inserted for $\eta(s)\phi(\sigma;s)$ its asymptotics given through the growth rate $\nu$, cf.~\eqref{eq:growth_rate_s}. The phase $\omega$ (let us take $\omega>0$) has been added in order to have a regular function $W$ for $\tau>0$. Note that the integral can only be performed for $\tau>\nu$, but the result is defined on the entire complex $\tau$-plane with single poles at $\nu\pm i\omega$.

Now, the integrand of the Bromwich integral \eqref{eq:BromInt} is given by $\hat{V}(\sigma;s)e^{s\tau}$. Consequently, we have to discuss here $\hat{W}(s)e^{s\tau}$ with
\bea\label{eq:hat_W}\hat{W}(s)={\cal L}\big[W(\tau)\big](s) &=&\frac{1}{2i}\left(e^{-s(\nu+i\omega)}E_1\big[-s(\nu+i\omega)\big]\right. 
\nn\\  &&\left.\nn\,\,\,\,-\,e^{-s(\nu-i\omega)}E_1\big[-s(\nu-i\omega)\big]\right),
\eea
where $E_1$ is an exponential integral. The function 
\beq\label{eq:func_f} f(z)=e^z E_1(z)\eeq
has a branch cut along the negative real axis in the complex $z$-plane. Let us discuss this function as being defined on a Riemannian surface with infinitely many sheets. The crossing of the negative real axis means the analytic transition into a neighbouring sheet. The function $f$ is best described by another function 
\beq\label{eq:func_g} g:[0,\infty)\times(-\infty,\infty)\to\mathbb{C},\qquad  g(r,\varphi)=f(re^{i\varphi}),\eeq
where $z=re^{i\varphi}$ is assumed to be located in the $k$-th sheet with $k=\lfloor(\varphi+\pi)/(2\pi)\rfloor$.\footnote{Here we used the notation $\lfloor\cdot\rfloor$ for the floor function.}. It turns out that:
\beq\label{eq:prop_g}
\begin{array}{lllll}
\lim\limits_{r\to\infty}\dfrac{\ln|g(r,\varphi)|}{r}&=&\sin\left(\varphi+\frac{5\pi}{2}\right) 
						&\mbox{for} & \varphi\in\left(-\frac{5\pi}{2},-\frac{3\pi}{2}\right) \\[3mm]
\lim\limits_{r\to\infty}|g(r,\varphi)|              &=&0                 
            &\mbox{for} & \varphi\in\left(-\frac{3\pi}{2},\frac{3\pi}{2}\right) \\[3mm]
\lim\limits_{r\to\infty}\dfrac{\ln|g(r,\varphi)|}{r}&=&\sin\left(\varphi-\frac{3\pi}{2}\right) 
						&\mbox{for} & \varphi\in\left(\frac{3\pi}{2},\frac{5\pi}{2}\right).
\end{array}
\eeq
Now, \eqref{eq:hat_W} means:
\beq\label{eq:hat_W_2}
\hat W(s)=\frac{1}{2i}\left[g(r,\varphi_+)-g(r,\varphi_-)\right]
\eeq
with ($\omega>0$)
\[r=|s|\sqrt{\nu^2+\omega^2},\quad \varphi_\pm=\arg(s)\pm[\arg(\nu+i\omega)-\pi].\]
Let us discuss $\nu<0$ first. We have $-\pi<\arg(s)<\pi$ and $\pi/2<\arg(\nu+i\omega)<\pi$. We conclude that
\[-\frac{3\pi}{2}<\varphi_+<\pi,\qquad -\pi<\varphi_-<\frac{3\pi}{2},\]
and hence $|\hat W(s)|$ vanishes whenever $s\to\infty$ with $-\pi<\arg(s)<\pi$.

The case $\nu>0$ is different. Now we have $\arg(\nu+i\omega)\in(0,\pi/2)$, and we get for $\arg(s)>\frac{\pi}{2}+\arg(\nu+i\omega)$:
\beq\nn
|g(r,\varphi_-)|\sim e^{r\sin\epsilon_-},\quad\epsilon_-=\arg(s)-\arg(\nu+i\omega)-\frac{\pi}{2}.
\eeq
Likewise, for $\arg(s)<-\arg(\nu+i\omega)-\frac{\pi}{2}$ we obtain:
\beq\nn
|g(r,\varphi_+)|\sim e^{r\sin\epsilon_+},\quad\epsilon_+=-\arg(s)-\arg(\nu+i\omega)-\frac{\pi}{2}.
\eeq
We conclude that for $\nu>0$ the function $\hat W(s)$ diverges exponentially as $|s|\to\infty$ when the limit is performed in the circular sector $|\arg(s)|>\frac{\pi}{2}+\arg(\nu+i\omega)$ \footnote{Outside this sector, in particular for $|\arg(s)|<\pi/2$, $|\hat W|$ tends to zero when $|s|\to\infty$, and hence the Bromwich integral \eqref{eq:BromInt} exists.}. The strongest divergence is obtained for $|\arg(s)|\to\pi$; there we have $|\hat W|\sim e^{|s|\nu}$. 

In fig.~\ref{fig:HalfCircle_s0l0_AnlID_020_sig050} we display the behaviour of the Laplace transform $\hat{V}(\sigma;s)$ in the half-plane $\Re(s)<0$ for meromorphic initial data $V_0(\sigma) = 2\,\Re\left[ (\sigma - \sigma_{0})^{-1}\right]$, $\dot V_0(\sigma)=0$ with single poles at $\sigma_0$ and $\sigma_0^*$. We take $\sigma_0=0.5+iy$ and considered, in particular, the parameter $y=0.7$. For these initial data, we obtain at $\sigma = 0.5$ a positive growth rate $\nu \approx 0.13$ (see figs.~\ref{fig:Amplitude_s0l0_Inset} and \ref{fig:BranchAmplitude_s0l0}). In fig.~\ref{fig:HalfCircle_s0l0_AnlID_020_sig050} the existence of a circular sector $|\arg{s}|>0.99 \pi$ becomes apparent within which $\hat{V}(\sigma;s)$ diverges exponentially. As expected, the strongest divergence occurs for $|\arg{s}| =\pi$ with the rate $\nu$.  

We now turn to the deformation of the Bromwich integration path. Let us again consider the case $\nu<0$ first. Although $|\hat W(s)|\to 0$ for $|s|\to\infty, \,|\arg(s)|<\pi$, we cannot simply apply Jordan's lemma in order to obtain vanishing contribution of the integral over the semicircular portion of $\Gamma_2$ in the complex $s$-plane. The reason is that $\hat W$ possesses a branch cut along the negative $s$-axis, a fact which is not included in the formulation of Jordan's Lemma. Moreover, when considering $\hat V$ we see that it possesses poles at the QNMs $s=s_n$, accumulating at infinity. One might be reminded of the function $1/\sin(\pi s)$ with poles at real integer values. However, for $1/\sin(\pi s)$ the residues of the poles remain of finite magnitude as $s$ tends to infinity. In contrast, 
the residues of $\hat V$ die out at the rate $e^{-\Re(s_n)\nu}$. Likewise, also the jump along the negative $s$-axis falls off at the rate $e^{-s\nu}$ when $s\to -\infty$. One may say that asymptotically the singular structures of $\hat V$ and $\hat W$ disappear exponentially and play merely a sub-dominant role. 

Let us illustrate this issue by a representative example. The function
\beq\label{eq:Example_h} \hat h(s)=\sum_{k=1}^\infty \frac{2^{-k}}{s+k}=\frac{1}{2}\Phi\left(\frac{1}{2},1,1+s\right),\eeq
where $\Phi$ is the so-called Lerch transcendent (see e.g.~\cite{gradshteyn2007}), resembles the property of $\hat V$ of having infinitely many first-order poles accumulating at infinity with rapidly decreasing residuals. We have $\hat h\to 0$ for $|s|\to\infty,\,\,|\arg(s)|<\pi$. Let us now discuss the Bromwich integral (see (\ref{eq:BromInt}), (\ref{eq:Gamma_1}) and figure~\ref{fig:BromwichInt}) which yields the associated inverse Laplace transform $h$:
\beq
\label{eq:h_BromInt}
h(\tau) = {\cal L}^{-1}[\hat h(s)](\tau)=\frac{1}{2\pi i}\int_{\Gamma_1}\hat{h}(s)e^{s\tau} ds.
\eeq
The inverse Laplace transformation can be applied separately to each addend in (\ref{eq:Example_h}), giving thus
\beq
h(\tau)=\sum_{k=1}^\infty 2^{-k}e^{-k\tau}=\frac{1}{2e^{\tau}-1},
\eeq
which is just the sum of the residues of $\hat h(s) e^{s\tau}$ in the left half-plane:
\beq
	\sum_{k=1}^\infty {\rm Res}_{-k}\left[\hat h(s)e^{s\tau}\right]=\sum_{k=1}^\infty 2^{-k}e^{-k\tau}=\frac{1}{2e^{\tau}-1}.
\eeq
It follows that 
\bea
	\nn\frac{1}{2\pi i}\int_{\Gamma_2}\hat{h}(s)e^{s\tau} ds&=&\frac{1}{2\pi i}\int_{\Gamma_1}\hat{h}(s)e^{s\tau} ds\\ \nn&=&\sum_{k=1}^\infty {\rm Res}_{-k}\left[\hat h(s)e^{s\tau}\right],
	\eea
i.e.~the integral over the semicircular portion of $\Gamma_2$ vanishes in the limit of infinite radius. That is to say, that Jordan's Lemma applies, even though the function $\hat h$ possesses poles accumulating at infinity (and therefore does not satisfy the prerequisites required for a strict application of Jordan's Lemma). 

From the preceding considerations we find that for $\nu<0$ the integral over the semicircular portion of $\Gamma_2$ in the complex $s$-plane vanishes. In contrast, for $\nu>0$ the exponential divergence in the circular sector $|\arg(s)|>\frac{\pi}{2}+\arg(\nu+i\omega)$ ruins the applicability of Jordan's Lemma. However, for times $\tau>\nu$ the functions $\hat{V}(\sigma;s)e^{s\tau}$ as well as $\hat{W}(s)e^{s\tau}$ possess the desired fall-off at infinity. This results from the fact that the strongest divergence in the circular sector is given by $e^{|s|\nu}$ in the limit $|\arg(s)|\to\pi$. We thus may finally conclude that the integral over the semicircular portion of $\Gamma_2$ does not contribute for (i) $\tau>0$ when $\nu<0$, and (ii) $\tau>\nu$ when $\nu>0$.

\section{Spectral decomposition}\label{sec:Spec_Decomp}

\begin{figure*}[t!]
\begin{center}
\includegraphics[width=8.5cm]{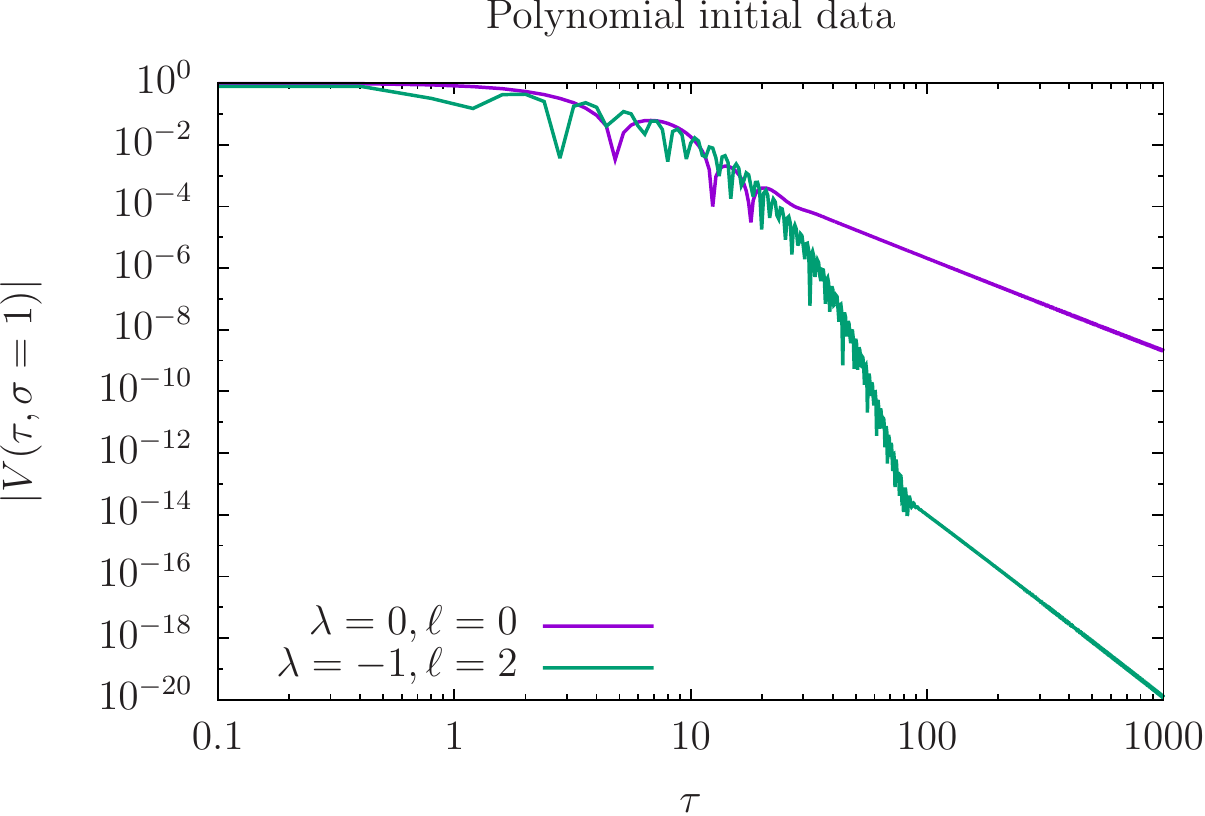}
\includegraphics[width=8.5cm]{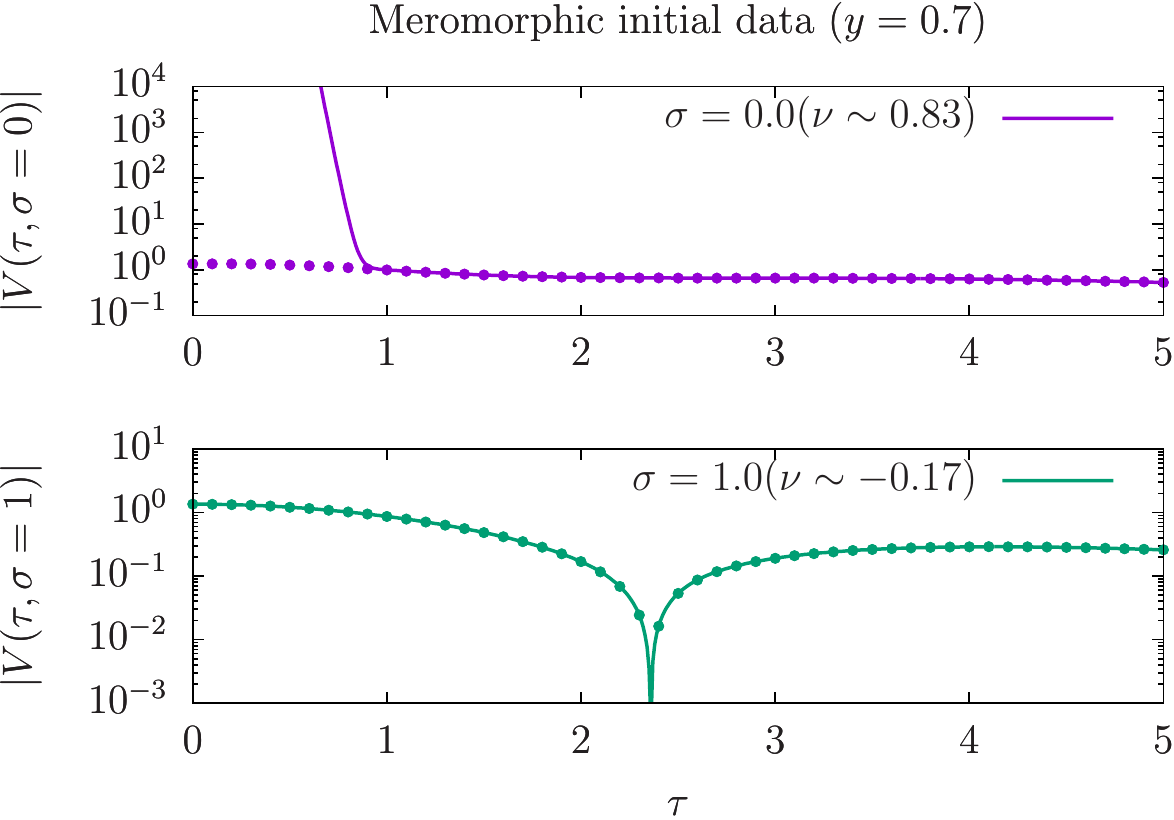}
\end{center}
\caption{Time evolution of the field $V(\tau, \sigma)$ according to the spectral decomposition \eqref{eq:wavefield_2}. Left panel: polynomial initial data with $V_0(\sigma)=1, \dot{V}_0(\sigma)=0$, for $\lambda=0, \ell=0$ and $\label = -1, \ell =2$. The spectral decomposition provides an efficient and stable method for a long time evolution (to be compared with the inverse Laplace transformation method in fig~\ref{fig:BromwitchIntegral}). Right panel: meromorphic initial data $V_0(\sigma) =2\,\Re\left[ (\sigma - \sigma_{0})^{-1}\right]$, $\dot V_0(\sigma)=0$, with $\sigma_{0} = 0.5 + i\, 0.7$ and $\lambda=0, \ell=0$. The continuous lines correspond to an evolution according to \eqref{eq:wavefield_2} while the dots are the results of the explicit time evolution with the code~\cite{Macedo2014}. The spectral decomposition \eqref{eq:wavefield_2} is valid for time $\tau \gtrsim \nu(\sigma)$.}
\label{fig:SpecDecTimeEvol}
\end{figure*}

It is now an easy task to put the pieces together that we collected in the previous sections.
Starting point is the representation (\ref{eq:BromInt}) of the wave field $V$. If we deform the Bromwich integration path from $\Gamma_1$ to $\Gamma_2$ (see fig.~\ref{fig:BromwichInt}), we gather
\ben
\item QNM contributions:
	Writing according to \eqref{eq:Series_hat_V}, \eqref{eq:ak_Pole}, \eqref{eq:hk=eta_Hk} and \eqref{eq:Series_phi} the Laplace transform $\hat V$ in the vicinity of the QNM $s_n$ as 
	\beq \label{eq:Vhat_QNM}\hat V(\sigma;s) = \frac{\eta_n\phi_n(\sigma)}{s-s_n}+\sum_{k=0}^\infty g_k(s)(1-\sigma)^k \eeq
	with coefficients $g_k(s)$ that are analytic in the vicinity of $s_n$ (cf.~discussion in Sec.~\ref{Sec:QNM_amplitudes}), we find
	\beq\label{eq:QNM_Contribution}
	\hspace*{6mm}\frac{1}{2\pi i}\sum_{n=0}^\infty 
	\oint_{C_n} \hat V(\sigma;s) e^{\tau s} ds=\sum_{n=0}^\infty  \eta_n\phi_n(\sigma) e^{\tau s_n},
	\eeq
	where $C_n$ denotes a sufficiently small circle which encompasses in a counterclockwise fashion the QNM $s_n$ (and only this one) and does not touch or cross the negative real $s$-axis.
	\item The branch cut contribution amounts to
		\bea
			\hspace*{10mm}\frac{1}{2\pi i}\int\limits_{-\infty}^0 \left[\hat V^-(\sigma;s)-
					\hat V^+(\sigma;s)\right] e^{\tau s} ds \nn && \\ \nn &&
	\hspace*{-55mm}=-\frac{1}{\pi}\int\limits_{-\infty}^0 \Im[\hat V^+(\sigma;s)] e^{\tau s} ds 
	=\int\limits_{-\infty}^0 \eta(s)\phi(\sigma; s) e^{\tau s} ds,\nn  
	\eea
	where the integral is to be performed along the negative real $s$-axis.
	\item No contribution from the semicircle for (i) $\tau>0$ when $\nu<0$, and (ii) $\tau>\nu$ when $\nu>0$, see discussion in 
	Sec.~\ref{sec:Asymptotic}. We recall that $\nu$ resembles the mutual growth rate of QNM and branch cut amplitudes, cf.~\eqref{eq:growth_rate_snk} and \eqref{eq:growth_rate_s}, obtained for analytic initial data.
\een
In sum, we arrive at the following {\em spectral decomposition} of the wave field satisfying the dissipative wave equation (\ref{eq:ReqTeukEq}):
\beq\label{eq:wavefield_1}
V(\tau, \sigma) = \sum_{n=0}^{\infty} \eta_{n} \phi_n(\sigma) e^{\tau s_{n}} + \int\limits_{-\infty}^{0} \eta(s)\phi(\sigma;s) e^{\tau s} ds.
\eeq
We recap that $\{s_n\}$ are the QNMs of our wave equation in question. The functions $\phi_n(\sigma)$ and $\phi(\sigma;s)$ are the solutions to the homogeneous equation \eqref{eq:HomogLaplaceTransfTeukEq_phi} taken at the QNMs and the branch cut $s\in\mathbb{R}^-$ respectively. The initial data are analytic for all $\sigma\in[0,1]$ and imply corresponding QNM amplitudes $\{\eta_n\}$ and the branch cut amplitude $\eta(s)\big|_{s\in\mathbb{R}^-}$ with characteristic mutual growth rate $\nu=\nu(\sigma)$. For $\nu<0$, the formula \eqref{eq:wavefield_1} was derived from the Bromwich integral for coordinate times $\tau>0$. However, having established \eqref{eq:wavefield_1}, we see that the expressions actually make sense for any $\tau>\nu$, meaning that we may analytically expand the solution to such times. In sum, we may conclude that \eqref{eq:wavefield_1} is a valid representation of the wave field solution for all times $\tau>\nu(\sigma)$.

A reformulated version is given by
\beq\label{eq:wavefield_2}
V(\tau, \sigma) = 2\sum_{n=0}^{\infty} \Re\big(\eta_{n} \phi_n(\sigma) e^{\tau s_{n}}\big) + \int\limits_{-\infty}^{0} \eta(s)\phi(\sigma;s) e^{\tau s} ds,
\eeq
where we utilized the condition (\ref{eq:Symmetry_hat_V}) that allows us to count merely the quasi-normal-modes $s_n$ with $\Im(s_n)>0$, as is done in fig.~\ref{fig:BromwichInt}. If algebraically special QNMs need to be taken into account, then the formula gets slightly modified, see Sec.~\ref{Sec:AlgSpecial}, eqn.~(\ref{eq:wavefield_3}) therein.

Fig.~\ref{fig:SpecDecTimeEvol} displays the time evolution according to the spectral decomposition \eqref{eq:wavefield_2} for the two types of initial data discussed in this work. The left panel brings the results of a polynomial initial data $V_0(\sigma)=1, \dot{V}_0(\sigma)=0$ and, in order to compare with the results from fig.~\ref{fig:BromwitchIntegral} with chose $\lambda=0, \ell=0$ and $\label = -1, \ell =2$. Moreover, the right panel compares the evolution according to \eqref{eq:wavefield_2} with the dynamics obtained numerically with the fully spectral code~\cite{Macedo2014} for the meromorphic initial data $V_0(\sigma) =2\,\Re\left[ (\sigma - \sigma_{0})^{-1}\right]$, $\dot V_0(\sigma)=0$ ($\sigma_{0} = 0.5 + i\, y$) and $\lambda=0, \ell=0$. According to the growth rate $\nu(\sigma)$ showed in the previous figs.~\ref{fig:Amplitude_s0l0_Inset} and \ref{fig:BranchAmplitude_s0l0}, for $y=0.7$, the spectral decomposition is valid only for $\tau\gtrsim 0.83$ at $\scri$ ($\sigma =0$), whereas eq.~\eqref{eq:wavefield_2} is valid for all time $\tau \gtrsim -0.17$ (in particular $\tau \ge 0$) at the horizon ($\sigma=1$).

All in all, we have found that the formula (\ref{eq:wavefield_2}) describes the wave field for all times $\tau>\nu(\sigma)$ and is, in contrast to the Bromwich integral method presented in Sec.~\ref{sec:BromIntResults} and Sec.~\ref{App:BromwichInt}, particularly well suited to resolve highly accurately the late time tail behaviour. More concretely, if $\tau$ is very large, then only the amplitude $\eta(s)$ within a tiny vicinity of the origin $s=0$ in the complex $s$-plane is important. Although the computation gets harder as $s\to 0$ (note that the coefficients $\mu_j$ diverge in this limit, cf.~(\ref{eq:A_expansion}) and (\ref{eq:mu_coeffs})), $\eta(s)$ can be computed independently for each $s$ directly from the initial data. This is a striking advantage compared to a time evolution code for which the tail behaviour results from a successive marching forward in time and depends therefore on all steps computed previously. A sophisticated study of tails in the more general Kerr spacetime is planned to be presented in a forthcoming article.

\section{Discussion}\label{sec:Discussion} 

In this article we have numerically constructed the ingredients $\{s_n,\phi_n, \eta_n\}_{n=1}^\infty$ as well as  $\{\phi(s), \eta(s)\}_{s\in\mathbb{R}^-}$ of the spectral decomposition (\ref{eq:wavefield_2}) that describes solutions $V$ to the initial value problem of the dissipative wave equation (\ref{eq:ReqTeukEq}) with initial data  that are analytic in terms of the compactified spatial hyperboloidal coordinate $\sigma\in[0,1]$. The spectral form (\ref{eq:wavefield_2}) arises through the study of the corresponding Laplace-transformed equation and an appropriate deformation of the associated Bromwich integration path. The ingredients in question were established in a sophisticated analysis of Taylor coefficients of relevant functions appearing in this context. 

In the course of the aforementioned steps we have discussed in detail that the characterization of QNMs in terms of the vanishing of the Wronskian determinant formed of specifically normalized solutions to the homogeneous Laplace transformed equation as described in \cite{Kokkotas99a} implies severe technical problems when attempting a straight-forward numerical computation.

In contrast, a well-functioning definition, which provides the justification of Leaver's continued fraction method \cite{Leaver85} to determine the QNMs, is given through the vanishing of an appropriate {\em discrete} Wronskian determinant. 

The form (\ref{eq:wavefield_2}) has been numerically confirmed in a number of tests in which a selection of different analytical initial data were chosen. After the determination of $\{s_n,\phi_n, \eta_n\}_{n=1}^\infty$ and $\{\phi(s), \eta(s)\}_{s\in\mathbb{R}^-}$, formula (\ref{eq:wavefield_2}) represents the desired solution for all coordinate times $\tau$ for which (\ref{eq:wavefield_2}) makes sense, that is for $\tau>\nu(\sigma)$ where $\nu(\sigma)$ is the mutual growth rate of QNM and branch cut excitation coefficients. 
The test was performed through a comparison with the fully-pseudo spectral time evolution algorithm described in \cite{Macedo2014}. 

Before moving on, we elaborate on striking similarities with a specific dissipative wave equation on hyperbolic slices in Minkowski space that can be handled explicitly. 

\subsection{A dissipative wave equation with obstacle in Minkowski space}\label{sec:Mink}
The following example is often being used in the so-called Lax-Phillips scattering theory \cite{Lax67}. Here, we aim not at a discussion within that framework but rather investigate the equation along the lines developed in this article. 

Consider the ordinary wave equation on Minkowski space,
\[\frac{\partial^2 U}{\partial x^2} + \frac{\partial^2 U}{\partial y^2} 
	+ \frac{\partial^2 U}{\partial z^2}- \frac{\partial^2 U}{\partial t^2}=0,\]
written in the Cartesian coordinates $(x,y,z,t)$ of some inertial frame. We introduce the specific hyperbolic coordinates $(\varrho,\theta,\varphi,u)$ given through
\bea
	x&=& r_{\rm O}\left(\varrho+1\right)\sin\theta\cos\varphi \nn \\
	y&=& r_{\rm O}\left(\varrho+1\right)\sin\theta\sin\varphi \nn \\
	z&=& r_{\rm O}\left(\varrho+1\right)\cos\theta \nn \\
	t&=& r_{\rm O}\left(2u+\varrho+1\right)\nn,
\eea
where $r_{\rm O}$ denotes the coordinate radius of a given spherical obstacle at which we require the wave field $U$ to vanish at all times. Assuming that $U$ also vanishes at $\scri$ (described by $\varrho\to\infty$), we rewrite it as
\[V(u,\varrho,\theta,\phi)=\left(\varrho+1\right) U(u,\varrho,\theta,\phi)\]
and expand the auxiliary field $V(u,\varrho,\theta,\phi)$ into the spherical harmonics $Y_{\ell m}(\theta, \varphi)$ basis
\[
V(u,\varrho,\theta,\phi)  = \sum_{\ell = 0}^{\infty} \sum_{m=-\ell}^{\ell} V_{\ell m}(u, \varrho)\,
		Y_{\ell m}(\theta, \varphi).
\]
We thus obtain a specific wave equation for each mode $V_{\ell m}(u, \varrho)$ (again we omit the indices $\ell m$  from now on): 
  \beq\label{eq:wave_Mink} 
  		V_{,\varrho\varrho}  - V_{,u\varrho} -\frac{l(l+1)}{(\varrho+1)^2}V= 0, \quad V(u, \varrho=0)=0.
  \eeq
The second equation represents the boundary condition that the wave field be always zero at the obstacle. We consider (\ref{eq:wave_Mink}) specifically for $l=0$, and with the initial data $V(0,\varrho)=V_0(\varrho)$ the solution
  \beq\label{eq:sol_wave_Mink} 
		V(u, \varrho)=V_0\left(\varrho+u\right)-V_0\left(u\right)
	\eeq
arises. 

We now formulate the following question: For which initial data $V_0$ can the solution (\ref{eq:sol_wave_Mink}) be written in the form (\ref{eq:wavefield_2})? The answer is given by the {\em inverse Laplace transform} applied to $\big(V_0-V_0|_{\varrho\to\infty}\big)$ in terms of the spatial coordinate $\varrho$. In fact, if $V_0$ can be written as
\beq\label{eq:LaplTransf_eta}
	\begin{array}{cll}
	 V_0(\varrho)&=&\int\limits_{-\infty}^{0}\eta(s) \left(e^{\varrho s}-1\right) ds={\cal L}\{\eta(-s)\}(\varrho) + V_\infty, \\
     V_\infty &=& V_0|_{\varrho\to\infty}=-\int\limits_{-\infty}^{0}\eta(s)ds
	\end{array}
\eeq
for some function $\eta(s)$, then (\ref{eq:sol_wave_Mink}) turns into
  \beq\label{eq:sol_wave_Mink_2} 
		V(u, \varrho)=\int\limits_{-\infty}^{0} \eta(s)  \underbrace{\left(e^{\varrho s}-1\right)}_{=\phi(\varrho;s)} e^{su}ds,
	\eeq
i.e.~we obtain the form (\ref{eq:wavefield_2}) with a purely continuous part. Now, if we expand $V_0$ into the entire complex $\varrho$-plane and find some finite value
\beq\label{eq:ximax}
\varrho_{\rm max}:=\max\{\Re(\varrho_{\rm S}):\quad\mbox{$\varrho_{\rm S}$ is singularity of $V_0$}\},
\eeq
i.e.~all singularities of $V_0$ are located to the left of the line $\Re(\varrho)=\varrho_{\rm max}$, then we have for $s\to-\infty$:
\beq\label{eq:eta_s_Mink}
	|\eta(s)|= \left|{\cal L}^{-1}\left\{V_0(\varrho)-V_\infty\right\}(-s)\right|\sim e^{-s\varrho_{\rm max}}.
\eeq
We thus conclude: The spectral decomposition (\ref{eq:sol_wave_Mink_2}) holds for $ u>\varrho_{\rm max}$ if the complex continuation of the initial data $V_0(\varrho)$ onto the complex $\varrho$-plane reveals singular	structures located entirely to the left of the line $\Re(\varrho)=\varrho_{\rm max}$ away from infinity. 

This is a very similar situation as the one that we encountered in the Schwarzschild case. In this Minkowski example, however, we can identify an explicit relation between the {\em spatially constant} growth rate and the location of singular	structures of the initial data $V_0(\varrho)$, namely $\nu=\varrho_{\rm max}$.

Going further and considering now singularities located at infinity in the complex $\varrho$-plane, we find that the matter gets more subtle. 
The following examples provide an impression (take $\alpha,\omega\in\mathbb{R}^+$ in the examples 2 and 4):
\beq\label{eq:V0_eta_examples}
	\begin{array}{ccc}
	{\rm No.}&V_0(\varrho) &  \eta(s) \\
	\hline\hline\\
	1&e^{-\varrho}-1 &\delta(s+1)\\
	\hline \\
    2&e^{-\alpha\varrho}\sin(\omega\varrho) & \mbox{nonexistent}\\
	\hline\\
	3&e^{-\sqrt{\varrho+1}} - 1 & \dfrac{e^{s+1/(4s)}}{2\sqrt{\pi}(-s)^{3/2}}\\
	\hline\hline \\
	4&(\varrho+1)^{-\alpha}-1 & \dfrac{e^s(-s)^{\alpha-1}}{\Gamma\left(\alpha\right)}
	\end{array}
\eeq

The first example describes a purely exponential fall-off of the solution, while the second one is a ring down oscillation with arbitrarily chosen frequency $\omega$ and decay rate $\alpha$. 

If we discuss the several choices $V_0$ in terms of a compactified spatial coordinate $\sigma=1/(\varrho+1)$, then we find for the first three examples in (\ref{eq:V0_eta_examples}) that the corresponding $V_0(\sigma)$ is $C^\infty$ at $\sigma=0$. In the fourth example, however, we have $V_0(\sigma)=\sigma^\alpha-1$ which, for $\alpha<1$, is not differentiable at $\sigma=0$. We conclude that $C^k$-regularity, and in  particular $C^\infty$-smoothness of the initial data $V_0(\sigma)$ at $\sigma=0$, is neither sufficient nor necessary for the existence of a corresponding function $\eta(s)$. This point deserves further clarification, to be conducted elsewhere.

\subsection{Conclusion and outlook}\label{Sec:Conclusion}
The final result of our analytical considerations combined with numerous numerical examples can be formulated as the following conjecture:

{\em 
Given analytical initial data $V_0(\sigma)$ and $\dot V_0(\sigma)$ for the wave equation \eqref{eq:ReqTeukEq}. Then the spectral decomposition (\ref{eq:wavefield_2}) holds for all $\tau>\nu(\sigma)$ where $\nu(\sigma)$ is the mutual growth rate of quasinormal mode and branch cut excitation coefficients defined by \eqref{eq:growth_rate_snk} and \eqref{eq:growth_rate_s}. 
}

A strict mathematical proof of this conjecture remains a challenging task which is far outside the scope of this paper. 

Clearly, it would be desirable to relax the conditions imposed on the initial data to allow for more generic configurations (for instance, data with compact support). From our experiences gathered in this paper as well as through numerous dynamical computations performed by many authors we surmise that the conjecture would still hold for generic initial data that are analytic in a vicinity of $\sigma=0$, i.e.~at $\scri$. More precisely, we expect that there are for each such initial data individual QNM as well as branch cut amplitudes with characteristic growth rates $\nu_{\rm QNM}(\sigma)$ and $\nu_{\rm cut}(\sigma)$ (maybe different for nonanalytical data) such that the spectral decomposition (\ref{eq:wavefield_2}) holds for all $\tau>\nu(\sigma)$ where $\nu(\sigma)=\max\{\nu_{\rm QNM}(\sigma),\, \nu_{\rm cut}(\sigma)\}$. These amplitudes, however, cannot be determined by the methods described in this paper, as they rely on an analysis of Taylor expansions. Note that we had to exclude initial data that are not analytic at $\scri$. Again a look at the dissipative wave equation in Minkowski space illuminates the situation, specifically example 2 in table \eqref{eq:V0_eta_examples} for which a branch cut amplitude and hence a corresponding growth rate does not exist. Consequently, for such initial data the spectral decomposition never holds. Going back to the Schwarzschild case, we expect that likewise for generic initial data of the form \eqref{eq:Example_ID} we cannot identify QNM and branch cut amplitudes, i.e., (\ref{eq:wavefield_2}) never holds ($\nu=\infty$), unless the Laplace parameter in \eqref{eq:Example_ID} is chosen to be some QNM, $s=s_n$, in which case (\ref{eq:wavefield_2}) holds for all times $\tau$ ($\nu=-\infty$).  

At the end, a final remark seems to be in place. In \cite{Berti:2006wq} it has been argued that {\em ``the integral over the quarter circles at infinite frequency produces the early time response of the black hole''}. Similar comments can be found in other papers (see e.g.~\cite{Harms:2013ib}), and they all seem to be reformulated versions of Leaver's statement {\em ``It is $G_F$ that propagates the high-frequency response, and which reduces to the free-space Green's function in the limit as the mass of the black hole
goes to zero.''} made in \cite{Leaver86c}. Our considerations lead us, however, to a different interpretation. We have seen that, for generic analytical initial data, the Laplace transform $\hat V$ {\em diverges exponentially} in some circular sector $|\arg(s)|\in(\pi-\varepsilon,\pi)$ in the complex $s$-plane when $|s|\to\infty$. This means, that the integral over the quarter circles at infinite frequency mentioned before cannot be evaluated. However, in the course of time, concretely for $\tau>\nu(\sigma)$, the associated $\hat V$ for initial data taken on such time slices vanishes when $|s|\to\infty,\,|\arg(s)|<\pi$, and hence the integral in question does not contribute by virtue of Jordan's Lemma. Therefore, we encounter the impression that the proposed early time contribution of the integral over the quarter circles at infinity is a misinterpretation. This integral cannot be evaluated for small times (and thus cannot reasonably be discussed physically) and vanishes for data at larger times $\tau>\nu(\sigma)$. 

We conclude this article with the observation that, due to our experiences gathered in the Minkowski example, the case in which the singularity is located at $\scri$, i.e.~at $\sigma=0$, is expected to require a more sophisticated investigation. While it might appear as a minor remaining uncertainty, this issue plays a fundamental role when attempting to identify an appropriate linear operator acting on an associated function space, whose spectrum is $\{s_n\}\cup\mathbb{R}^-$ with corresponding proper and improper eigenvectors $\phi_n$ and $\phi(s)$. Again, 
the treatment of this interesting functional-analytical question is far beyond the scope of this paper.

\section*{Acknowledgements}
The authors are deeply indebted to Bernd Schmidt for bringing our attention to this interesting topic and for numerous insightful discussions. This work was supported by the DFG-grants SFB/Transregio 7 ``Gravitational Wave Astronomy' and GRK 1523/2. Rodrigo P. Macedo was supported by CNPq under the programme "Ci\^encia sem Fronteiras".

\section{Appendix}\label{App}
\subsection{The Bromwich Integral}\label{App:BromwichInt}

\begin{figure*}[t!]
\begin{center}
\includegraphics[width=8.5cm]{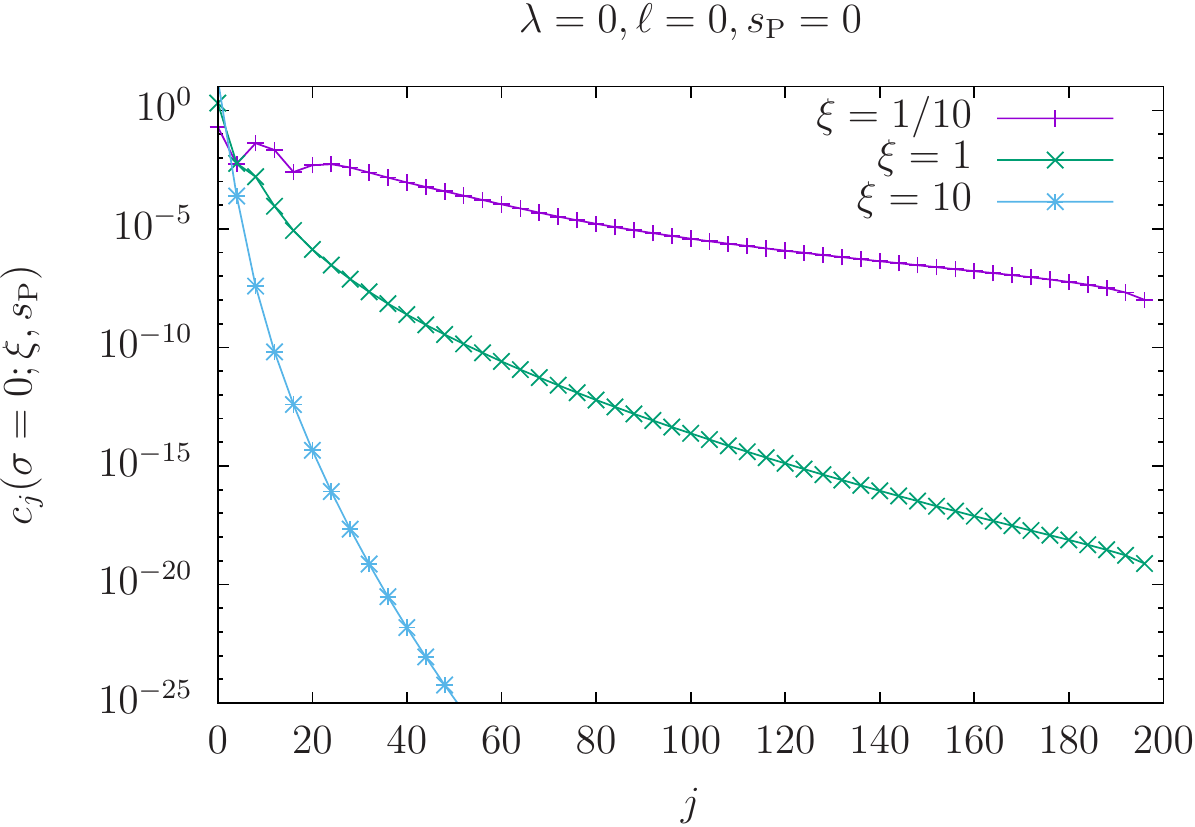}
\includegraphics[width=8.5cm]{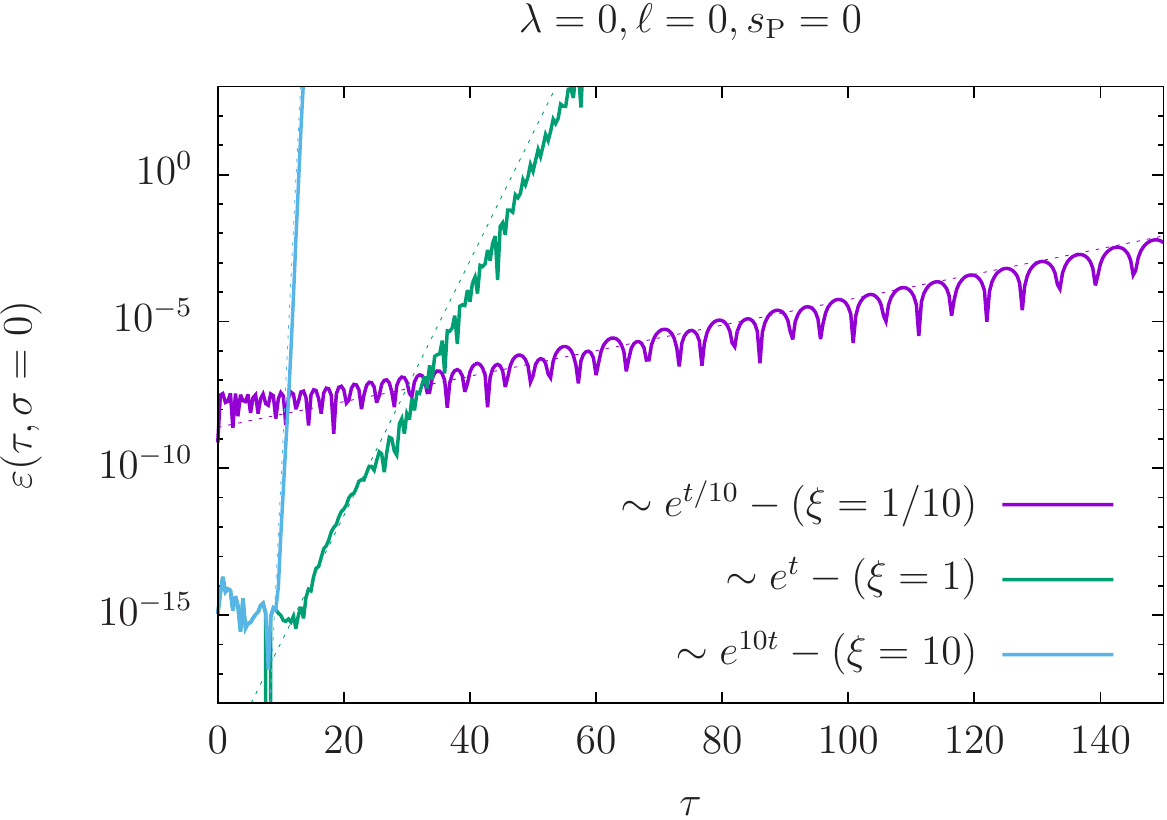}
\end{center}
\caption{Numerical accuracy of the Bromwich integral method with resolution $n_{\chi}=200$, applied to the initial data $V_0(\sigma)=1, \dot{V}_0(\sigma)=0$, with parameter $s_{\rm P}=0$ (cf.~(\ref{eq:Compact_chi})) along the Bromwich integration paths $\Re(s)=\xi\in\{1/10, 1,10\}$. Left panel: Chebyshev coefficients $c_j(0;\xi,s_{\rm P})$, cf.~(\ref{eq:ChebDec_ReV}) and their dependence on $\xi$. Right panel: Corresponding dynamical error $\varepsilon(\tau, \sigma=0)$ which diverges as $e^{\xi \tau}$. Results were obtained with the fixed Taylor resolutions $k_{\rm max} = 500$, $J_{\rm max} = 10$, cf. Sec.~\ref{sec:TaylorExpansions}.}
\label{fig:ChebBromwichInt}
\end{figure*}
The Bromwich integral solution method considers the integration path $\Gamma_1$ (see (\ref{eq:Gamma_1})), that is, values $\hat V(s)$ for 
$s = \xi + i \chi$ with some prescribed $\xi>0$ and $\chi\in(-\infty, \infty)$ are collected in order to evaluate (\ref{fig:BromwichInt}), i.e.~:
\bea\label{eq:V_of_Vhat}
V(\tau, \sigma) &=& \frac{e^{\xi\tau}}{2\pi}\int_{-\infty}^{+\infty} \hat{V}(\sigma;\xi + i\chi)e^{i\chi\tau} d\chi.
\eea
Since $V(\tau, \sigma)$ is bounded, its Laplace transform $\hat V$ vanishes for $|s|\to\infty,\, \arg(s)\in(-\pi/2,\pi/2)$, see (\ref{eq:LaplaceTransf}). Consequently, we have for all $s$ with $\Re(s)>0$ via Jordan's Lemma:
\beq\label{eq:int_CR}\lim_{R\to\infty}\int_{C_R(s)}\frac{\hat V(\sigma;\tilde s)d\tilde s}{\tilde s-s}=0\eeq
where $C_R(s)$ is a semi-circle in the right half-plane about the point $s$, i.e.~:
\[   C_R(s)=\{\tilde s\in\mathbb{C}\mid \tilde s =s+Re^{i\varphi},\,\varphi\in(-\pi/2,\pi/2)\}.\]
Consider now Cauchy's integral theorem,
\beq\label{eq:Cauchy_P} P\oint\limits_{\Gamma_R(s)}\frac{\hat V(\sigma;\tilde s)d s}{\tilde s-s}=\pi i \hat V(\sigma;s),\eeq
where
\[   \Gamma_R(s)=C_R(s)\cup\{\tilde s\in\mathbb{C}\mid \tilde s =s+i\tilde\chi,\,\tilde\chi\in(-R,R)\},\]
with $P$ denoting the {\em Cauchy principal value}
and the integration evaluated in a counter-clockwise fashion along the closed curve $\Gamma_R(s)$. By virtue of (\ref{eq:int_CR}) and (\ref{eq:Cauchy_P}) it follows that the imaginary part of $\hat V$ is the {\em Hilbert transform} of the real part:
\beq\label{eq:Hilbert_transforms}
	\Im\left[\hat V(\sigma;\xi+i\chi)\right]=\frac{1}{\pi}P\int_{-\infty}^\infty \frac{\Re\left[\hat V(\sigma;\xi+i\bar\chi)\right]
	d\bar\chi}{\bar\chi-\chi}.
\eeq
We insert this expression into (\ref{eq:V_of_Vhat}) and obtain via
\[P\int_{-\infty}^\infty \frac{e^{-i\omega}}{\omega}d\omega=-\pi i\]
that 
\bea
\nn
V(\tau, \sigma) &=&\frac{e^{\xi\tau}}{\pi}\int_{-\infty}^{+\infty} 
\Re[\hat{V}(\sigma;\xi + i\chi)] \, e^{i\chi\tau} d\chi\\ 
\label{eq:BromwInt2}
 &=&\frac{2e^{\xi\tau}}{\pi}\int_{0}^{\infty} \Re[\hat{V}(\sigma;\xi + i\chi)] \, \cos(\chi\tau) d\chi,\qquad
\eea
where the latter expression arises through the symmetry condition (\ref{eq:Symmetry_hat_V}).
Now, for some prescribed auxiliary parameter $s_{\rm P}\in {\mathbb R^-_0}$, we compactify the integration interval with the help of the new variable $x\in(0, x_{\rm P}]$ via
\bea
\label{eq:Compact_chi}
x = \frac{x_{\rm P}}{1+x_{\rm P}\chi^2}\quad\mbox{where}\quad x_{\rm P}=(\xi-s_{\rm P})^{-2}.
\eea
Taking into account that $\Re(\hat V)$ is an even function in the variable $\chi$ and vanishes when $\chi\to\infty$, we may approximate $x^{-1}\Re(\hat V)$ in terms of a Chebyshev series:
\beq
\label{eq:ChebDec_ReV}
\Re[\hat{V}(\sigma; \xi + i\chi)] \approx x \sum_{j=0}^{n_\chi-1} c_j(\sigma;\xi, s_{\rm P}) \, T_j \left(\frac{2x}{x_{\rm P}} - 1\right),
\eeq  
where a particular numerical resolution $n_\chi$ was chosen. The Chebyshev coefficients $c_j(\sigma;\xi, s_{\rm P})$ are obtained through requiring equality in (\ref{eq:ChebDec_ReV}) at the Chebyshev-Gauss grid points 
\beq
\label{eq:BromInt_Grid}\nn
\chi_k = \sqrt{\frac{1}{x_k} - \frac{1}{x_{\rm P}}},\quad\mbox{with}\quad x_k = x_{\rm P}\cos\left(\frac{\pi(2k+1)}{2 n_\chi}\right)
\eeq 
and $k = 0, \ldots, n_\chi -1$. Here, the left hand sides in (\ref{eq:ChebDec_ReV}) are evaluated according to the procedure described in section \ref{sec:TaylorExpansions}. Note that $x^{-1}\Re(\hat{V})$ is $C^\infty$ for $x\in[0,x_{\rm P}]$ \footnote{Analyticity breaks down at $x=0$ when $\chi\to\infty$, see discussion in Sec.~\ref{sec:Asymptotic}.}, and hence a Chebyshev expansion provides an accurate numerical approximation.

Considering the polynomial structure of the $T_j$'s explicitly, we introduce
coefficients ${\cal P}_{jl}$ via the requirements
\beq
\label{eq:ChebPolExp}
x\,T_j \left(\frac{2x}{x_{\rm P}}  - 1\right) = \sum_{l=0}^{j} {\cal P}_{jl}\,x^{l+1}.
\eeq
Through this rearrangement we can explicitly evaluate the integral (\ref{eq:BromwInt2}):
\beq
V(\tau, \sigma) \approx \sum_{j=0}^{n_\chi-1} c_j(\sigma;\xi, s_{\rm P}) \sum_{l=0}^{j} {\cal P}_{jl}Q_l(\tau;\xi, s_{\rm P})
\eeq
with
\bea\label{eq:IntegralTerms}
\lefteqn{ Q_l(\tau;\xi, s_{\rm P})= \frac{2e^{\xi\tau}}{\pi}\int_{0}^{+\infty} x^{l+1} \cos(\chi\tau) d\chi}\\
&&= \nn\frac{2e^{s_{\rm P}\tau}}{l!}\sum_{k=0}^{l}\frac{(l+k)!}{(l-k)!\,k!}\left[2 (\xi-s_{\rm P})\right]^{-(l+k+1)}\tau^{l-k}.
\eea
We provide a few examples of the functions $Q_l(\tau;\xi, s_{\rm P})$:
\bea
Q_0(\tau;\xi, s_{\rm P}) &=& \frac{e^{s_{\rm P}\tau}}{\xi-s_{\rm P}} \nn \\
Q_1(\tau;\xi, s_{\rm P}) &=& e^{s_{\rm P}\tau}\frac{1+\tau(\xi-s_{\rm P})}{2(\xi-s_{\rm P})^3} \nn \\
Q_2(\tau;\xi, s_{\rm P}) &=& e^{s_{\rm P}\tau}\frac{3+3\tau(\xi-s_{\rm P}) + \tau^2(\xi-s_{\rm P})^2}{8(\xi-s_{\rm P})^5} \nn
\eea

\begin{figure*}[t!]
\begin{center}
\includegraphics[width=8.5cm]{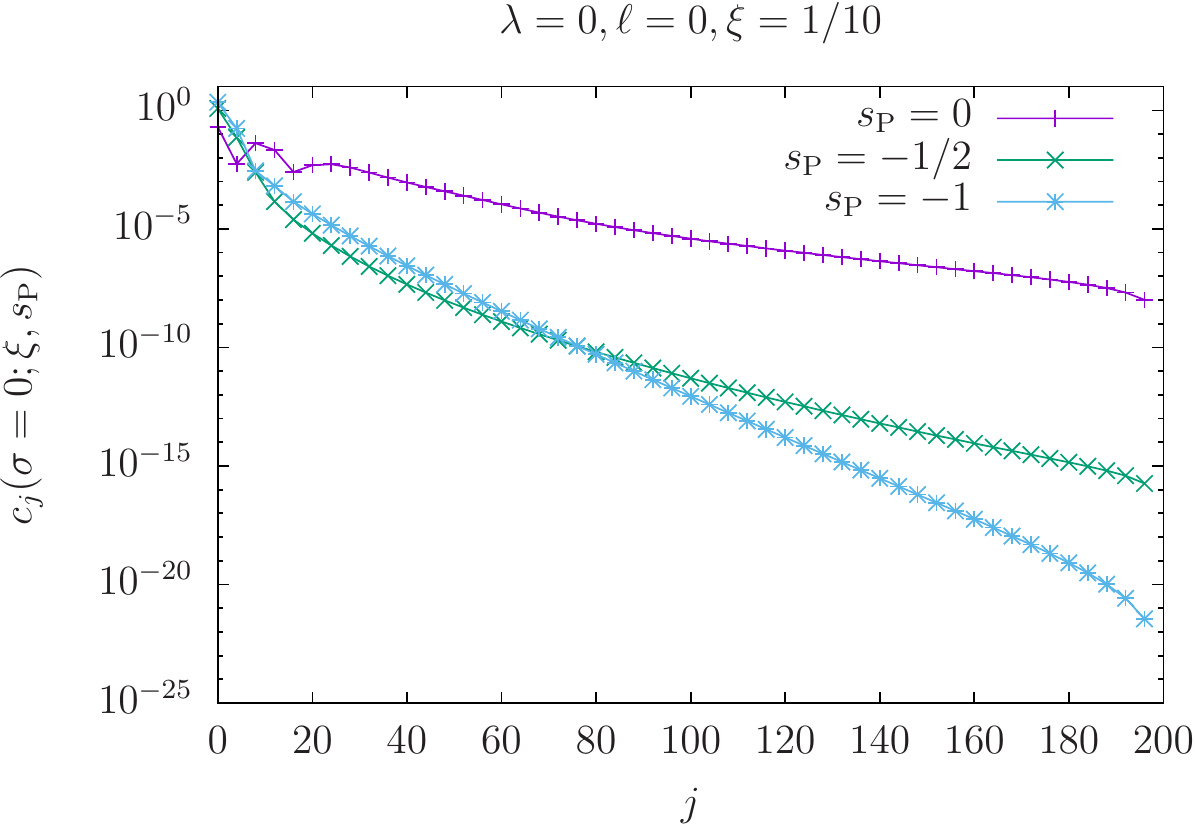}
\includegraphics[width=8.5cm]{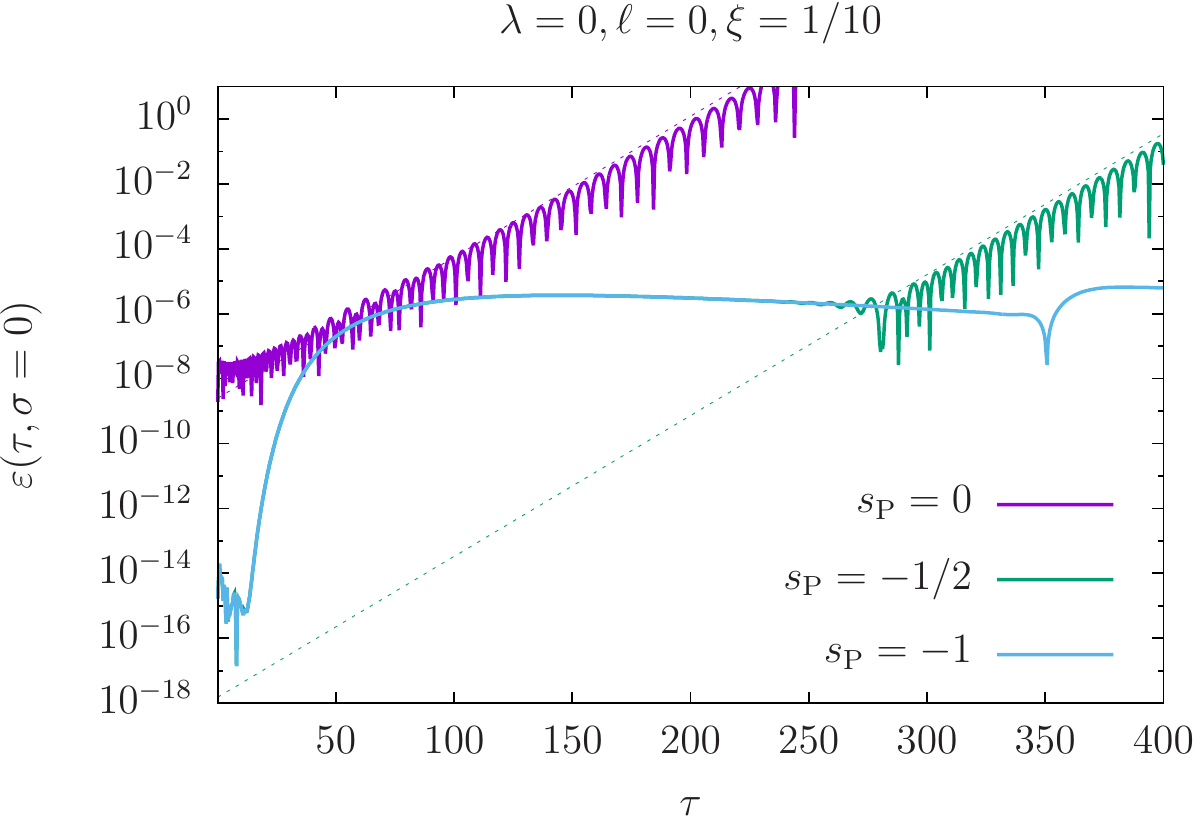}
\end{center}
\caption{Numerical accuracy of the Bromwich integral method with resolution $n_{\chi}=200$, as in fig.~\ref{fig:ChebBromwichInt} applied to the initial data $V_0(\sigma)=1, \dot{V}_0(\sigma)=0$, along the fixed Bromwich integration path $\Re(s)=\xi=1/10$ for different parameters $s_{\rm P}$ (cf.~(\ref{eq:Compact_chi})). Left panel: Chebyshev coefficients $c_j(0;\xi,s_{\rm P})$, cf.~(\ref{eq:ChebDec_ReV}) and their dependence on $s_{\rm P}$. Right panel: Corresponding dynamical error $\varepsilon(\tau, \sigma=0)$.
Results were obtained with the fixed Taylor resolutions $k_{\rm max} = 500$, $J_{\rm max} = 10$, cf. Sec.~\ref{sec:TaylorExpansions}.}
\label{fig:ChebBromwichInt_2}
\end{figure*}

The systematic error of this method is a consequence of the truncation of the Chebyshev series (\ref{eq:ChebDec_ReV}) at the resolution order $n_\chi$.
For $\tau\sim 0$, the accuracy is determined by the behaviour of the Chebyshev coefficients $c_j(\sigma;\xi, s_{\rm P})$, whereas the time propagation of this initial numerical error is related to the behaviour of the functions $Q_l(\tau;\xi, s_{\rm P})$. As a representative we consider the coordinate location $\sigma=0$ and study how the choice of the two parameters  $\xi>0$ and $s_{\rm P}\le 0$ affects the numerical performance.

Fig.~\ref{fig:ChebBromwichInt} compares the dependency of $c_j(0;\xi, s_{\rm P})$ on the parameter $\xi$ (left panel) for a fixed value $s_{\rm P}=0$. Note that the coefficients  fall off stronger and stronger as $\xi$ increases. However, the right panel of this figure reveals that the {\em dynamical} error grows in the course of time exponentially as $e^{\xi \tau}$ \footnote{The dynamical error is defined as
$
\varepsilon(\tau, \sigma) = |V_{\rm Bromwich}(\tau, \sigma) - V_{\rm HighAcc}(\tau, \sigma)|,
$
where $V_{\rm Bromwich}(\tau, \sigma)$ is  the solution obtained via the Bromwich integral method  and $V_{\rm HighAcc}(\tau, \sigma)$ being a highly accurate numerical solution obtained with a time-marching scheme based on a fully spectral code~\cite{Macedo2014}.}.

For $\xi \sim 0$, we can improve the efficiency of the method by shifting the parameter $s_{\rm P}$ accordingly. Fixing now $\xi = 1/10$, the left panel of fig.~\ref{fig:ChebBromwichInt_2} depicts $c_j(0;\xi, s_{\rm P})$ for different values of $s_{\rm P}$. Note that an appropriate choice of $s_{\rm P}$ significantly enhances the decay rate of $c_j(0;\xi, s_{\rm P})$. As in figure~\ref{fig:ChebBromwichInt}, the right panel of fig.~\ref{fig:ChebBromwichInt_2} displays the corresponding dynamical errors, and it becomes apparent that a stable solution can be obtained for a much longer time period. In the long run, however, the time behaviour of the numerical Bromwich solution is dominated by the decay rate $\sim e^{s_{\rm P}\tau}\,\tau^{l}$ of the integral terms $Q_l(\tau; \xi, s_{\rm P})$, cf.~(\ref{eq:IntegralTerms}) and fig.~\ref{fig:BromwitchIntegral}. Consequently, for very late times, the dynamical error corresponds to the inverse power law behaviour of the tail decay.

We conclude that in contrast to the spectral decomposition (\ref{eq:wavefield_2}), the Bromwich integral method, as realized in this work, is not suitable for describing the wave field's long term tail behaviour. It does, however, provide a neat test and justification of the Taylor coefficient techniques which present the core of this work.

\subsection{Examples for relevant Taylor asymptotics}\label{App:Examples_Taylor_Asymptotics}
In this section we provide an explicit example that resembles the properties of a function $f$,
\beq\label{eq:Taylor_f}
	f=\sum_{k=0}^\infty H_k (1-\sigma)^k,
\eeq
whose Taylor coefficients posses an asymptotics of the kind given in \eqref{eq:Hk_Asymptotics} and \eqref{eq:kappa_zeta}. The explicit example is given by the function
\beq
	\label{eq:example_f}
	f=e^{s/\sigma}.
\eeq
It can easily be verified that $f$ satisfies the ordinary differential equation 
\beq
	\label{eq:eqn_f}
	[\sigma^2\partial_\sigma+s]\,f=0.
\eeq
Inserting the Taylor expansion (\ref{eq:Taylor_f}) into (\ref{eq:eqn_f}) yields the recurrence relation
\beq
	\label{eq:recurrence_uk}
	(k+1)H_{k+1}-(2k+s)H_k+(k-1)H_{k-1}=0.
\eeq
Treating this recurrence relation in exactly the same manner as described in Sec.~\ref{sec:Homogeneous_Tayl_asymptotics}, we obtain the following asymptotics of the Taylor coefficients:
\beq
	\label{eq:asympt_uk}
	H_k\sim A_\infty^\pm\, \underbrace{k^{-3/4}e^{\pm 2\sqrt{sk}} Y\left(\pm\frac{1}{\sqrt{k}}\right)}_{\displaystyle =:a^\pm_k},\quad k\to \infty
\eeq
with
\bea
	Y(x)&=&1+\sum_{j=1}^\infty y_j x^j,\\[2mm]
	y_1 &=& \frac{4 s^2-9}{48 \sqrt{s}} \\
	y_2 &=& \frac{s^3}{288}-\frac{5 s}{64}-\frac{15}{512 s}\qquad \ldots\quad\mbox{etc.}
\eea
Clearly, for $s\notin\mathbb{R}^-$, the term in (\ref{eq:asympt_uk}) with the positive sign (meaning the square root in the exponent with positive real part) is dominant in comparison to the term with the negative sign. Hence we have:
\beq\label{eq:Uinf_limit}
	 A_\infty^+ = \lim_{k\to\infty}\left(H_k k^{3/4}e^{-2\sqrt{sk}}\right), \quad s\notin\mathbb{R}^-.
\eeq	 
If we compute the coefficients $H_k$ with the help of (\ref{eq:recurrence_uk}), utilized here as upwards recurrence relation with $H_{-1}=0$ and $H_0=e^s$, we can determine the limit (\ref{eq:Uinf_limit}) numerically. Examples are provided in (\ref{eq:Table_Ainf}).

For $s\in\mathbb{R}^-$, however, both asymptotics in (\ref{eq:asympt_uk}) need to be considered. Then the two unknowns $A_\infty^\pm$ can be obtained through the conditions 
\beq\label{eq:a_1a_2} H_1=s\, e^s, \quad H_2=\frac{s}{2} (2 + s)\,e^s, \eeq
which result from (\ref{eq:recurrence_uk}) for $k=1$ and $k=0$, taking into account that $H_{-1}=0$ and $H_0=e^s$. Indeed, if we write in accordance with (\ref{eq:asympt_uk})
\beq\label{eq:Hk_a}
	H_k=A_\infty^+ a^+_k + A_\infty^- a^-_k
\eeq
for two coefficients $H_{K+1}$ and $H_{K}$, where $K$ is some large number, and climb down to $H_2$ and $H_1$ utilizing (\ref{eq:recurrence_uk}) now as downwards recurrence relation, the conditions (\ref{eq:a_1a_2})  determine uniquely the two unknowns $A_\infty^\pm$. We generally find that $A_\infty^+=(A_\infty^-)^*\propto e^{\pi i/4}$ for $s\in\mathbb{R}^-$, see   examples in (\ref{eq:Table_Ainf}).

\beq\label{eq:Table_Ainf}
	\begin{array}{cccccccccc}
		s &=& 1            &:& \qquad & A_\infty^+ & \approx & 0.46509 & & \\
		s &=& e^{\pi i/4}  &:& \qquad & A_\infty^+ & \approx & 0.34251 &+& 0.20995\, i \\
		s &=& i            &:& \qquad & A_\infty^+ & \approx & 0.17696 &+& 0.21969\, i \\
		s &=& e^{3\pi i/4} &:& \qquad & A_\infty^+ & \approx & 0.11641 &+& 0.16027\, i \\[3mm]
		s &=& -\frac{1}{2} &:& \qquad & A_\infty^+ & \approx & 0.13063 &+& 0.13063\, i\\
		s &=& -1          &:& \qquad & A_\infty^+ & \approx & 0.12099 &+& 0.12099\, i\\
		s &=& -2          &:& \qquad & A_\infty^+ & \approx & 0.08727 &+& 0.08727\, i\\
		s &=& -10         &:& \qquad & A_\infty^+ & \approx & 0.00239 &+& 0.00239\, i
	\end{array}
\eeq

The preceding detailed study provides us with an understanding of regularity properties of solutions $\phi(s)$ 
to the homogeneous equation
\beq\label{eq:A_times_phi_s_2}
	{\mathbf{A}}(s)\phi(s)=0,\qquad\mbox{$\phi(s)$ analytic at $\sigma=1$,}
\eeq
when $s$ is not a quasinormal mode, $s\notin\{s_n\}$. Just like $f=e^{s/\sigma}$, the  functions $\phi(s)$ diverge at $\sigma=0$ if $\Re(s)>0$. However, they are $C^\infty$ for all $\sigma\in[0,1]$ if $\Re(s)<0$, despite the fact that the corresponding Taylor coefficients' magnitudes $|H_k|$ grow faster in $k$ than any polynomial (for $s\notin\mathbb{R}^-$). It is interesting to note that apparently the relation between the growth rate of $|H_k|$ and the frequency in the coefficients' oscillations determines whether the function is $C^\infty$ or diverges at $\sigma=0$:
\bea\nn
\phi&=&\sum_{k=0}^\infty H_k(1-\sigma)^k,\qquad H_k\sim e^{(g+i\omega)\sqrt{k}}\mbox{\, as $k\to\infty$:} \\[3mm]
		&& \label{eq:phi_conv_div} \mbox{$g<|\omega|:$\, $\phi$ is $C^\infty$,\quad $g>|\omega|:$\, $\phi$ diverges.} 
		\eea
As depicted in Sec.~\ref{sec:Homogeneous_Tayl_asymptotics}, this has the consequence that for $\Re(s)<0$ there are 
$C^\infty$-initial data \eqref{eq:Example_ID} which imply the regular $C^\infty$-solution \eqref{eq:Example_V} to our wave equation (\ref{eq:ReqTeukEq}). The situation is very similar to the Minkowskian wave obtained for the initial data No.~2 in (\ref{eq:V0_eta_examples}). Note that for this example the spectral representation (\ref{eq:sol_wave_Mink_2}) does not hold, and hence we presume that likewise (\ref{eq:wavefield_2}) is not applicable for the initial data \eqref{eq:Example_ID}, see discussion in section \ref{Sec:Conclusion}. The particular case $s\in\mathbb{R}^-$ provides us with a solution with an exponential fall-off without oscillations, to be compared with the Minkowskian wave associated with the initial data No.~1 in (\ref{eq:V0_eta_examples}). Just like there, the situation can be described marginally with the formula  (\ref{eq:wavefield_2}) if we accept Dirac-delta-type amplitudes $\eta(s)$.

Let us now take a closer look at the function $\Phi$ introduced in \eqref{eq:Phi} which satisfies the ODE
\bea\nn
0&=&\sigma ^2(1-\sigma) \Phi''
+[(\lambda -1) \sigma  (3 \sigma -2)-s] \Phi ' \\ &&
-\left[(\lambda -1) (2 \lambda  \sigma -\lambda -\sigma )+\ell(\ell+1)\right]\Phi.\label{eq:ODE_Phi}
\eea
Writing $\Phi$ as $f$ in \eqref{eq:Taylor_f} in terms of a Taylor expansion about $\sigma=1$, we find the recurrence relation
\beq\label{eq:RecRel_Phi}
\alpha_k H_{k+1} + \tilde\beta_k H_{k} + \tilde\gamma_k H_{k-1}=0 
\eeq
with
\bea
\tilde{\beta}_k&=&-\lambda ^2+4 k \lambda +2 \lambda -2 k (k+1)-\ell (\ell+1)-1 \nn \\
\tilde\gamma_k &=&(k-2 \lambda ) (k-\lambda ).\label{eq:coeff}
\eea
The analysis of the relation \eqref{eq:RecRel_Phi} along the lines presented in Sec.~\ref{sec:Homogeneous_Tayl_asymptotics} reveals the asymptotics
\beq\label{eq:asympt_Hk_Phi}
	H_k=U_\infty^+ u^+_k + U_\infty^- u^-_k
\eeq
with
\beq
	u_k^\pm \sim k^{-\frac{1}{4}(4\lambda+2s-3)}e^{\pm 2\sqrt{-sk}} Z\left(\pm\frac{1}{\sqrt{k}}\right),\quad (k\to \infty),
\eeq
which by virtue of \eqref{eq:phi_conv_div} describes for $\Re(s)>0$ a function $\Phi$ that is $C^\infty$ for all $\sigma\in[0,1]$ but blows up exponentially at $\sigma=0$ for $\Re(s)<0$.

We conclude this section by providing an example resembling the properties of a function $f$  with Taylor coefficients $H_k$ which can be written as in  \eqref{eq:Hk_a} with $A_\infty^+=0$, i.e.~with an asymptotics similar to the one described in \eqref{eq:RecRel_psi}. This example is meant to demonstrate the analytical properties of the solutions $\phi_n$ to the homogeneous problem \eqref{eq:HomogLaplaceTransfTeukEq_phi} at the QNMs $s_n\notin\mathbb{Z}^-$.

Looking for some $j\in\mathbb{N}$ at the inhomogeneous equation
\beq\label{eq:A_times_phi_s_3}
	[\sigma^2\partial_\sigma+s]\,f=\sigma^j ,
\eeq
where $s\in\mathbb{C}\backslash\mathbb{R}^-_0$, we find the solution
\[f(\sigma)=\sigma^{j-1} e^{s/\sigma}E_j\left(\frac{s}{\sigma}\right),\]
which is analytic for $\sigma\in(0,1]$ and still $C^\infty$ at $\sigma=0$. Its Taylor coefficients obey the asymptotics	$H_k\sim A_\infty^- a^-_k $ as $k\to \infty$ (see (\ref{eq:asympt_uk})). Taking again this expression
for two coefficients $H_{K+1}$ and $H_{K}$, where $K$ is some large number, and climbing down to $H_j$ utilizing (\ref{eq:recurrence_uk}) once more as downwards recurrence relation\footnote{With the inhomogeneity $\sigma^j$ in (\ref{eq:A_times_phi_s_3}), the recurrence relation (\ref{eq:recurrence_uk}) remains the same for $k>j$.}, 
the condition $H_j=(-1)^jf^{(j)}(\sigma=1)/j!$ implies the numerical value of $A_\infty^-$. Exemplary cases for $j=1$ are provided in (\ref{eq:Table_Ainf_minus}).
\beq\label{eq:Table_Ainf_minus}
	\begin{array}{cccccccccc}
		s &=& 1            &:& \qquad & A_\infty^- & \approx & -2.92228 & & \\
		s &=& e^{\pi i/4}  &:& \qquad & A_\infty^- & \approx & -2.15206 &-&  1.31915\, i \\
		s &=& i            &:& \qquad & A_\infty^- & \approx & -1.11188 &-&  1.38033\, i \\
		s &=& e^{3\pi i/4} &:& \qquad & A_\infty^- & \approx & -0.73143 &-&  1.00699\, i 
	\end{array}
\eeq

\subsection{Negative integer Laplace parameters}\label{App:Neg_Lapl_Par}
When the Laplace parameters $s$ is a negative integer then $\alpha_{-(s+1-\lambda)}$ as well as $\gamma_{-s}$ and $\gamma_{-(s+\lambda)}$ vanish. Consequently, neither the forward recurrence relation \eqref{eq:UpRecRel_phi} nor the backward recurrence relation \eqref{eq:DownRecRel_psi} can be completely carried out for the sequences $\{ H_k\}$ and $\{ I_k\}$, respectively. Yet, the calculation of the jump function for values in the neighbourhood of the negative integers indicates that the solution $\hat{V}(\sigma;s)$ is well behaved at $s \in \mathbb{Z}^-$.     

Here we modify the calculation of the Taylor coefficients $a_k$ (cf.~\eqref{eq:Series_hat_V}) for $s \in \mathbb{R}^-\backslash\mathbb{Z}^-$ and consider subsequently a smooth limit of $s$ towards the neighbouring negative integer.\footnote{Throughout this section, we assume that the negative real axis is approached from above. For simplicity, we omit the index $\bullet^{+}$ in all the relevant quantities.}. To this end, we start by defining the critical indices 
\bea
\tilde{k} &=&\left\lfloor -s-1  + \lambda \right\rceil, \nn \\
\hat{k} &=& \left\lfloor\min( -s, -s-\lambda ) \right\rceil, \label{eq:CricIndx}\\
 k^* &=&\left\lfloor \max( -s, -s-\lambda ) \right\rceil, \nn
\eea
where $\lfloor \bullet \rceil$ denotes the round function which provides the nearest integer. We further introduce the new coefficients $K_k$ and $J_k$ via
\bea
\label{eq:Kk}
K_k &=& \left\{ 
		  \begin{array}{ccc}
		     H_k & {\rm for} & k \leq \tilde{k} \\
		     \alpha_{\tilde k}\,H_k & {\rm for} & k > \tilde{k}
		   \end{array}
		\right., \\
\label{eq:Jk}
J_k &=& \left\{ 
		  \begin{array}{ccc}
		     \gamma_{\hat{k}}\,\gamma_{k^*}\,I_k & {\rm for} & k < \hat{k} \\
		     \gamma_{k^*}\,I_k & {\rm for} &  \hat{k} \leq k < k^*  \\
		     I_k & {\rm for} & k \ge k^*
		   \end{array}
		\right. .
\eea
Now, for $k<\tilde{k}$ and for $k>\tilde{k}+1$, the coefficients $K_k$ satisfy the same forward recurrence relation as $H_k$, cf.~\eqref{eq:UpRecRel_phi}. Besides, combining the definition \eqref{eq:Kk} with the recurrence relation \eqref{eq:UpRecRel_phi} for $k=\tilde{k}$ and $k=\tilde{k}+1$ one gets
\bea
\label{eq:Kp1}
K_{\tilde{k} + 1} &=& -\left( \beta_{\tilde{k}} K_{\tilde{k}} +  \gamma_{\tilde{k}} K_{\tilde{k}-1}\right), \\
\label{eq:Kp2}
K_{\tilde{k} + 2} &=& -\left( \beta_{\tilde{k}} K_{\tilde{k}+1} +  \alpha_{\tilde{k}}\,\gamma_{\tilde{k}} K_{\tilde{k}}\right)/\alpha_{\tilde{k}+1}.
\eea
In the limit $s\to\lfloor s \rceil\in \mathbb{Z}^-$, both $K_{\tilde{k} + 1}$ and $K_{\tilde{k} + 2}$ are well defined, despite the fact that  $\alpha_{\tilde{k}}\to 0$. This can be followed from the relations \eqref{eq:UpRecRel_phi}, \eqref{eq:Kp1} and \eqref{eq:Kp2}.

The analysis of the coefficients $J_k$ follows a similar route. They satisfy the same backward recurrence relation \eqref{eq:DownRecRel_psi} for the coefficients $I_k$, except for the values $k=k^*$, $k=k^*-1$, $k=\hat{k}$ and $k=\hat{k}-1$. Combining the definition \eqref{eq:Jk} with the backward recurrence relation \eqref{eq:DownRecRel_psi} for $k=k^*$ and $k=k^*-1$, we obtain
\bea
J_{k^*-1} &=& -\left( \alpha_{k^*} J_{k*+1} + \beta_{k^*}J_{k^*}\right),  \label{eq:Jktilde-1}\\
J_{k^*-2} &=& -\left( \gamma_{k^*}\,\alpha_{k^*-1} J_{k*} + \beta_{k^*-1}J_{k^*-1}\right)/\gamma_{k^*-1}. \label{eq:Jktilde-2} \qquad
\eea 
Likewise, if we replace here $k^*$ by $\hat k$, then we obtain expressions which provide $J_{\hat k-1}$ and $J_{\hat k-1}$.
\footnote{Care must be taken if $|\lambda|=1$, for then $k^*-1=\hat{k}$ and $k^*-2=\hat{k}-1$.}. Again, eqs.~\eqref{eq:Jktilde-1} and \eqref{eq:Jktilde-2} (together with the corresponding case $k^*$ replaced by $\hat{k}$) are well defined in the limit $s\to\lfloor s \rceil\in \mathbb{Z}^-$ where $\gamma_{k^*}$ and $\gamma_{\hat{k}}$ tend to zero.

We now turn our attention to the computation of the coefficients $a_k$. Given $a_0$ we can invoke the forward recurrence relation \eqref{eq:RecRel} to compute $a_k$ for $1\le k\le \tilde k$. However, since $\alpha_{\tilde k}\to 0$ for $s\to\lfloor s \rceil$, the forward recurrence calculation of $a_{\tilde k+1}$ breaks down in this limit. Then for larger indices $k\ge \tilde k+2$, the forward recurrence relation \eqref{eq:RecRel} can again be used. That is to say that for the computation of the entirety $\{a_k\}$ in the limit $s\to\lfloor s \rceil$ via the forward recurrence relation it is necessary to provide the two coefficients $a_0$ and $a_{\tilde k+1}$. Indeed, they can be obtained explicitly by rewriting eq.~\eqref{eq:sol_ak} with the help of \eqref{eq:Kk} and \eqref{eq:Jk}. In terms of the abbreviations
\bea
\tilde\Pi^{(\alpha)}_j&=&\prod\limits_{m=0}^{j-1}\alpha_m \label{eq:tilde_Pi_alpha} \\
      \Pi^{(\alpha)}_j&=&\prod\limits_{\substack{m=0 \\[0.5mm] \{ m\neq \tilde{k}\}}}^{j-1}\alpha_m  \label{eq:Pi_alpha} \\
      \Pi^{(\gamma)}_j&=&\prod\limits_{\substack{m=0 \\[0.5mm] \{ m\neq k^*\} \\ \{m\neq \hat{k} \} }}^{j}\gamma_m\label{eq:Pi_gamma}
\eea
we find:
\beq
\label{eq:a0_NegInt}
a_0 = -\frac{1}{J_{-1}} \sum_{j=0}^{\infty} \frac{J_j B_j \tilde\Pi^{(\alpha)}_j}{\Pi^{(\gamma)}_j}
\eeq
and
\bea
\label{eq:a_ktilde+1_NegInt}
a_{\tilde{k}+1} &=& -\frac{1}{J_{-1}}\left( K_{\tilde{k}+1}\sum_{j=\tilde{k}+1}^{\infty} \frac{J_j B_j\Pi^{(\alpha)}_j}{ \Pi^{(\gamma)}_j}\right. \nn \\
&&\hspace*{1cm} + \left. J_{\tilde{k}+1}\sum_{j=0}^{\tilde{k}} \frac{\Theta_j K_j B_j \Pi^{(\alpha)}_j}{ \Pi^{(\gamma)}_j} \right).
\eea
The factors $\Theta_j$ assume different values, depending on $\lambda$:
\bea
\lambda <0:\,  \Theta_j &=& 1,  \nn  \\
\lambda = 0:\, \Theta_j &=& \gamma_{k^*} \nn  \\
\lambda >0:\, \Theta_j &=&
\left\{ 
	\begin{array}{ccc}
		\gamma_{k^*}\gamma_{\hat{k}} & \rm{for} & j<\hat{k} \\
		\gamma_{k^*} & \rm{for} & \hat{k} \leq j <k^* \\
		1                     & \rm{for} & k^* \leq j \leq \tilde{k}
	\end{array}
\right. \nn
\eea

Note that in these expression no divisions by  $\gamma_{k^*}$ and $\gamma_{\hat{k}}$ occur and hence both coefficients $a_0$ as well as $a_{\tilde{k}+1}$ assume well defined values in the limit $s\to\lfloor s \rceil$. We finish this section by noticing that for $|\lambda|=2$ there are specific $s$-values for which this construction fails since then $J_{-1}$ vanishes. Similar to the discussion in sec.~\ref{Sec:QNMs_from_discr_Wronskian} (see in particular eq.~\eqref{eq:QNM_I_m1_=0}) we conclude that these values have to be considered as QNMs, to be treated in Sec.~\ref{Sec:AlgSpecial}.

\subsection{Algebraically special QNMs}\label{Sec:AlgSpecial}
\subsubsection{Polynomial solutions to the homogeneous Laplace transformed equation}\label{Sec:Polynomial_Solutions}
For gravitational perturbations $|\lambda|=2$ specific QNMs arise, the so-called algebraically special $s_{(\ell)}$-values which are negative integers. As for each $s\in\mathbb{Z}^-$, the corresponding coefficients $H_k$ and $I_k$ (introduced in Sec.~\ref{sec:Homogeneous_Tayl_asymptotics}) cannot be defined for all $k$ (see Sec.~\ref{App:Neg_Lapl_Par}; especially, there is no sensible value $I_{-1}$), the characterization of QNMs in terms of the points 1.~and 2.~in Sec.~\ref{Sec:QNMs_from_discr_Wronskian} fails. However, it turns out that for $s=s_{(\ell)}$ a {\em polynomial} solution $\phi_{(\ell)}$ to the homogeneous problem (\ref{eq:HomogLaplaceTransfTeukEq_phi}) can be found. This means that the characterization of QNMs in terms of the point 3.~in Sec.~\ref{Sec:QNMs_from_discr_Wronskian} still holds and should therefore be regarded as generically valid definition of QNMs for perturbations in the asymptotically flat Schwarzschild spacetime.

The solution $\phi_{(\ell)}$ emerges through the following considerations. Let us write  $\phi_{(\ell)}$ as in \eqref{eq:Series_phi} as Taylor expansion about $\sigma=1$ with coefficients $H_k$ that are {\em not} subject to the scaling condition in \eqref{eq:UpRecRel_phi} but are normalized in the sequel through a different requirement. We look at the corresponding recurrence relation (\ref{eq:RecRel_phi}), denoted for the indexes $k\in\{j-1, j, j+1, j+2\}$:
\bea
	\alpha_{j-1} H_j     + \beta_{j-1} H_{j-1} + \gamma_{j-1} H_{j-2} &=& 0 \label{jm1}\\
	\alpha_{j}   H_{j+1} + \beta_{j}   H_{j}   + \gamma_{j}   H_{j-1} &=& 0 \label{j}\\
	\alpha_{j+1} H_{j+2} + \beta_{j+1} H_{j+1} + \gamma_{j+1} H_{j}   &=& 0 \label{jp1}\\
	\alpha_{j+2} H_{j+3} + \beta_{j+2} H_{j+2} + \gamma_{j+2} H_{j+1} &=& 0 \label{jp2}
\eea
It can be verified explicitly that for \footnote{In~\cite{Chandrasekhar:1984} the algebraically special Laplace parameters are given by $\bar s_{(\ell)}=\frac{1}{2}s_{(\ell)}$, cf.~\eqref{eq:s_bar}.}
\[ |\lambda|=2,\quad s=s_{(\ell)}=-\frac{1}{3}(\ell-1)\ell(\ell+1)(\ell+2),\quad j=-s-2\]
the following properties arise:
\ben
	\item The coefficient $\gamma_{-s}$ vanishes, $\gamma_{-s}=\gamma_{j+2}=0$.
	\item The $2\times 2$-matrix
	\[\hat M=\left(\begin{array}{cc}\beta_j & \alpha_j \\ \gamma_{j+1} & \beta_{j+1} \end{array}\right)\]
	is singular and possesses the nontrivial null eigenvector
	\[\vec v =\left(\begin{array}{c}\ell^2+\ell-3  \\ 1-\lambda \end{array}\right)\]
	satisfying $\hat M\vec v =0$.
	\item Additionally, we have for
		\ben 
			\item $\lambda=-2$ \quad that \quad $\alpha_{j-1}=0$.
			\item $\lambda=+2$ \quad that \quad $\gamma_{j}=0$.
		\een
\een
Now, we may write (\ref{j}), (\ref{jp1}) in the form
\beq\label{eq:matrix_M}
	\hat M \left(\begin{array}{c} H_j  \\ H_{j+1} \end{array}\right) 
		= - \left(\begin{array}{c} \gamma_j H_{j-1}  \\ \alpha_{j+1} H_{j+2} \end{array}\right)
\eeq
The corresponding algebraically special solutions $\phi_{(\ell)}$ are characterized by the following properties:
\ben
	\item[(A)] $H_k=0$ for $k\ge j+2$.
	\item[(B)] $H_k=0$ for $k<0$.\\
	With statement (A) we conclude that $\phi_{(\ell)}$ is a polynomial of order $\frac{1}{3}(\ell-1)\ell(\ell+1)(\ell+2)-1$.
	\item[(C)] The coefficients $(H_j, H_{j+1})$ form a null eigenvector of $\hat M$, i.e.~
		\beq\label{eq:pk_nulleigvector}\left(\begin{array}{c} H_j  \\ H_{j+1} \end{array}\right) 
		= K \,\vec v=K \left(\begin{array}{c}\ell^2+\ell-3  \\ 1-\lambda \end{array}\right)  \eeq
		with some constant $K$.
\een
The property (\ref{eq:pk_nulleigvector}) means, that (\ref{eq:matrix_M}) is realized for vanishing right hand side, i.e.~we have
\bea	
	 \label{eq:eig1=0}\gamma_j H_{j-1}&=&0  \\
	 \label{eq:eig2=0}\alpha_{j+1} H_{j+2}&=&0
\eea
Let us first look at (\ref{eq:eig2=0}). As 
$\alpha_{j+1}=s_{(\ell)}\lambda\ne 0$
we have that $H_{j+2}=0$. Looking now at (\ref{jp2}) it follows with $\gamma_{j+2}=0$ (see point 1.~above) and $\alpha_{j+2}=(\lambda-1)(s-1)\ne 0$ that $H_{j+3}=0$. With the two vanishing coefficients $H_{j+2}=0=H_{j+3}$, the upwards recurrence relation
\[ H_{k+1}=-\frac{1}{\alpha_{k}}\left(\beta_{k}   H_{k}   + \gamma_{k}   H_{k-1}\right),\qquad k\ge j+3 \]
tells us that all $H_k=0$ for $k\ge j+2$, i.e.~we obtain consistency with the above statement (A). 

The first property (\ref{eq:eig1=0}) is realized differently for the two cases $\lambda=-2$ and $\lambda=2$. Let us start with the discussion of the situation $\lambda=-2$. As $\gamma_j=4-2\lambda=8\ne 0$, we have $H_{j-1}=0$. With the property 3.(a) above and $\gamma_{j-1}=9-3\lambda=15\ne 0$, we obtain from (\ref{jm1}) that $H_{j-2}=0$. Now, with $H_{j-1}=0=H_{j-2}$, the downwards  recurrence relation 
\[ H_{k-1}=-\frac{1}{\gamma_{k}}\left(\alpha_kH_{k+1} + \beta_{k}   H_{k}\right),\qquad k\le j-2\]
 tells us that all $H_k=0$ for $k\le j-1$, thus realizing in particular the above statement (B). Taking \eqref{eq:pk_nulleigvector} into account, the algebraically special polynomial solution amounts thus to
\[\phi_{(\ell)}=K(1-\sigma)^j [(\ell^2+\ell-3)+3(1-\sigma)]\]
for 
\[ \lambda=-2,\quad s=s_{(\ell)},\quad j=-s_{(\ell)}-2. \]
For the remaining case $\lambda=2$ we have $\gamma_j=0$ (cf.~point 3.(b)) and hence (\ref{eq:eig1=0}) is realized trivially. Let us now look at the recurrence relation for the indexes $k=0,\ldots, j-1$, where we assume that $H_{-1}=0$ in order to satisfy the above statement (B), i.e.~
\beq\label{eq:AlgSp_linSys}
\begin{array}{ccccccc}
	\alpha_0     H_1    &   + & \beta_0 H_0    &       &                     &=& 0 \\
	\alpha_{1}   H_{2}  &   + &\beta_{1}    H_1&     + &\gamma_1   H_0       &=& 0 \\
	                    &     &                &       &                     &\vdots &  \\
	\alpha_{j-2} H_{j-1}&   + &\beta_{j-2} H_{j-2}& +  &\gamma_{j-2} H_{j-3} &=& 0 \\
	                    &     &\beta_{j-1} H_{j-1}& +  &\gamma_{j-1} H_{j-2} &=& -\alpha_{j-1} H_j\\[2mm]
	                    &     &                &       &                     && \hspace*{-2cm} 
	                    =-K \alpha_{j-1} (\ell^2+\ell-3).   
\end{array}	                    
\eeq
Here we have rewritten the very last relation for $k=j-1$ such that the quantity $H_j$ known from (\ref{eq:pk_nulleigvector})
appears as inhomogeneity on the right hand side. Now, (\ref{eq:AlgSp_linSys}) forms a system of $j$ linear equations to determine uniquely the coefficients $\{H_0,\ldots,H_{j-1}\}$, thereby realizing $H_{-1}=0$ through which the solution becomes analytic at $\sigma=1$. We thus obtain for 
\[ \lambda=2,\quad s=s_{(\ell)},\quad j=-s_{(\ell)}-2 \]
the algebraically special polynomial solution as
\[\phi_{(\ell)}=\sum_{k=0}^{j+1} H_k(1-\sigma)^k\]
with the coefficients $\{H_0,\ldots,H_{j-1}\}$ fixed by (\ref{eq:AlgSp_linSys}) and $\{H_j,H_{j+1}\}$ given through (\ref{eq:pk_nulleigvector}). In the following we use polynomial solutions $\phi_{(\ell)}$ which are normalized by the requirement $K=1$. Note that the scaling in \eqref{eq:UpRecRel_phi} would not work in the case $\lambda=-2$.

\begin{figure}[t!]
\begin{center}
\includegraphics[width=7.0cm]{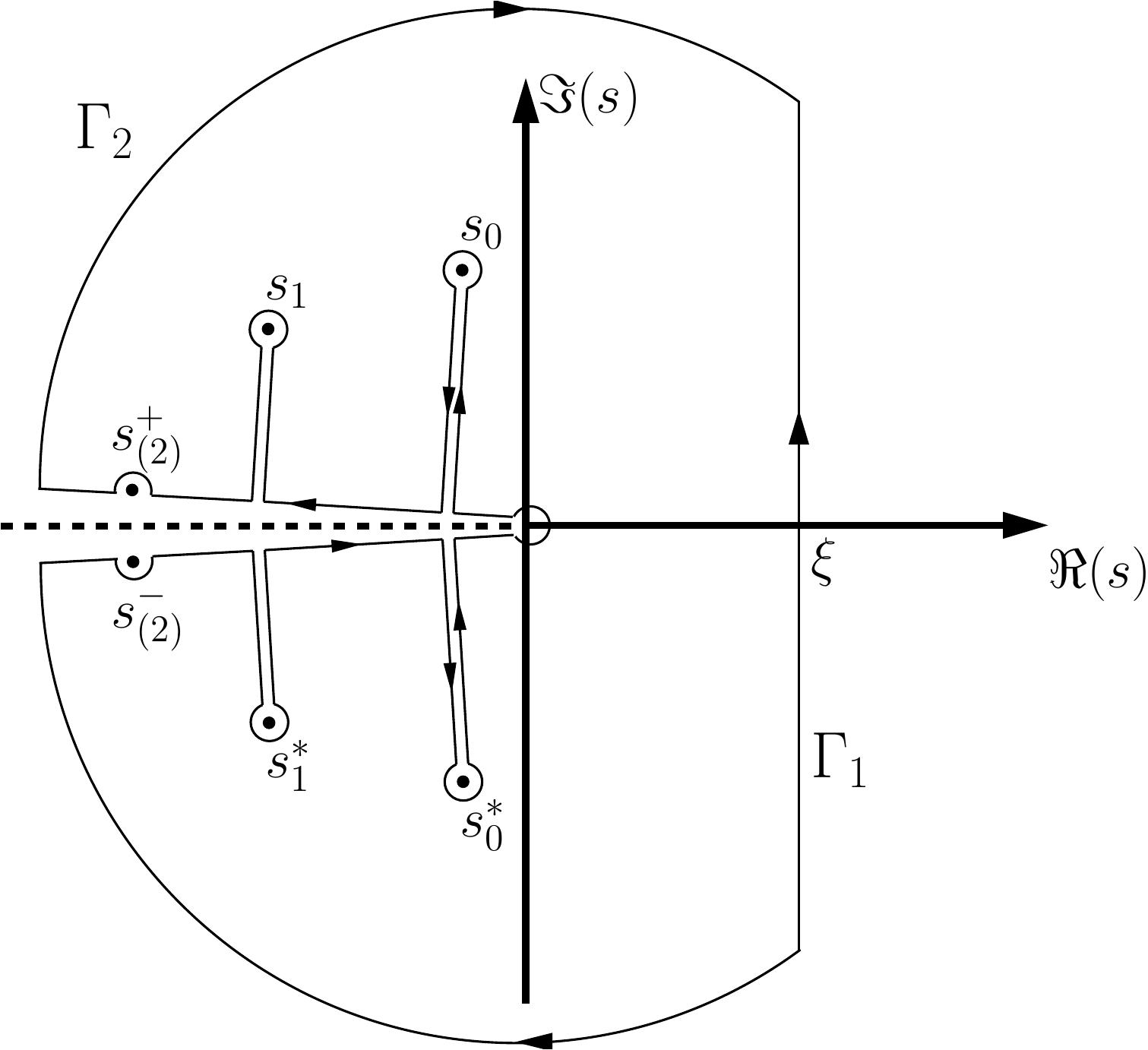}
\end{center}
\caption{Deformed integration path $\Gamma_2$ for the inverse Laplace transformation in the case $|\lambda|=2$ where algebraically special QNMs $s_{(\ell)}^\pm\in\mathbb{R}^-$ are present. The Laplace transform $\hat V$ is defined on multiple sheets where the negative $s$-axis presents the transition between the sheets. At $s=s_{(\ell)}^\pm$, $\hat V$ possesses single poles with residues $\eta^\pm_{(\ell)}\phi_{(\ell)}$, and the integration is to be performed along corresponding infinitesimal semicircles about $s_{(\ell)}^\pm$.}
\label{fig:BromwichInt_AlgSpc}
\end{figure}

\subsubsection{Algebraically special QNM amplitudes}\label{Sec:AlgSpecial_amplitudes}

A characterization that works well for the entirety of QNMs is given by the condition $J_{-1}=0$ with $J_k$ introduced in \eqref{eq:Jk} (see also \eqref{eq:RecRel_psi}). Clearly, then the computation of $a_0$ and $a_{\tilde k+1}$ according to \eqref{eq:a0_NegInt} and \eqref{eq:a_ktilde+1_NegInt} will fail, which leads us to the conclusion that in the vicinity of the algebraically special QNM $s_{(\ell)}$ the Laplace transform $\hat V$ is of the form \eqref{eq:Vhat_QNM}, i.e.~it possesses a single pole there. As $s_{(\ell)}$ is located on the branch cut, we need to consider separately the approach $s\to s_{(\ell)}$ from above ($\arg(s)\to \pi$) and from below ($\arg(s)\to-\pi$), i.e. $s\to s_{(\ell)}^\pm$. Alternatively, $\hat V$ can be regarded as being defined on multiple sheets in the complex $s$-plane where the negative $s$-axis presents the transition between the sheets. At the two $s$-values $s=s_{(\ell)}^\pm$ the Laplace transform $\hat V$ possesses single poles with residues $\eta^\pm_{(\ell)}\phi_{(\ell)}$, which by virtue of (\ref{eq:Symmetry_hat_V}) satisfy $\eta^-_{\rm (\ell)}=[\eta^+_{\rm (\ell)}]^*$. Considering the representation (\ref{eq:BromInt}) of the wave field $V$ which we now evaluate along the deformed integration path $\Gamma_2$ displayed in fig.~\ref{fig:BromwichInt_AlgSpc}, we find that in the limit of infinitesimal semicircles about the algebraically special QNMs $s_{(\ell)}^\pm$ the contributions 
\[\frac{1}{2\pi i}\cdot\pi i \,{\rm Res}_{s^\pm_{(\ell)}}\left( \hat V e^{\tau s_{(\ell)}}\right),\]
arise, i.e.~in sum:
\[\frac{1}{2}\left(\eta^-_{\rm (\ell)}\phi_{(\ell)} e^{\tau s_{(\ell)}} + 
\eta^+_{\rm (\ell)}\phi_{(\ell)} e^{\tau s_{(\ell)}}\right)=\Re(\eta^+_{\rm (\ell)})\phi_{(\ell)} e^{\tau s_{(\ell)}}.
	\]
	Thus, the corresponding spectral decomposition (see Sec.~\ref{sec:Spec_Decomp}) of the wave field satisfying the dissipative wave equation (\ref{eq:ReqTeukEq}) assumes the form:
	\bea\nn\label{eq:wavefield_3}
V(\tau, \sigma) &=&\nn 2\sum_{n=0}^{\infty} \Re\big(\eta_{n} \phi_n(\sigma) e^{\tau s_{n}}\big) 
+ \Re\big(\eta^+_{(\ell)}\big) \phi_{(\ell)}(\sigma) e^{\tau s_{(\ell)}}\nn \\ 
&&+ \int\limits_{-\infty}^{0} \eta(s)\phi(\sigma;s) e^{\tau s} ds,
\eea	
where the set $\{s_n\}_{n=0}^{\infty}$ contains only those quasinormal modes for which $\Im(s_n)>0$.

In order to determine the amplitude $\eta_{(\ell)}^+$, we proceed along the same line of reasoning as in sec.~\ref{Sec:QNM_amplitudes}. In particular, the analogue of eq.~\eqref{eq:coeff_g0} for negative integer $s$-values can be derived by means of eqn.~\eqref{eq:a0_NegInt}, with the source coefficient $B_k$ being replaced by $(B_k - \eta_{(\ell)}^+C_k)$ where $C_k$ is given by \eqref{eq:coeffs_Ck} (see also \eqref{eq:Operator_C}) with the Taylor coefficients $H_k$ of $\phi_{(\ell)}$. Requiring the regularity of the corresponding coefficient $g_0^+(s_{(\ell)})$, we obtain
\beq
\eta_{(\ell)}^+ = \frac{
											\sum\limits_{k=0}^{\infty} J_k^+ B_k \dfrac{ \tilde\Pi^{(\alpha)}_k }{\Pi^{(\gamma)}_k}
									}{
											\sum\limits_{k=0}^{\infty} J_k^+ C_k \dfrac{ \tilde\Pi^{(\alpha)}_k }{\Pi^{(\gamma)}_k}
									}
 \quad {\rm for} \quad \lambda = +2.
\eeq
We emphasise, however, that this procedure works only for $\lambda = +2$. For $\lambda= -2$, we observe that 
$\tilde\Pi^{(\alpha)}_k=0$ for $k>\lambda-s$ and $J^+_k=0$ for $k\le\lambda-s$. Hence, every addend of the sum in \eqref{eq:a0_NegInt} vanishes, regardless of the source coefficient $B_k$. This means that this equation cannot be used to establish the amplitude $\eta_{(\ell)}^+$. Instead, the corresponding formula \eqref{eq:a_ktilde+1_NegInt} can be employed. Requiring a regular coefficient $g^+_{\tilde k+1}(s_{(\ell)})$ we find:
\beq
\eta_{(\ell)}^+ = \frac{
K_{\tilde{k}+1}\sum\limits_{k=\tilde{k}+1}^{\infty} J_k^+ B_k\dfrac{ \Pi^{(\alpha)}_k }{\Pi^{(\gamma)}_k}+ J_{\tilde{k}+1}^+\sum\limits_{k=0}^{\tilde{k}}        K_k B_k\dfrac{ \Pi^{(\alpha)}_k }{\Pi^{(\gamma)}_k}
									}{
K_{\tilde{k}+1}\sum\limits_{k=\tilde{k}+1}^{\infty} J_k^+ C_k\dfrac{ \Pi^{(\alpha)}_k }{\Pi^{(\gamma)}_k}+ J_{\tilde{k}+1}^+\sum\limits_{k=0}^{\tilde{k}}        K_k C_k\dfrac{ \Pi^{(\alpha)}_k }{\Pi^{(\gamma)}_k}
									}
\eeq
for $\lambda=-2$.
\bibliographystyle{apsrev4-1-noeprint.bst}
\bibliography{bibitems}

\end{document}